# Subsurface Vacancy Engineering Enables Atomically Clean and Oxidation-Resistant Copper Interfaces for Anode-Free Lithium Metal Batteries

*Yue Li[1, 2]†, Xuanguang Ren[1, 3]†, Xueting Feng[3]†, Lingcheng Kong[1]†, Fengping Luo[2], Yang Xu[2], Liu Qian[1], Yusheng Ye[4]*, Ziqiang Zhao[1, 2]*, Xin Gao[1]*, Jin Zhang[1, 3]**

[1]School of Materials Science and Engineering, Peking University, Beijing, 100871, China.

[2]State Key Laboratory of Nuclear Physics and Technology, School of Physics, Peking University, Beijing, 100871, China.

[3]Beijing National Laboratory for Molecular Sciences, College of Chemistry and Molecular Engineering, Peking University, Beijing, 100871, China.

[4]Beijing Key Laboratory of Environmental Science and Engineering, School of Materials Science and Engineering, Beijing Institute of Technology, Beijing, 100081, China.

†These authors contributed equally to this work.

*Correspondence to: Y. Y., ysye@bit.edu.cn; Z. Z., zqzhao@pku.edu.cn; X. G., gaoxin-cnc@pku.edu.cn; J. Z., jinzhang@pku.edu.cn.

## Abstract

Interfaces govern reaction pathways and stability in electrochemical systems, yet creating clean, well-defined metal interfaces at scale remains challenging. In anode-free lithium metal batteries (AFLMBs), the current-collector interface is decisive for lithium nucleation and solid electrolyte interphase (SEI) formation, and ideally should support efficient charge transport, uniform reaction distribution, and long-term chemical and structural stability. Here we report an ion-implantation strategy that produces an atomically clean and oxidation-resistant copper interface. Implanting copper ions into commercial foils removes the native oxide while generating subsurface vacancy clusters directly beneath the surface—an atomic-scale modification that does not increase the collector thickness but fundamentally alters interfacial chemistry. Experiments and multiscale simulations reveal that these vacancies act as strong oxygen traps, preventing re-oxidation, enhancing interfacial conductivity, and guiding the formation of an ultrathin, $Li_2O$-enriched SEI that promotes uniform lithium deposition and suppresses parasitic reactions. Applied in AFLMBs, the engineered current collectors deliver long-term stability with a Coulombic efficiency of 98.8% over 600 cycles under lean-electrolyte conditions. These findings demonstrate atomic-scale interface control of copper current collectors as a route toward stable and practical lithium metal batteries.

Interfaces play a critical role in determining the performance and longevity of electrochemical systems, from catalysis to energy storage. At these boundaries, charge transfer occurs, reaction selectivity is defined, and degradation process initiate.[1] Achieving an ideal interface, characterized by efficient charge transport, uniform reaction distribution, and long-term chemical and structural stability, is essential for advancing next-generation electrochemical technologies. To this end, interface engineering strategies that modulate surface wettability, nucleation dynamics, and interfacial kinetics have been widely explored.[2,3]

Anode-free lithium metal batteries (AFLMBs) have recently emerged as a compelling architecture for high-energy-density storage. By dispensing with a conventional anode and plating lithium directly onto a bare copper current collector (CuCC) during initial charging, AFLMBs offer simplified construction, reduced cost, and high gravimetric energy densities.[4–6] However, the absence of a host anode renders lithium plating highly sensitive to interfacial conditions at the CuCC, where dendritic growth, unstable solid electrolyte interphase (SEI) formation, and irreversible lithium loss often lead to rapid performance degradation and poor cycle life.[7–11]

While considerable efforts have been devoted to stabilizing this interface via electrolyte optimization,[12] three-dimensional host architectures,[13] artificial SEI layers,[14] interfacial defect engineering,[15] and surface modifications.[16] These strategies typically address interfacial manifestations rather than their fundamental origin. In particular, commercial Cu foils[17,18] develop a native oxide layer upon air exposure, which becomes the effective lithium plating interface.[19] This amorphous oxide is electronically insulating and chemically unstable, promoting heterogeneous nucleation and uncontrolled SEI evolution. Although treatments such as acid etching or thermal reduction can partially remove the oxide,[20–22] rapid reoxidation under ambient conditions remains a persistent barrier. Coating-based approaches provide some protection but often compromise conductivity or scalability.

Here, we introduce a substrate-intrinsic atomic interface engineering (AIE) strategy that employs homo-elemental ion implantation to address the fundamental instabilities of anode-free lithium metal batteries. By implanting Cu ions into commercial Cu current collectors (CuCCs), this approach avoids extrinsic dopants, isolates the intrinsic energetic effects of implantation, and preserves the interfacial thickness, thereby providing an ideal platform for probing lithium–collector interactions. Ion implantation concurrently removes the native oxide and generates subsurface vacancy clusters (Figure 1a), which act as potent oxygen traps that suppress re-oxidation and endow the collector with >90-day ambient stability. Multiscale simulations and spectroscopy show that implantation drives substantial lattice reconfiguration, producing an atomically clean and chemically robust surface. Integrated into anode-free cells (cells without lithium metal at the initial state; Figure 1b), the engineered interface regulates lithium nucleation, promotes the preferential formation of a $Li_2O$-rich SEI, and mitigates electrolyte corrosion. This leads to an average of 98.8% Coulombic efficiency over 600 cycles under lean-electrolyte conditions. Our findings establish a scalable and robust route for engineering CuCC interfaces, resolving key degradation pathways in AFLMBs and offering a generalizable strategy for stable, high-performance anode-free architectures.

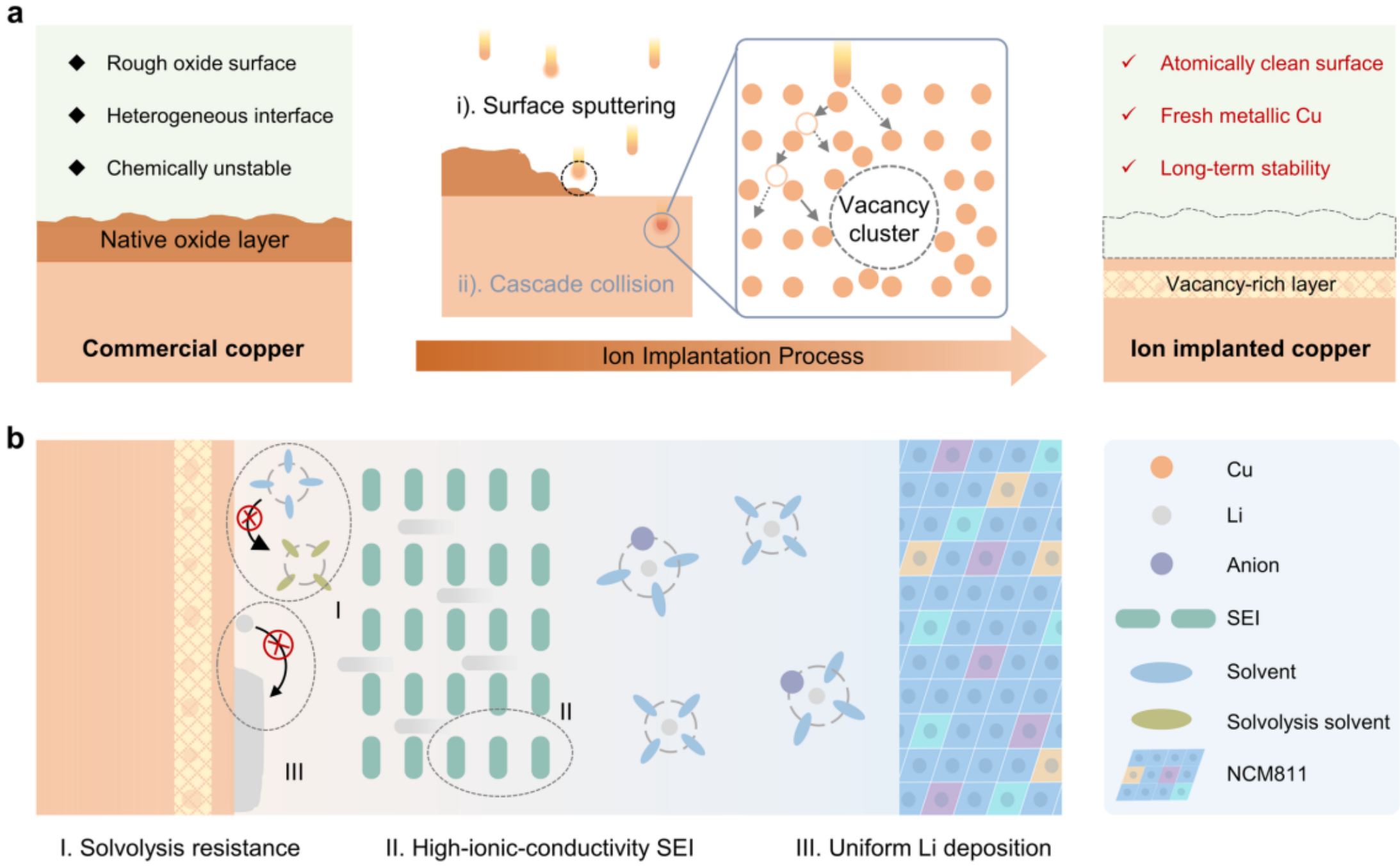

**Figure 1.** Schematic of the overall concept. **a** Formation mechanism of anti-oxidation Cu current collectors (CuCCs) with superior interfacial properties via the ion implantation process. **b** Key advantages of ion-implanted CuCCs for achieving high-performance AFLMBs.

## *Results and Discussion*

## Interfacial regulation of CuCCs via ion implantation

The CuCC plays a critical role in AFLMBs, as it directly governs electron transport and interfacial reactions with lithium, thereby profoundly influencing Li nucleation behavior, interfacial stability, and overall electrochemical performance. Here, we employ Cu ion implantation to treat commercial Cu foils for precise interfacial regulation, achieving controlled Li nucleation and enhanced interfacial stability. To the best of our knowledge, this work represents the first demonstration of homo-elemental ion-implanted CuCCs for high-performance battery applications, as schematically illustrated in Figure 1a. Although ion implantation has been widely used to enhance corrosion and wear resistance in metals and to modify the surfaces of ceramics and semiconductors,[23,24] its potential for constructing battery interfaces remains largely unexplored.

During the ion implantation process, accelerators impart high kinetic energy to ions, which bombard the material surface and induce complex physical and chemical transformations. In our system, accelerated Cu ions strike the Cu foil surface, sputter away the native oxide layer, penetrate into the near-surface region, and generate vacancy defects beneath the surface (Figure 1a). Distinct from conventional additive deposition techniques,[25] such as physical vapor deposition (PVD) or atomic layer deposition (ALD), this ion-implantation approach constructs a vacancy-enriched interfacial structure without altering the foil thickness or compromising the intrinsic electrical conductivity and mechanical integrity of Cu. Instead, it maintains a chemically "fresh" Cu surface by trapping oxygen via vacancy sites, thereby suppressing surface reoxidation.

Ion implantation primarily generates two effects: the mass effect and the energy effect.[24,26] To eliminate doping-related complications associated with the mass effect, we employed Cu ion implantation, ensuring that the implanted ions were chemically identical to the substrate. Benefiting from the milliampere-range beam current provided by a metal vapor vacuum arc (MEVVA) source,[27] the entire implantation process was completed within 25 min. Notably, the implantation area reached 8 inches in diameter, enabling the simultaneous treatment of Cu foils sufficient for fabricating approximately 300 coin-sized cell current collectors (Figure 2a), highlighting the strong potential of this strategy for scalable and industrially relevant applications.

To comprehensively elucidate the effects of ion implantation on Cu current collectors, we systematically examined the morphology, crystallography, and surface chemistry of Cu foils subjected to different treatments, namely pristine Cu (Pristine-Cu), acid-treated Cu, and ion-implanted Cu (Ion-Cu). Scanning electron microscopy (SEM) images reveal that Ion-Cu exhibits a smoother and more uniform surface morphology compared to Pristine-Cu (Figures 2b and 2c). Electron backscattered diffraction (EBSD) orientation maps show similar grain size, shape, and crystallographic orientation between the two foils (Supplementary Figure 1). X-ray diffraction (XRD) patterns confirm that the overall crystallographic texture remains unchanged. However, the increased full width at half maximum (FWHM) of the diffraction peaks indicates a higher density of lattice defects in Ion-Cu (Supplementary Figure 2). Cross-sectional high-resolution transmission electron microscopy (HRTEM) further highlights these structural differences: Pristine-Cu displays a three-layer structure comprising an amorphous carbon layer, an oxide layer (~2-3 nm, $CuO/Cu_2O$), and a bulk Cu layer, whereas Ion-Cu presents a two-layer structure consisting only of carbon and Cu (Figures 2d-g and Supplementary Figure 3). The carbon layer was introduced as a protective coating during focused ion beam (FIB) milling. The absence of surface oxide in Ion-Cu was further corroborated by time-of-flight secondary ion mass spectrometry (TOF-SIMS) measurements (Supplementary Figure 4).

X-ray photoelectron spectroscopy (XPS) was used to quantitatively compare the oxidation levels among Pristine-Cu, acid-treated Cu, and Ion-Cu, based on the percentage area of the divalent Cu peaks at 935.44 eV and 955.24 eV[28,29] (Supplementary Figure 5). Ion-Cu exhibited a substantially lower oxidation level than the pristine sample, and slightly lower than the acid-treated counterpart. Four-point probe measurements further demonstrate that Ion-Cu has significantly lower and more uniform surface resistance across the test area (Supplementary Figure 6), with an average resistance of $1.59 \pm 0.04\ \Omega$ (coefficient of variation [C.V.]: 2.70%) compared to $2.28 \pm 0.35\ \Omega$ (C.V.: 15.31%) for Pristine-Cu.

Collectively, these results demonstrate that Ion-Cu achieves smoother surface morphology, higher defect density, enhanced electronic conductivity, and long-term chemical stability relative to Pristine-Cu. The primary driver for the enhanced performance is the elimination of chemical and electronic heterogeneity. The "roughness" observed on Pristine-Cu is largely symptomatic of the non-uniform, insulating amorphous oxide layer. By removing this layer, Ion-Cu achieves a chemically homogeneous metallic surface with uniform electronic conductivity, which is the dominant factor in suppressing localized high-current-density hotspots and facilitating uniform Li nucleation. While acid etching can similarly remove the native oxide, it cannot prevent rapid re-oxidation and generates chemical waste. By contrast, the MEVVA-based ion implantation process operates at room temperature without chemical reagents, treats large areas in a single batch, and modifies the existing substrate without altering its dimensions.

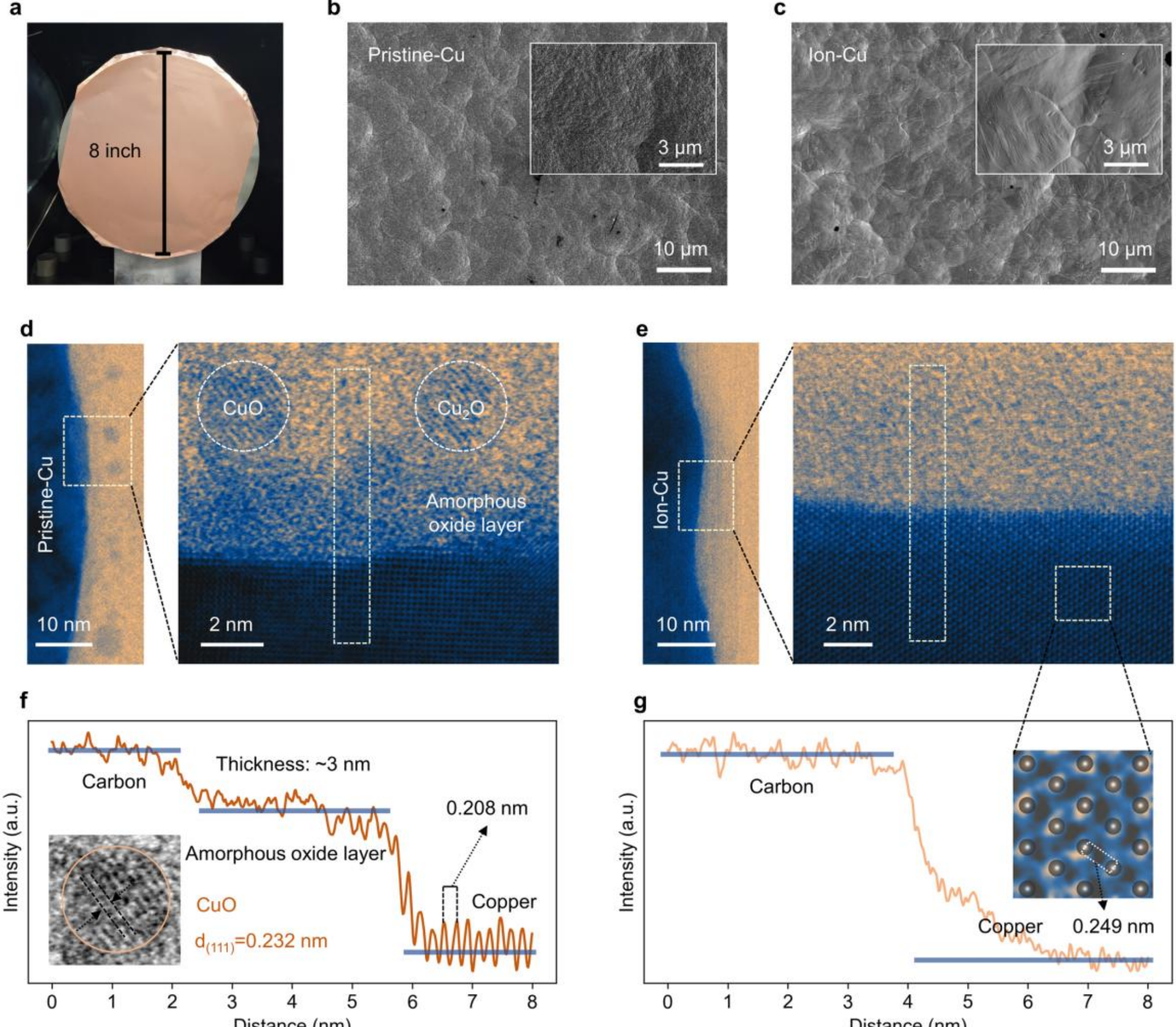

**Figure 2.** Surface cleaning via ion implantation on Cu foils. **a** Photograph of the ion implantation process on Cu foils. **b-c** Top-view SEM images of Pristine-Cu (**b**) and Ion-Cu (**c**), with insets showing magnified regions that highlight surface morphology differences. **d-e** Cross-sectional HRTEM images of Pristine-Cu (**d**) and Ion-Cu (**e**), showing the presence and absence of a native oxide layer, respectively. **f-g** Magnified HRTEM images and intensity profiles of regions highlighted in (**d**) and (**e**).

The stability of the interface in designed CuCCs is another critical parameter for practical battery applications. We examined the long-term interfacial stability of acid-treated Cu and Ion-Cu, as shown in Supplementary Figure 5. Ion-Cu exhibited minimal degradation even after 90 days, whereas the acid-treated sample rapidly re-oxidized upon air exposure, reverting to its initial state.

To elucidate the mechanism underlying the enhanced oxidation resistance of Ion-Cu, we conducted dark-field TEM measurements and multiscale simulations. Figures 3a and 3b present low-magnification dark-field TEM images of Pristine-Cu and Ion-Cu, respectively, while Figure 3c displays intensity profiles corresponding to the highlighted regions (Boxes A and B). A notable increase in contrast roughness is observed in Ion-Cu, indicating that ion implantation introduces substantial lattice damage. HRTEM images (Figures 3d and 3e) further reveal atomic-scale details: Pristine-Cu exhibits a highly ordered lattice structure, whereas Ion-Cu contains numerous bright spots and regions of reduced contrast. The bright spots may correspond to interstitial Cu atoms, while the circular regions likely represent nanoscale vacancy clusters. Figure 3f presents intensity profiles of Box C (Pristine-Cu) and Boxes D and E (Ion-Cu). The periodic oscillations reflect the underlying Cu lattice. Box C shows a gradual intensity increase due to FIB-induced thickness variations, whereas Boxes D and E reveal distinct features corresponding to vacancy clusters, with measured diameters of 1.52 nm and 1.45 nm, equivalent to clusters containing approximately eight Cu atoms. These vacancy clusters are identified as localized regions where multiple adjacent atomic columns exhibit uniformly diminished intensity, and the observed dimensions are consistent with the size distribution predicted by CD simulations.

To quantitatively characterize the ion implantation-induced defects, Monte Carlo (MC) simulations were performed to assess damage accumulation, and density functional theory (DFT) calculations were used to evaluate the binding energies between vacancies and oxygen atoms. These results were incorporated into a cluster dynamics (CD) model to simulate the evolution of vacancy clusters and the corresponding lattice oxygen concentrations (Supplementary Figure 7).

Figure 3g illustrates the MC simulation results for ion implantation under the experimental conditions, obtained using the Transport of Ions in Matter (TRIM) module. The simulated target consisted of a 3 nm oxide layer atop a 27 nm Cu layer according to the commercial Cu foil case. The blue curve (left axis) shows the depth distribution of implanted Cu ions (normalized count), reflecting the ion concentration profile along the depth direction, with a mean stopping depth of approximately 7 nm. The red curve (right axis) displays the displacement-per-atom rate (dpa $s^{-1}$), quantifying the frequency of atomic displacements caused by collisions between incident ions and lattice atoms. The dpa $s^{-1}$ profile peaks at ~5 nm beneath the surface. After subtracting the oxide layer thickness, the peak displacement depth shifts to ~2 nm, consistent with the defect-rich region observed in Figure 3e. These results confirm that the vacancy cluster formation is beneath the Cu surface but confined to a shallow depth ($<0.2\%$ of the total foil thickness), thereby preserving the macroscopic dimensions of the battery. The Bragg peak profile inherent to ion implantation concentrates lattice displacement at subsurface depths, while the topmost surface retains a relatively well-ordered crystalline structure (Figure 2e), suggesting that the vacancy-mediated oxygen-trapping function is spatially confined beneath the electrode–electrolyte interface.

To assess this effect, DFT calculations were performed on 20 representative structures containing varying numbers of vacancies and oxygen atoms (Supplementary Figure 8). We systematically quantified the binding energetics for both oxygen capture (Supplementary Figure 9a) and vacancy accumulation (Supplementary Figure 9b). These thermodynamic parameters were fitted to analytical functions (Equations 1 and 2, see Methods). Finally, with the TRIM and DFT results serving as critical inputs, CD simulations were conducted under experimental conditions. These simulations reveal that vacancy defects tend to aggregate into nanoscale clusters (Supplementary Figure 9c). The resulting cluster size distribution peaks at approximately 50 atoms, corresponding to a diameter of ~1.0 nm. This is slightly smaller than the experimentally observed clusters (~1.4 nm in diameter), possibly due to limited resolution of HRTEM imaging or thermally enhanced vacancy migration during ion implantation. Further CD simulations examined the evolution of lattice oxygen concentrations. Assuming an initial lattice oxygen concentration of $10^{-5}$ and oxygen production rates set ($P_O$) of $1\times10^{-9}$ $s^{-1}$, Figure 3h depicts a three-stage process. In Stage 1, residual lattice oxygen is rapidly captured by vacancy clusters, reducing its concentration below $10^{-12}$ within one second. Stage 2 represents the anti-oxidation phase, during which newly generated lattice oxygen (from surface adsorption) is efficiently trapped by vacancies,[30,31] maintaining a near-zero oxygen concentration. In Stage 3, the saturation of vacancy clusters leads to the gradual reappearance of lattice oxygen. It should be noted that varying $P_O$ affects the duration of each stage but does not alter the underlying physical processes. The selected $P_O$ value aligns with experimental observations showing that Ion-Cu begins to exhibit slight oxidation after 90 days. This time scale is corroborated by the acid-treated Cu control, which shares an initially oxide-free surface but lacks subsurface vacancies and re-oxidizes within days (Supplementary Figure 5b), confirming that the sustained anti-oxidation behavior of Ion-Cu originates from the vacancy-mediated oxygen trapping rather than oxide removal alone.

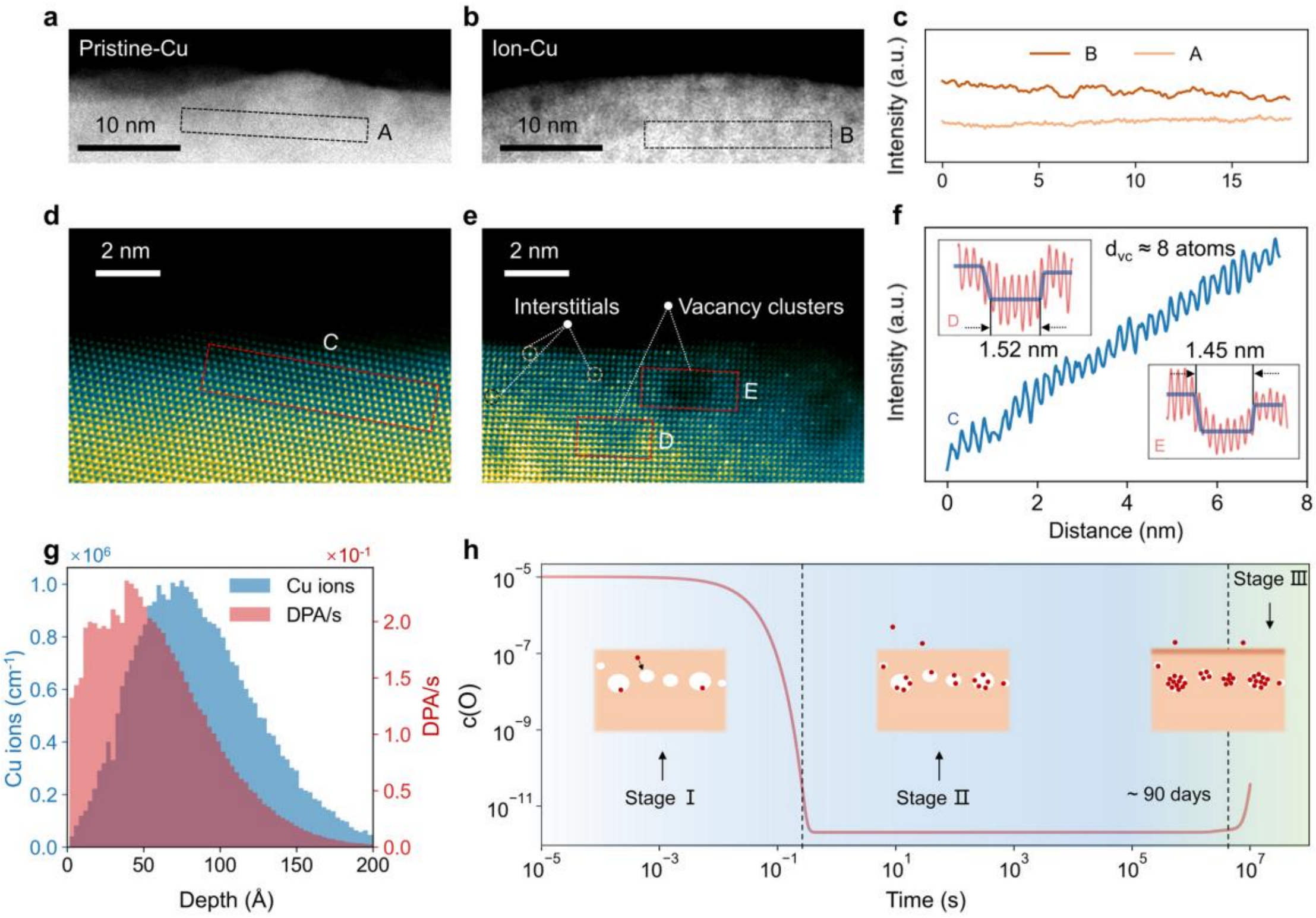

**Figure 3.** Subsurface vacancy formation via ion implantation on Cu foils and its effect on oxidation resistance. **a-b** Dark-field HRTEM images of Pristine-Cu (**a**) and Ion-Cu (**b**). **c** Intensity profiles of regions highlighted in (**a**) and (**b**). **d-e** Atomic-scale HRTEM images of Pristine-Cu (**d**) and Ion-Cu (**e**), showing nanoscale vacancy clusters and interstitial defects in Ion-Cu. **f** Intensity profiles of selected regions: Box C in Pristine-Cu, and Boxes D and E in Ion-Cu. **g** TRIM simulation results showing depth distribution of implanted Cu ions (blue curve) and displacements per atom (dpa $s^{-1}$, red curve). **h** Simulated evolution of lattice oxygen concentration, showing three distinct stages.

## SEI Formation on Ion-implanted CuCCs

The formation and evolution of the solid electrolyte interphase (SEI) on CuCCs are critical determinants of interfacial stability and cycling reversibility in AFLMBs. In conventional CuCCs, nonuniform Li nucleation and continuous surface reoxidation often lead to inhomogeneous SEI growth, repeated SEI rupture, and excessive electrolyte consumption. By contrast, the vacancy-enriched surface created via ion implantation fundamentally alters the interfacial chemistry and nucleation behavior, thereby regulating SEI formation from the initial plating stage.

TOF-SIMS and XPS measurements were performed to compare the SEI evolution on Ion-Cu and Pristine-Cu surfaces (Figure 4). As shown in Figures 4a and 4b, the SEI formed on Ion-Cu is significantly thinner than that on Pristine-Cu. Additionally, the Ion-Cu SEI exhibits markedly reduced signals from organic species (e.g., $C_2H^-$ and $C_2HO^-$), indicative of suppressed solvent decomposition. XPS analysis reveals the atomic concentrations at various depths for SEI formed on different CuCCs (Supplementary Figure 10). At every etching stage, the SEI on Ion-Cu exhibits lower carbon (C) and oxygen (O) contents and a higher proportion of inorganic Li species compared to that on Pristine-Cu, showing the formation of a thinner, denser SEI layer. Detailed analysis of the O 1s spectra shows contributions from $Li_2CO_3$, C-O-C, and $Li_2O$ species. With increasing etching depth, a greater abundance of $Li_2O$ is detected on Ion-Cu, whereas $Li_2CO_3$ remains dominant on Pristine-Cu (Figures 4c and 4d). Given that $Li_2CO_3$ is unstable in direct contact with Li metal and prone to react, forming amorphous $LiC_x$,[32] its accumulation leads to SEI thickening, continuous electrolyte depletion, and degradation of electrochemical performance. Moreover, $Li_2CO_3$ imposes the highest diffusion barrier among SEI inorganic components, hindering $Li^+$ transport.[33] In contrast, on Ion-Cu, the $Li_2O$ layer envelops residual $Li_2CO_3$ (Figures 4e and 4f), effectively isolating it from direct Li-metal contact and thus mitigating side reactions. A higher presence of N-$SO_x$ and $LiN_xO_y$ is observed on Pristine-Cu surfaces, whereas $LiN_x$ is more prominent on Ion-Cu surfaces, suggesting distinct decomposition pathways of electrolyte components (Supplementary Figure 11).

To uncover the origin of the distinct SEI compositions between Ion-Cu and Pristine-Cu, cyclic voltammetry (CV) was conducted on the three types of Cu foils (Supplementary Figure 5c-f). CV results demonstrate that SEI formation proceeds through two potential-dependent stages. At 1.0 to 2.0 V vs. $Li/Li^+$, SEI initiates via direct electron transfer (ET). In this region, Ion-Cu exhibits a higher exchange current density than Acid-Treated Cu, and both are substantially larger than that of Pristine-Cu, indicating that Ion-Cu and Acid-Treated Cu form a more robust and inorganic-rich direct-ET SEI. This trend aligns well with previous reports that high-CE electrolytes also promote enhanced direct-ET current in this potential window.

At potentials below 0.5 V, since both Acid-Treated Cu and Ion-Cu already possess a stable and compact direct-ET SEI formed at 1.0 to 2.0 V, Li underpotential deposition (UPD) does not induce excessive additional SEI formation. In contrast, Pristine-Cu forms only a sparse and low-quality direct-ET SEI. Coupled with the strong affinity of surface copper oxides toward solvent molecules, Li UPD triggers massive parasitic solvent decomposition, producing a highly organic-rich and unstable SEI.

For Pristine-Cu, persistent UPD/UPS peaks in subsequent cycles reflect continuous SEI formation and dissolution, which causes sustained Li corrosion and gradual thickening of a resistive and unstable SEI layer (Supplementary Figure 5e). By comparison, the complete absence of UPD/UPS peaks for both Acid-Treated Cu and Ion-Cu confirms that their SEI layers are stable and fully established in the initial cycle, effectively protecting the Li metal anode without repeated side reactions (Supplementary Figure 5f).

Clearly, oxide removal plays a vital role in regulating SEI formation. Furthermore, Ion-Cu delivers a more favorable SEI composition relative to Acid-Treated Cu (Supplementary Figure 5c), which agrees well with its larger exchange current in the 1.0 to 2.0 V region. The fundamental origin lies in the lower residual oxide content of Ion-Cu. In addition, the subsurface vacancy clusters, which are located beneath the surface without direct electrolyte contact, provide a durable oxygen-trapping effect that sustains an oxide-free state over long period. This conclusion is further supported by direct immersion tests (Supplementary Figure 11c), in which the Pristine-Cu electrolyte solution shows visible discoloration and significantly higher interfacial impedance after 12 h, whereas the Ion-Cu solution remains colorless with minimal impedance increase.

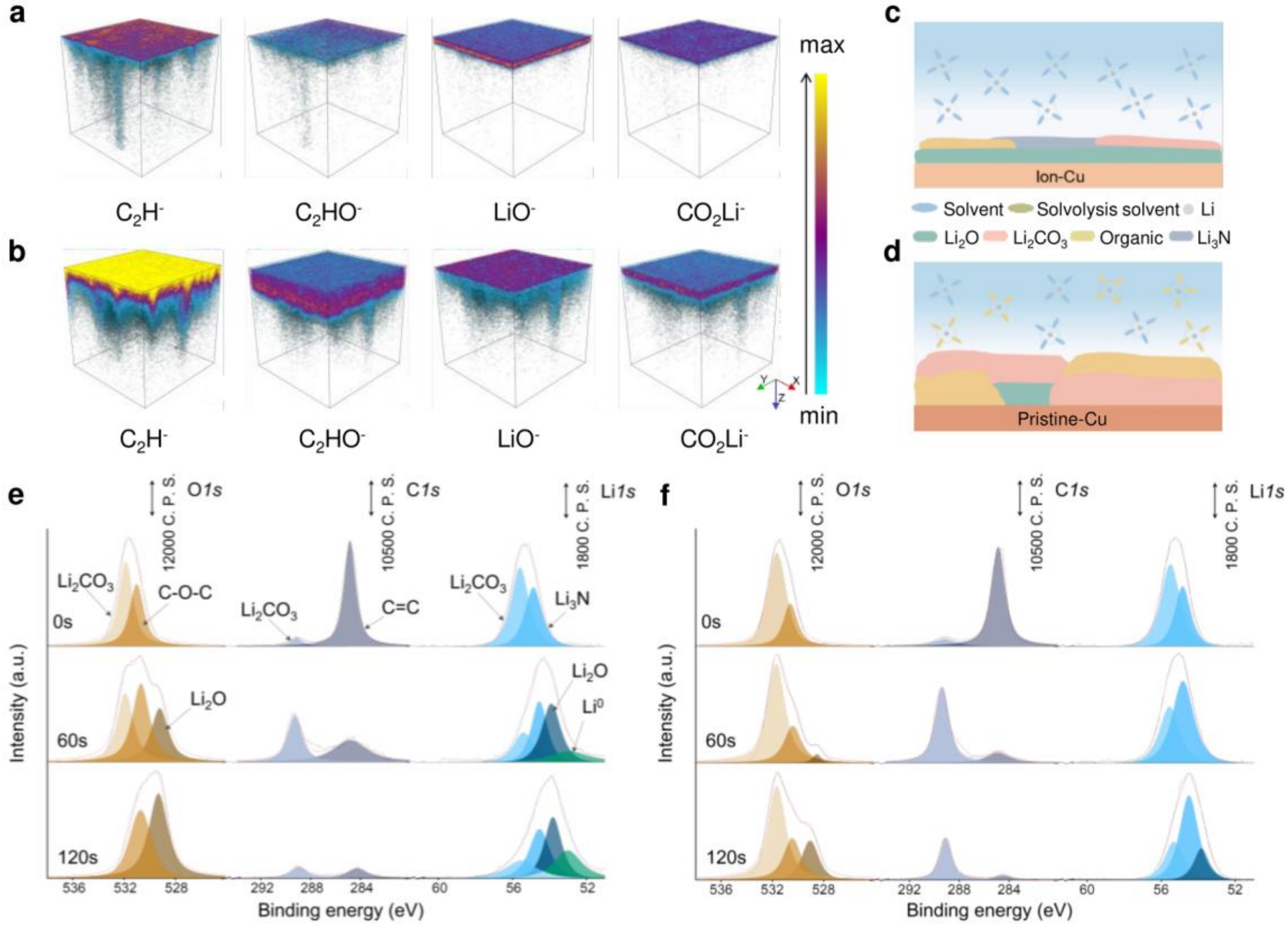

**Figure 4.** SEI composition on different CuCCs. **a-b** 3D-rendered TOF-SIMS images of the SEI on Ion-Cu (**a**) and Pristine-Cu (**b**) at applied potentials approaching the $Li^0$-metal potential (0 V vs. $Li/Li^+$) under $Cs^+$ sputtering. **c-d** Depth-resolved XPS spectra (O1s, C1s, and Li1s) of the SEI on the deposited Li on Ion-Cu (**c**) and Pristine-Cu (**d**). The SEI evolution on Ion-Cu and Pristine-Cu surfaces under potentials approaching $Li^0$-metal (0 V vs. $Li/Li^+$). **e-f** Schematic diagrams of SEI structures on Ion-Cu (**e**) and Pristine-Cu (**f**).

## Li plating/stripping behavior

Li||Cu half-cells pairing Li foils with different CuCCs were assembled using an ether-based electrolyte (1.0 M LiTFSI in a 1:1 vol/vol mixture of DOL/DME with 5% $LiNO_3$) to evaluate Li plating/stripping behavior. For a comprehensive comparison, oxidized Cu (Oxi-Cu) was also prepared by soaking commercial Cu foils in 1 M NaOH solution for 24 hours to achieve complete surface oxidation. Electrochemical measurements, SEM characterizations, and DFT calculations were employed to study the Li plating/stripping behavior. Benefiting from the $Li_2O$-rich SEI induced by ion implantation, Ion-Cu exhibits the smallest interfacial impedance (25.39 Ω) and lowest nucleation overpotential (38.4 mV) in Li||Cu half-cells (Supplementary Figures 12-14). Ion-Cu promoted nucleation sites with larger radii compared to Pristine-Cu (Supplementary Figures 15-17). This optimized nucleation behavior directly translates to improved lithium deposition morphology. As the degree of surface oxidation decreased from Oxi-Cu to Ion-Cu, the deposited Li becomes progressively larger, denser, and less granular (Figures 5a-c). Li deposited on Ion-Cu shows the dense structure with a thickness of approximately 2.5 μm at 0.5 mAh $cm^{-2}$ (Supplementary

Figure 18). Even under high current density, Ion-Cu enables densely packed, film-like Li deposition, while Oxi-Cu and Pristine-Cu exhibit pronounced dendritic Li growth (Supplementary Figure 19). After stripping, Ion-Cu displays negligible dead Li, indicating its highly reversible Li plating/stripping behavior (Supplementary Figure 20).

Li plating/stripping behavior was also studied by DFT calculations. Li adsorption and migration on representative metallic Cu, including (100), (110), and (111) facets, as well as on amorphous $CuO_x$ surfaces, were investigated (Figure 5d). The amorphous $CuO_x$ surfaces were modeled by randomly incorporating oxygen atoms into crystalline Cu slab models at a Cu:O atomic ratio of 1:1. The resulting structures were subjected to three sequential relaxation steps—geometry optimization with fixed in-plane dimensions, molecular dynamics equilibration at 300 K using the density-functional tight-binding (DFTB) method, and final geometry optimization—to obtain equilibrated amorphous configurations (see Methods in Supplementary Information for full details). The Li adsorption energies were found to be significantly lower (i.e., more thermodynamically favorable) on $CuO_x$ surfaces compared to Cu surfaces, indicating stronger Li binding and more effective trapping at initial nucleation sites. Transition state (TS) analysis further revealed that the surface diffusion barrier for Li on $CuO_x$ (~1.74 eV) is more than four times higher than that on metallic Cu (~0.38 eV), suggesting a substantial reduction in Li migration rates on oxidized surfaces (Figure 5e). This dramatic decrease in surface mobility promotes localized Li deposition, consistent with experimental observations (Supplementary Figure 21). To assess whether deposited Li atoms preferentially migrate laterally or vertically onto existing Li clusters, surface models incorporating Li steps were constructed to evaluate the corresponding energy barriers. The calculated barrier for vertical climbing (~1.32 eV) was much higher than for lateral surface diffusion (~0.38 eV), indicating a strong preference for Li to spread laterally across the Cu surface rather than to vertically accumulate (Supplementary Figure 22). This lateral diffusion behavior contributes to the formation of flat and dense Li morphologies on Ion-Cu substrates. Furthermore, the energetic stability of two distinct Li deposition configurations—dispersed Li layers (M1) versus clustered Li islands (M2)—was compared on both Cu and $CuO_x$ surfaces. We calculated the energy difference ($\Delta E$) between clustered and dispersed states (Figure 5f). On metallic Cu, dispersed Li were energetically more favorable, while on $CuO_x$ surfaces, clustered Li was preferred (Supplementary Figure 23). These trends were further validated through ab initio molecular dynamics (AIMD) simulations (Supplementary Figure 24). Interfacial formation energy calculations confirmed that dispersed Li configurations are thermodynamically favored on Cu, whereas clustering is promoted on $CuO_x$ surfaces, except for $CuO_x$ (111), where the upward migration of O atoms during relaxation induced anomalous behavior.

## Li-Cu half-cell cycling performance

To test the half-cell performance for different CuCCs (Pristine-Cu, Oxi-Cu, and Ion-Cu), Li||Cu half-cells pairing Li foils were assembled using an ether-based electrolyte

(1.0 M LiTFSI in a 1:1 vol/vol mixture of DOL/DME with 5% $LiNO_3$). Ion-Cu achieves a CE of 98.8% over 600 cycles, outperforming Pristine-Cu (98.5% over 300 cycles) and Oxi-Cu (short-circuits after only 50 cycles at 0.5 mA $cm^{-2}$ and 1 mAh $cm^{-2}$) (Figure 5g). At a higher current density of 2 mA $cm^{-2}$, Ion-Cu maintains a CE of 97.9% over 200 cycles (Figure 5h), while Pristine-Cu and Oxi-Cu fail after 100 and 30 cycles, respectively. When the current density is further increased to 5 mA $cm^{-2}$, the Ion-Cu cell reaches a high CE of 96.43% after only 2 activation cycles, and remains stable above 95% over 50 cycles, whereas the CE of Pristine-Cu failed after 42 cycles with a lower average CE of 87% (Supplementary Figure 25).

Moreover, Ion-Cu demonstrates consistently high average CEs of 99.23%, 99.15%, and 98.57% across different electrolyte systems with improved Li plating/stripping behaviors, underscoring the robustness of the interfacial engineering strategy (Figure 5i and Supplementary Figures 26-28). The voltage profiles are presented in Supplementary Figure 29, Ion-Cu exhibits the smallest polarization voltage in all electrolytes. Impressively, even in a fluorine-free ester-based electrolyte (1.0 M $LiPF_6$ in a 1:1 vol/vol mixture of EC/DEC)—which typically yields poor CE—Ion-Cu significantly improves lithium plating/stripping behavior and long-term cycling stability, demonstrating its broad applicability (Supplementary Figure 30).

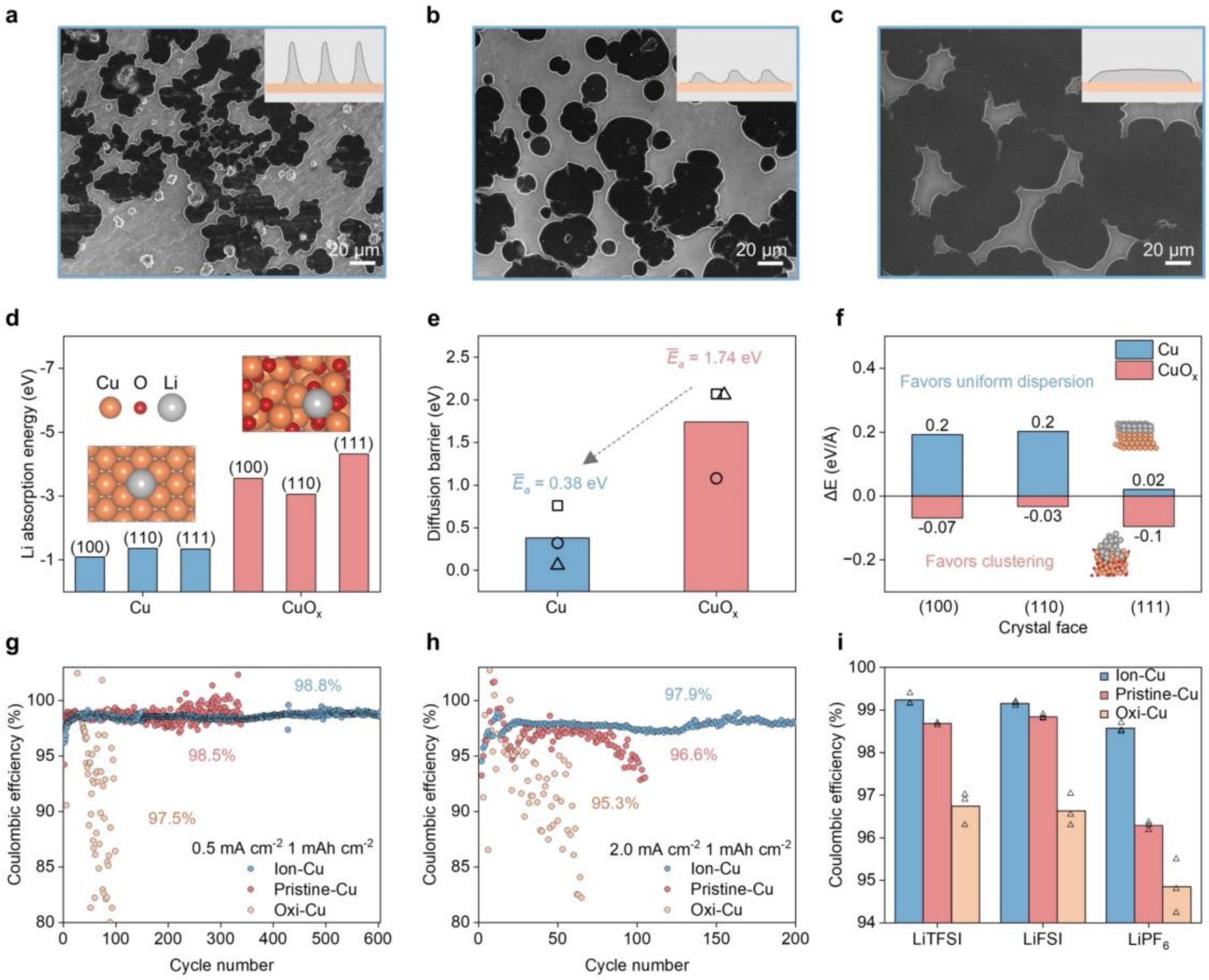

**Figure 5.** Li deposition morphology and electrochemical performance of Li-Cu half-cells. **a-c** SEM images of Li deposited on Oxi-Cu (**a**), Pristine-Cu (**b**), and Ion-Cu (**c**) with a capacity of 0.5 mAh $cm^{-2}$ under a current density of 0.5 mA $cm^{-2}$. **d** Li adsorption energies on Cu and amorphous $CuO_x$ surfaces for different facets ((100), (110), and (111)). **e** Li diffusion barriers ($E_a$) on Cu and $CuO_x$ surfaces, showing significantly hindered mobility on oxidized surfaces. **f** Thermodynamic energy difference (ΔE) between clustered and dispersed Li configurations on different crystal facets. **g-h** Cycling performance of Li||Cu half-cells using different CuCCs at 0.5 mA $cm^{-2}$ (**g**) and 2 mA $cm^{-2}$ (**h**), both with an areal capacity of 1.0 mAh $cm^{-2}$. **i** Coulombic efficiency (CE) of Li||Cu half-cells measured with the Aurbach protocol using different electrolytes: 1.0 M LiTFSI in a 1:1 vol/vol mixture of DOL/DME with 5% $LiNO_3$; 1:1.8:2 mol/mol/mol LiFSI/DME/TTE; and 1.0 M $LiPF_6$ in a 1:1 vol/vol EC/DEC with 10% FEC and 1% VC additives.

## Full cell performance

To further validate the effectiveness of the vacancy-engineered interface created by ion implantation in practical battery systems, AFLMBs were fabricated by pairing Ion-Cu with various commercial cathodes, including NCM523, NCM811, and $LiFePO_4$ (LFP). These full-cell configurations provide a stringent assessment of Li reversibility and interfacial stability under realistic operating conditions.

The Ion-Cu||NCM523 full cell delivers an impressive initial discharge capacity of 163.95 mAh $g^{-1}$ at 0.2 C (Figure 6a), while maintaining 70.0% capacity retention over 60 cycles. This performance corresponds to a substantial 43.4% reduction in the capacity decay rate compared with cells employing Pristine-Cu (47.0%) and Oxi-Cu (40.9%) under identical conditions. The notably lower initial capacity of the Oxi-Cu cell originates from the electronically insulating nature of the intentionally thickened oxide layer, which increases interfacial charge-transfer resistance and polarization, while simultaneously consuming a larger fraction of the limited cathode-sourced lithium inventory through irreversible parasitic reactions during the first cycle (Supplementary Figure 13). Voltage-capacity profiles recorded at the 60th cycle reveal significantly reduced polarization and more stable electrochemical behavior for the Ion-Cu-based cell relative to the Pristine-Cu and Oxi-Cu counterparts (Figure 6b and Supplementary Figure 31). When the cycling rate is increased to 0.6 C, Ion-Cu-based AFLMBs with NCM523 cathodes still retain 42.3% of their capacity after 100 cycles (Supplementary Figure 32).

When paired with a high-energy-density NCM811 cathode, the Ion-Cu AFLMB maintains a capacity retention of 59.2% after 100 cycles, accompanied by stable voltage profiles throughout cycling (Figure 6c and 6d). Under high mass-loading conditions (33.1 mg $cm^{-2}$), the performance advantage of Ion-Cu becomes even more pronounced: the Ion-Cu electrode preserves 40.2% capacity retention after 80 cycles, whereas the Pristine-Cu-based cell rapidly degrades to only 11.16% retention (Supplementary Figure 33). The universality of Ion-Cu as high-performance anode current collector was further demonstrated using commercial LFP cathodes in anode-free cells. The LFP full cells exhibit highly stable cycling behavior, maintaining 85.7% of their initial capacity over 140 cycles at a charge/discharge rate of 0.5 C/0.6 C (Figure 6f). The corresponding charge-discharge profiles show negligible potential decay over prolonged cycling (Figure 6g), indicative of a highly stable interfacial structure enabled by ion implantation. Benchmark comparisons with previously reported AFLMB systems suggest that the Ion-Cu configuration exhibits competitive robustness among reported anode-free battery architectures (Figured 6e and 6h).

Beyond coin-cell demonstrations, the practical viability of Ion-Cu was further validated in pouch-cell configurations. A 50 mAh prototype Ion-Cu||NCM523 AFLMB pouch cell was assembled and operated stably for 60 cycles under lean electrolyte conditions (2.4 g $Ah^{-1}$) (Supplementary Figures 34-35). The pouch cell achieves a high specific energy density of approximately 320 Wh $kg^{-1}$, highlighting the exceptional effectiveness of Ion-Cu in enhancing Li reversibility in practical AFLMBs.

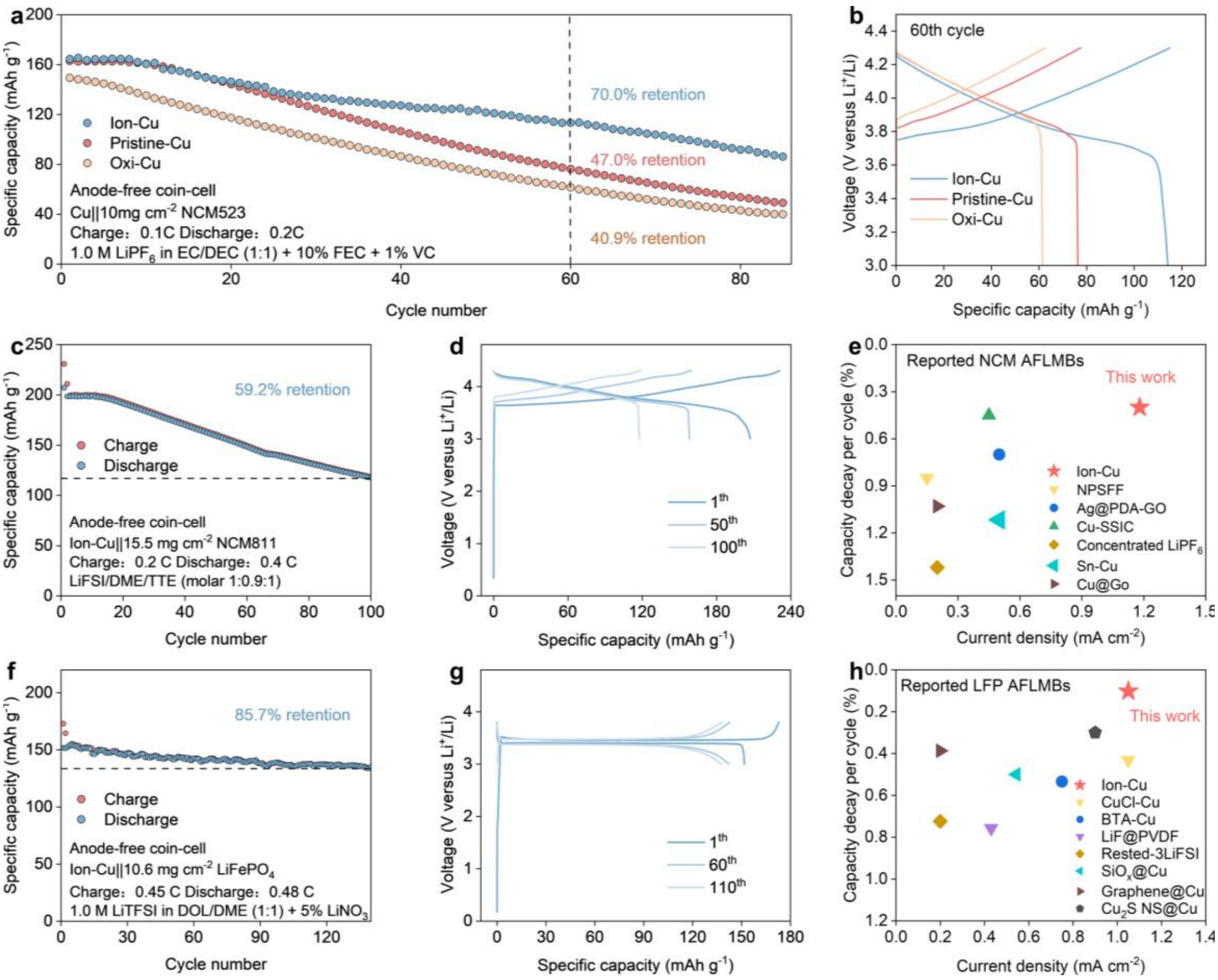

**Figure 6. Electrochemical performance of full cells with different CuCCs. a** Long-term cycling performance of anode-free Cu||NCM523 coin cells. **b** Corresponding charge/discharge profiles of the Cu||NCM523 coin cells at the 60th cycle. **c** Long-term cycling performance of anode-free Ion-Cu||NMC811 coin-cell. **d** Corresponding charge/discharge profiles of the Ion-Cu||NCM811 coin-cell at $1^{th}$, $50^{th}$, and $100^{th}$ cycles. **e** Electrochemical performance comparison of AFLMBs with NCM cathode between this work and previously reported work.[34–39] **f** Long-term cycling performance of anode-free Ion-Cu||LFP coin-cell. **g** Corresponding charge/discharge profiles of the Ion-Cu||LFP coin-cell at $1^{th}$, $60^{th}$, and $110^{th}$ cycles. **h** Electrochemical performance comparison of AFLMBs with LFP cathode between this work and previously reported work.[40–46]

## *Conclusions*

In summary, we developed an AIE strategy to construct anti-oxidation CuCCs for AFLMBs, starting from an anode-free cell configuration. Through Cu ion implantation, the native amorphous oxide layer on commercial Cu foils can be effectively removed, while subsurface vacancy clusters are introduced that are capable of capturing residual oxygen atoms, as supported by multiscale experimental characterizations and simulations. The resulting anti-oxidation Cu surface exhibits improved interfacial stability, mitigates electrolyte solvolysis, reduces and homogenizes surface resistance, and facilitates more uniform lateral Li deposition. These interfacial modifications favor the formation of a thin, $Li_2O$-enriched SEI, which is associated with reduced interfacial charge-transfer resistance and enhances $Li^+$ transport kinetics. As a consequence, the modified CuCCs show improved electrochemical performance, including a CE of 98.8% over 600 cycles in Li-Cu half-cells, 85.7% capacity retention after 140 cycles in full-cell configurations with LFP cathodes. This work suggests that ion implantation is a versatile approach for interfacial engineering of CuCCs. Owing to its compatibility with large-area processing, short treatment time, and direct applicability to commercial Cu foils without altering the bulk structure, the proposed strategy may offer a practical route toward improving the performance and reliability of AFLMBs.

## *Methods*

***Preparation of Ion-Cu.*** Ion-Cu was prepared by implanting Cu ions into commercial Cu foils (MSK-WTMC-CU2G) at room temperature using a FAD-MEVVA ion implanter, whose beam spot exceeds 8 inches in diameter and allowed simultaneous treatment of foils sufficient for more than 300 current collectors. The ion fluence was varied from $10^{13}$ to $10^{16}$ ions $cm^{-2}$ and the ion energy from 5 to 20 keV; unless otherwise stated, all electrochemical measurements used the optimized condition of 9 keV and $1\times10^{16}$ ions $cm^{-2}$. Full details are provided in the Supporting Information.

***Preparation of Acid-Treated Cu.*** Commercial Cu foils were immersed in 1.0 M $CH_3COOH$ for 10 min at room temperature, rinsed with deionized water and ethanol, dried under argon, and transferred to an argon-filled glovebox before use.

***Electrochemical measurements.*** Cells were assembled as 2032-type coin cells with a Celgard 2325 separator in an argon-filled glovebox ($O_2$ and $H_2O$ below 0.01 ppm). Li||Cu half-cells used Li foil as the counter electrode, and Coulombic efficiencies were determined by the Aurbach method. Anode-free full cells (Ion-Cu||NCM523, ≈10 mg $cm^{-2}$) and pouch cells were activated at 0.05 C between 3.0 and 4.3 V (1 C = 165 mA $g^{-1}$) and subsequently cycled at 0.1–0.2 C. Unless otherwise stated, the electrolyte was 1 M LiTFSI in DOL/DME (1:1 by volume) with 5 wt% $LiNO_3$. Additional electrolytes and protocols are described in the Supporting Information.

***Characterizations.*** Surface morphology was examined by scanning electron microscopy (SEM, Hitachi S-4800). X-ray photoelectron spectroscopy (XPS) was performed on a Kratos AXIS Ultra spectrometer with monochromatic Al Kα radiation, with binding energies calibrated to the C 1s peak at 284.6 eV. Crystallography was assessed by X-ray diffraction (XRD, Rigaku SmartLab, Cu Kα) and electron backscattered diffraction (EBSD, eFlash HR). Samples for SEI analysis were rinsed with anhydrous DME and transferred under vacuum to avoid air exposure.

***Computational methods.*** Multiscale modeling combined Monte Carlo (TRIM) simulations of implantation damage, density functional theory (DFT) calculations of vacancy–oxygen energetics, and cluster-dynamics simulations of vacancy-cluster and lattice-oxygen evolution. Lithium adsorption and diffusion on crystalline and amorphous Cu and $CuO_x$ surfaces were evaluated by DFT, complemented by ab initio and density-functional tight-binding molecular dynamics. Full parameters, codes, and associated references are provided in the Supporting Information.

## *Associated Content*

***Data Availability Statement.*** The data that support the findings of this study are available in the main text and Supporting Information. All relevant data are available from the corresponding authors upon request.

***Supporting Information.*** The Supporting Information is available free of charge at https://pubs.acs.org/doi/XXXXX.

Experimental and computational methods; supplementary SEM, EBSD, GIXRD, and HRTEM characterization; TOF-SIMS and XPS depth-profiling of the SEI; multiscale simulation (TRIM, DFT, cluster dynamics) details; supplementary electrochemical performance; Supplementary Figures S1–S35 and Supplementary Tables S1–S2 (PDF).

***Preprint.*** A previous version of this manuscript was deposited as a preprint: *Li, Y.; Ren, X.; Feng, X.; Kong, L.; Luo, F.; Xu, Y.; Qian, L.; Ye, Y.; Zhao, Z.; Gao, X.; Zhang, J. Atomic Interface Engineering of Battery Current Collectors via Ion Implantation. 2025, arXiv:2508.00236. arXiv. https://doi.org/10.48550/arXiv.2508.00236 (accessed June 5, 2026).*

## *Conflict of Interest*

The authors declare no conflict of interest.

## *Acknowledgements*

This work was financially supported by the Ministry of Science and Technology of China (2022YFA1203302, 2022YFA1203304 and 2025YFE0211900), the National Natural Science Foundation of China (52302034, 52021006, 52102032, 52272033, 52402041), the Scientific Research Innovation Capability Support Project for Young Faculty (SRICSPYF-ZY2025066), the Strategic Priority Research Program of CAS (XDB36030100), the Beijing National Laboratory for Molecular Sciences (BNLMS-CXTD-202001) and the Shenzhen Science and Technology Innovation Commission (KQTD20221101115627004).

## *References*

(1) Magnussen, O. M.; Groß, A. Toward an Atomic-Scale Understanding of Electrochemical Interface Structure and Dynamics. *J. Am. Chem. Soc.* **2019**, *141*, 4777–4790.
(2) Irvine, J. T. S.; Neagu, D.; Verbraeken, M. C.; Chatzichristodoulou, C.; Graves, C.; Mogensen, M. B. Evolution of the Electrochemical Interface in High-Temperature Fuel Cells and Electrolysers. *Nat. Energy* **2016**, *1*, 15014.
(3) Li, C.; Shao, Q.-G.; Luo, K.; Gao, Y.-Q.; Zhao, W.-W.; Cao, N.; Du, S.-Y.; Jin, X.; Zou, P.-C.; Zang, X.-B. A Lean-Zinc and Zincophilic Anode for Highly Reversible Zinc Metal Batteries. *Adv. Funct. Mater.* **2023**, *33*, 2305204.
(4) Nanda, S.; Gupta, A.; Manthiram, A. Anode-Free Full Cells: A Pathway to High-Energy Density Lithium-Metal Batteries. *Adv. Energy Mater.* **2021**, *11*, 2000804.
(5) Weber, R.; Genovese, M.; Louli, A. J.; Hames, S.; Martin, C.; Hill, I. G.; Dahn, J. R. Long Cycle Life and Dendrite-Free Lithium Morphology in Anode-Free Lithium Pouch Cells Enabled by a Dual-Salt Liquid Electrolyte. *Nat. Energy* **2019**, *4*, 683–689.
(6) Heubner, C.; Maletti, S.; Auer, H.; Hüttl, J.; Voigt, K.; Lohrberg, O.; Nikolowski, K.; Partsch, M.; Michaelis, A. From Lithium-Metal toward Anode-Free Solid-State Batteries: Current Developments, Issues, and Challenges. *Adv. Funct. Mater.* **2021**, *31*, 2106608.
(7) Huang, W.; Zhao, C.; Wu, P.; Yuan, H.; Feng, W.; Liu, Z.; Lu, Y.; Sun, S.; Fu, Z.; Hu, J.; Yang, S.; Huang, J.; Zhang, Q. Anode-Free Solid-State Lithium Batteries: A Review. *Adv. Energy Mater.* **2022**, *12*, 2201044.
(8) Fang, C.; Lu, B.; Pawar, G.; Zhang, M.; Cheng, D.; Chen, S.; Ceja, M.; Doux, J.-M.; Musrock, H.; Cai, M.; Liaw, B.; Meng, Y. S. Pressure-Tailored Lithium Deposition and Dissolution in Lithium Metal Batteries. *Nat. Energy* **2021**, *6*, 987–994.
(9) Lin, D.; Liu, Y.; Cui, Y. Reviving the Lithium Metal Anode for High-Energy Batteries. *Nat. Nanotechnol.* **2017**, *12*, 194–206.
(10) Wang, H.; Yu, Z.; Kong, X.; Kim, S. C.; Boyle, D. T.; Qin, J.; Bao, Z.; Cui, Y. Liquid Electrolyte: The Nexus of Practical Lithium Metal Batteries. *Joule* **2022**, *6*, 588–616.
(11) Yuan, X.; Liu, B.; Mecklenburg, M.; Li, Y. Ultrafast Deposition of Faceted Lithium Polyhedra by Outpacing SEI Formation. *Nature* **2023**, *620*, 86–91.
(12) Zhang, E.; Chen, Y.; Holoubek, J.; Yu, Z.; Zhang, W.; Lyu, H.; Choi, I. R.; Kim, S. C.; Serrao, C.; Cui, Y.; Bao, Z. Monofluorinated Acetal Electrolyte for High-Performance Lithium Metal Batteries. *Proc. Natl. Acad. Sci. U.S.A.* **2025**, *122*, e2418623122.
(13) Kwon, H.; Lee, J.-H.; Roh, Y.; Baek, J.; Shin, D. J.; Yoon, J. K.; Ha, H. J.; Kim, J. Y.; Kim, H.-T. An Electron-Deficient Carbon Current Collector for Anode-Free Li-Metal Batteries. *Nat. Commun.* **2021**, *12*, 5537.
(14) Zhai, P.; Wang, T.; Jiang, H.; Wan, J.; Wei, Y.; Wang, L.; Liu, W.; Chen, Q.; Yang, W.; Cui, Y.; Gong, Y. 3D Artificial Solid-Electrolyte Interphase for Lithium Metal Anodes Enabled by Insulator–Metal–Insulator Layered Heterostructures. *Adv. Mater.* **2021**, *33*, 2006247.

(15) Zhao, H.; Du, S.; Liu, Q.; Liu, X.; Lu, D.; Liu, T.; Yin, Y.; Li, Y.; Zou, P.; Wu, Z. Defective Interface Customizes Anion-Aggregated Helmholtz Plane for Anode-Less Lithium Metal Batteries. *ACS Nano* **2025**, *19*, 34942–34953.
(16) Lu, G.; Liu, W.; Yang, Z.; Wang, Y.; Zheng, W.; Deng, R.; Wang, R.; Lu, L.; Xu, C. Superlithiophilic, Ultrastable, and Ionic-Conductive Interface Enabled Long Lifespan All-Solid-State Lithium-Metal Batteries under High Mass Loading. *Adv. Funct. Mater.* **2023**, *33*, 2304407.
(17) Kim, S. J.; Kim, Y. I.; Lamichhane, B.; Kim, Y.-H.; Lee, Y.; Cho, C. R.; Cheon, M.; Kim, J. C.; Jeong, H. Y.; Ha, T.; Kim, J.; Lee, Y. H.; Kim, S.-G.; Kim, Y.-M.; Jeong, S.-Y. Flat-Surface-Assisted and Self-Regulated Oxidation Resistance of Cu(111). *Nature* **2022**, *603*, 434–438.
(18) Yang, J. C.; Kolasa, B.; Gibson, J. M.; Yeadon, M. Self-Limiting Oxidation of Copper. *Appl. Phys. Lett.* **1998**, *73*, 2841–2843.
(19) Yoon, J. S.; Liao, D. W.; Greene, S. M.; Cho, T. H.; Dasgupta, N. P.; Siegel, D. J. Thermodynamics, Adhesion, and Wetting at Li/Cu(-Oxide) Interfaces: Relevance for Anode-Free Lithium–Metal Batteries. *ACS Appl. Mater. Interfaces* **2024**, *16*, 18790–18799.
(20) Lee, S. Y.; Mettlach, N.; Nguyen, N.; Sun, Y. M.; White, J. M. Copper Oxide Reduction through Vacuum Annealing. *Appl. Surf. Sci.* **2003**, *206*, 102–109.
(21) Qu, G.; Vegunta, S. S. S.; Mai, K.; Weinman, C. J.; Ghosh, T.; Wu, W.; Flake, J. C. Copper Oxide Removal Activity in Nonaqueous Carboxylic Acid Solutions. *J. Electrochem. Soc.* **2013**, *160*, E49.
(22) Chavez, K. L.; Hess, D. W. A Novel Method of Etching Copper Oxide Using Acetic Acid. *J. Electrochem. Soc.* **2001**, *148*, G640.
(23) Li, W.; Zhan, X.; Song, X.; Si, S.; Chen, R.; Liu, J.; Wang, Z.; He, J.; Xiao, X. A Review of Recent Applications of Ion Beam Techniques on Nanomaterial Surface Modification: Design of Nanostructures and Energy Harvesting. *Small* **2019**, *15*, 1901820.
(24) Nastasi, M.; Mayer, J. W. *Ion Implantation and Synthesis of Materials*; Springer: Berlin, Heidelberg, 2006.
(25) Xu, R.; Cheng, X.-B.; Yan, C.; Zhang, X.-Q.; Xiao, Y.; Zhao, C.-Z.; Huang, J.-Q.; Zhang, Q. Artificial Interphases for Highly Stable Lithium Metal Anode. *Matter* **2019**, *1*, 317–344.
(26) Wright, G. M.; Barnard, H. A.; Kesler, L. A.; Peterson, E. E.; Stahle, P. W.; Sullivan, R. M.; Whyte, D. G.; Woller, K. B. An Experiment on the Dynamics of Ion Implantation and Sputtering of Surfaces. *Rev. Sci. Instrum.* **2014**, *85*, 023503.
(27) Zhang, T.; Wang, X.; Liang, H.; Zhang, H.; Zhou, G.; Sun, G.; Zhao, W.; Xue, J. Behaviour of MEVVA Metal Ion Implantation for Surface Modification of Materials. *Surf. Coat. Technol.* **1996**, *83*, 280–283.
(28) Iijima, J.; Lim, J.-W.; Hong, S.-H.; Suzuki, S.; Mimura, K.; Isshiki, M. Native Oxidation of Ultra High Purity Cu Bulk and Thin Films. *Appl. Surf. Sci.* **2006**, *253*, 2825–2829.

(29) Zuo, Z.-J.; Li, J.; Han, P.-D.; Huang, W. XPS and DFT Studies on the Autoxidation Process of Cu Sheet at Room Temperature. *J. Phys. Chem. C* **2014**, *118*, 20332–20345.
(30) McColl, K.; Coles, S. W.; Zarabadi-Poor, P.; Morgan, B. J.; Islam, M. S. Phase Segregation and Nanoconfined Fluid O2 in a Lithium-Rich Oxide Cathode. *Nat. Mater.* **2024**, *23*, 826–833.
(31) Kilic, M. E.; Lee, J.-H.; Lee, K.-R. Oxygen Ion Transport in Doped Ceria: Effect of Vacancy Trapping. *J. Mater. Chem. A* **2021**, *9*, 13883–13889.
(32) Han, B.; Zhang, Z.; Zou, Y.; Xu, K.; Xu, G.; Wang, H.; Meng, H.; Deng, Y.; Li, J.; Gu, M. Poor Stability of Li2CO3 in the Solid Electrolyte Interphase of a Lithium-Metal Anode Revealed by Cryo-Electron Microscopy. *Adv. Mater.* **2021**, *33*, 2100404.
(33) Zhao, Q.; Stalin, S.; Archer, L. A. Stabilizing Metal Battery Anodes through the Design of Solid Electrolyte Interphases. *Joule* **2021**, *5*, 1119–1142.
(34) Zhan, J.; Deng, L.; Liu, Y.; Hao, M.; Wang, Z.; Dong, L.; Yang, Y.; Song, K.; Qi, D.; Wang, J.; Wang, S.; Liu, H.; Zhou, W.; Chen, H. Self-Selective (220) Directional Grown Copper Current Collector Design for Cycling-Stable Anode-Less Lithium Metal Batteries. *Adv. Mater.* **2025**, *37*, 2413420.
(35) Wondimkun, Z. T.; Tegegne, W. A.; Jiang, S.-K.; Huang, C.-J.; Sahalie, N. A.; Weret, M. A.; Hsu, J.-Y.; Hsieh, P.-L.; Huang, Y.-S.; Wu, S.-H.; Su, W.-N.; Hwang, B. J. Highly-Lithiophilic Ag@PDA-GO Film to Suppress Dendrite Formation on Cu Substrate in Anode-Free Lithium Metal Batteries. *Energy Storage Mater.* **2021**, *35*, 334–344.
(36) Oyakhire, S. T.; Zhang, W.; Shin, A.; Xu, R.; Boyle, D. T.; Yu, Z.; Ye, Y.; Yang, Y.; Raiford, J. A.; Huang, W.; Schneider, J. R.; Cui, Y.; Bent, S. F. Electrical Resistance of the Current Collector Controls Lithium Morphology. *Nat. Commun.* **2022**, *13*, 3986.
(37) Hagos, T. T.; Thirumalraj, B.; Huang, C.-J.; Abrha, L. H.; Hagos, T. M.; Berhe, G. B.; Bezabh, H. K.; Cherng, J.; Chiu, S.-F.; Su, W.-N.; Hwang, B.-J. Locally Concentrated LiPF6 in a Carbonate-Based Electrolyte with Fluoroethylene Carbonate As a Diluent for Anode-Free Lithium Metal Batteries. *ACS Appl. Mater. Interfaces* **2019**, *11*, 9955–9963.
(38) Zhang, S. S.; Fan, X.; Wang, C. A Tin-Plated Copper Substrate for Efficient Cycling of Lithium Metal in an Anode-Free Rechargeable Lithium Battery. *Electrochim. Acta* **2017**, *258*, 1201–1207.
(39) Wondimkun, Z. T.; Beyene, T. T.; Weret, M. A.; Sahalie, N. A.; Huang, C.-J.; Thirumalraj, B.; Jote, B. A.; Wang, D.; Su, W.-N.; Wang, C.-H.; Brunklaus, G.; Winter, M.; Hwang, B.-J. Binder-Free Ultra-Thin Graphene Oxide As an Artificial Solid Electrolyte Interphase for Anode-Free Rechargeable Lithium Metal Batteries. *J. Power Sources* **2020**, *450*, 227589.
(40) Li, Z.; Huang, X.; Kong, L.; Qin, N.; Wang, Z.; Yin, L.; Li, Y.; Gan, Q.; Liao, K.; Gu, S.; Zhang, T.; Huang, H.; Wang, L.; Luo, G.; Cheng, X.; Lu, Z. Gradient Nano-Recipes to Guide Lithium Deposition in a Tunable Reservoir for Anode-Free Batteries. *Energy Storage Mater.* **2022**, *45*, 40–47.
(41) Kang, T.; Zhao, J.; Guo, F.; Zheng, L.; Mao, Y.; Wang, C.; Zhao, Y.; Zhu, J.; Qiu,

Y.; Shen, Y.; Chen, L. Dendrite-Free Lithium Anodes Enabled by a Commonly Used Copper Antirusting Agent. *ACS Appl. Mater. Interfaces* **2020**, *12*, 8168–8175.
(42) Tamwattana, O.; Park, H.; Kim, J.; Hwang, I.; Yoon, G.; Hwang, T.; Kang, Y.-S.; Park, J.; Meethong, N.; Kang, K. High-Dielectric Polymer Coating for Uniform Lithium Deposition in Anode-Free Lithium Batteries. *ACS Energy Lett.* **2021**, *6*, 4416–4425.
(43) Chen, W.; Salvatierra, R. V.; Ren, M.; Chen, J.; Stanford, M. G.; Tour, J. M. Laser-Induced Silicon Oxide for Anode-Free Lithium Metal Batteries. *Adv. Mater.* **2020**, *32*, 2002850.
(44) Assegie, A. A.; Chung, C.-C.; Tsai, M.-C.; Su, W.-N.; Chen, C.-W.; Hwang, B.-J. Multilayer-Graphene-Stabilized Lithium Deposition for Anode-Free Lithium-Metal Batteries. *Nanoscale* **2019**, *11*, 2710–2720.
(45) Yang, Z.; Liu, W.; Chen, Q.; Wang, X.; Zhang, W.; Zhang, Q.; Zuo, J.; Yao, Y.; Gu, X.; Si, K.; Liu, K.; Wang, J.; Gong, Y. Ultrasmooth and Dense Lithium Deposition toward High-Performance Lithium-Metal Batteries. *Adv. Mater.* **2023**, *35*, 2210130.
(46) Zhang, Q.-K.; Zhang, X.-Q.; Wan, J.; Yao, N.; Song, T.-L.; Xie, J.; Hou, L.-P.; Zhou, M.-Y.; Chen, X.; Li, B.-Q.; Wen, R.; Peng, H.-J.; Zhang, Q.; Huang, J.-Q. Homogeneous and Mechanically Stable Solid–Electrolyte Interphase Enabled by Trioxane-Modulated Electrolytes for Lithium Metal Batteries. *Nat. Energy* **2023**, *8*, 725–735.

# Supplementary Materials

# Subsurface Vacancy Engineering Enables Atomically Clean and Oxidation-Resistant Copper Interfaces for Anode-Free Lithium Metal Batteries

*Yue Li[1, 2]†, Xuanguang Ren[1, 3]†, Xueting Feng[3]†, Lingcheng Kong[1]†, Fengping Luo[2], Yang Xu[2], Liu Qian[1], Yusheng Ye[4]*, Ziqiang Zhao[1, 2]*, Xin Gao[1]*, Jin Zhang[1, 3]**

[1] School of Materials Science and Engineering, Peking University, Beijing, 100871, China.
[2] State Key Laboratory of Nuclear Physics and Technology, School of Physics, Peking University, Beijing, 100871, China.
[3] Beijing National Laboratory for Molecular Sciences, College of Chemistry and Molecular Engineering, Peking University, Beijing, 100871, China.
[4] Beijing Key Laboratory of Environmental Science and Engineering, School of Materials Science and Engineering, Beijing Institute of Technology, Beijing, 100081, China.

†These authors contributed equally to this work.

* Correspondence to: Y. Y., ysye@bit.edu.cn; Z. Z., zqzhao@pku.edu.cn; X. G., gaoxin-cnc@pku.edu.cn; J. Z., jinzhang@pku.edu.cn.

**The PDF file includes:**

## Methods

### Preparation of Ion-Cu

Ion-Cu was prepared by implanting Cu ions into commercial Cu foils (MSK-WTMC-CU2G) at room temperature. The implantation was carried out using a FAD-MEVVA ion implanter, which delivers a uniform beam spot exceeding 8 inches in diameter, as verified by Gafchromic EBT3 self-developing dosimetry film.[1] This beam uniformity allowed for the simultaneous implantation of Cu foils sufficient to fabricate over 300 battery current collectors used in this work. The ion fluence was controlled within a range from 1E13 to 1E16 ions $cm^{-2}$, and the ion energies were adjusted between 5 to 20 keV.

The optimized conditions used for all electrochemical measurements reported in this work are an ion energy of 9 keV and an ion fluence of $1\times10^{16}$ ions $cm^{-2}$. Lower fluences ($<1\times10^{14}$ ions $cm^{-2}$) resulted in incomplete removal of the native oxide, while further increases beyond the optimized fluence did not yield significant additional improvements, indicating effective saturation of the vacancy generation and oxide removal effects.

### Preparation of Acid-Treated Cu

The acid treatment was performed by immersing commercial Cu foils in 1.0 M $CH_3COOH$ solution for 10 minutes at room temperature, followed by rinsing with deionized water and ethanol, and drying under argon flow. The treated foils were immediately transferred to an argon-filled glovebox to minimize air exposure before subsequent characterization or cell assembly.

### Electrochemical performance test

All the batteries are assembled using the 2032-type coin cell with Celgard 2325 separator and disassembled in an argon-filled glovebox (the content of $O_2$ and $H_2O$ is maintained below 0.01 ppm). Half-cells are assembled with pure Li foil as counter electrode and Cu foil as working electrodes. In the experiment, a single layer of Celgard 2325 separator is used. A volume of 40 μL of electrolyte is subsequently injected. Following assembly, the battery is allowed to rest for a period of 2 hours. For EIS measurements, BioLogic is used, and the impedance curve is generated by scanning from a frequency of 100 mHz to 200 KHz. The CEs of the electrolytes were measured based on Aurbach method in Li||Cu cells, where 5 mAh $cm^{-2}$ of Li was first deposited onto the Cu foil as Li reservoir. This was followed by 10 subsequent cycles of plating and stripping at 0.5 mA $cm^{-2}$ for 1 mAh $cm^{-2}$. Finally, all deposited Li was stripped from Cu, and the total capacity recovered was divided by the amount deposited to obtain the CE. For the cycling process, the areal capacity for lithium plating is controlled at a constant rate, while the cut off potential for Li stripping is controlled at 1V. The SEI was formed on CuCCs using a constant current–constant-voltage procedure. The current was applied at 100 μA $cm^{-2}$ until the voltage reached 0 mV vs Li/$Li^+$. The AFLMB (Ion-Cu ||10 mg $cm^{-2}$ NCM523) and AFLMB pouch cells (Cu collector||10mg $cm^{-2}$ NCM523) are charged to 4.3 V at 0.05C and then discharged to 3.0 V at 0.05C (1C=165 mA $g^{-1}$) in the first cycles. After the above activation processes, the cell is cycled at 0.1C/0.2C for long cycling.

**Preparation of electrolytes**

1 M LiFSI in DME, 1 M LiTFSI in DOL/DME (1:1 by volume) with 5 wt.% $LiNO_3$, 1 M $LiPF_6$ in EC/DEC (1:1 by volume) with 10% FEC and 1% VC, 0.6 M LiDFOB ,0.6 M $LiBF_4$ in FEC/DEC (2:1 by volume) and 1 M $LiPF_6$ in EC/DEC (1:1 by volume) were purchased from DuoDuoChem. LiFSI was purchased from Sigma-Aldrich. DME and TTE were purchased from Alfa Aesar and they were dried by adequate 3 Å molecular sieve for more than 48 h before use. Unless otherwise stated, the electrolyte used is 1 M LiTFSI in DOL/DME (1:1 by volume) with 5 wt.% $LiNO_3$.

**Characterizations**

The morphology of the materials is examined by scanning electron microscopy (SEM, Hitachi S-4800). X-ray photoelectron spectroscopy (XPS) were measured on an AXIS Ultra photoelectron spectrometer using monochromatic Al Kα radiation (Kratos Analytical). CuCCs are rinsed with pure anhydrous DME solvent to remove residual LiTFSI, dried, and then sealed in the glovebox before being transferred using a vacuum transfer vessel for SEI characterizations. The sputtering process helped to remove the characterized surface from the material layer by layer and exhibited the inner layer. The binding energies are calibrated with respect to the C 1s peak at 284.6 eV. X-ray diffraction (XRD) patterns are examined by Rigaku SmartLab using Cu Kα radiation. Electron backscattered diffraction (EBSD) are examined by eFlash HR.

**Computational method**

**Task 1: Ion implantation effects on Cu foils.**

We conducted parameter-passing multiscale simulations, with the fundamental procedures illustrated in Supplementary Figure 7. Monte Carlo simulations (as implemented in the TRIM code)[2] were performed to quantify ion implantation-induced damage (displacements per atom, dpa/s), and density functional theory (DFT) calculations were used to obtain the energetics for the interaction between vacancies and oxygen atoms. These results were integrated into a cluster dynamics (CD) model to simulate the evolution of vacancy clusters and lattice oxygen concentrations. Details are as follows:

The fundamental defect parameters necessitated in cluster dynamics simulations include defect diffusion barriers and defect binding energies. The diffusion barriers of an interstitial, a vacancy, and an oxygen atom are set as 0.08 eV, 0.66 eV, and 0.50 eV.[3, 4] The binding energies of interstitial–interstitial and vacancy–vacancy clusters were obtained from literature,[4,5] with the latter extrapolated to larger sizes according to the capillary law.[6] To obtain the interaction energetics between vacancies and oxygen atoms, we performed DFT calculations on 20 structures varying in vacancy (1-4) and oxygen atom (0-4) to obtain the fitted functions. DFT calculations in this task were performed using the Vienna Ab initio Simulation Package (VASP).[7] The projector augmented wave (PAW)[8] pseudopotential with the PBE[9] generalized gradient approximation (GGA) exchange correlation function was utilized in the computations. The cutoff energy of the plane waves basis set was 500 eV and Monkhorst-Pack mesh of 3×3×3 was used in K-sampling in the adsorption energy calculation. All structures were spin polarized and all atoms and cell parameters were fully relaxed with the energy convergence tolerance of $10^{-5}$ eV per atom, and the final force on each atom was < 0.02 eV Å$^{-1}$. Using the above method, the binding energy of oxygen with existing vacancy clusters ($O_1$+$O_{x-1}V_y$=$O_xV_y$), with x=1-4 and y=1-4, as a function of the O/V ratio (eV) were obtained. The fitted relationship is expressed in Equation 1:

$$\Delta E_b^O(x, y) = A + B\left(\frac{x}{y}\right)^C, A = 2.535, B = -1.085, C = 0.581$$

Oxygen binding also enhances the stability of vacancy clusters. The binding energy increment for $V_1$+$O_xV_{y-1}$=$O_xV_y$ is similarly modeled as a function of the O/V ratio, expressed in Equation 2:

$$\Delta E_b^V(x, y) = D\left(\frac{x}{y}\right)^E, D = 0.785, E = 1.297$$

Finally, using results from TRIM and DFT, cluster dynamics simulations were conducted under experimental parameters.

In summary, the key simulation parameters are: initial lattice oxygen concentration of $10^{-5}$, surface oxygen permeation rate $P_O = 1\times10^{-9}$ s$^{-1}$ (calibrated to the experimentally observed ~90-day re-oxidation onset), and simulation temperature T = 300 K.

**Task 2: Li deposition behavior on modified Cu surfaces**

In this task, first-principles calculations based on the Density Functional Theory (DFT) has been performed using the code CASTEP.[10,11] During our calculations, Perdew-Burke-Ernzerhof (PBE) exchange-correlation functional[12] within the generalized gradient approximation (GGA) was employed, together with the use of ultrasoft pseudopotentials. During our calculations, the energy cut off for the plane wave basis expansion was set to 450 eV. The convergence criteria for total energy and maximum force were set as $10^{-4}$ eV and 0.03 eV/Å, respectively. The sampling in the Brillouin zone for all surface slabs was set actual spacing larger than 0.05 /Å, and 4×4×4 by the Monkhorst-Pack method[13] for bulk Li crystal. The van der Waals interaction has been considered using the Grimme dispersion scheme.[14] All surface calculations employed p(3×3) supercells with slab thickness of four or more atomic layers, with the bottom four layers fixed during geometry optimization.

To simulate Li-adsorption and diffusion on Cu surfaces, slab models for (110), (100) and (111) surfaces have been cleaved from ideal copper crystal and simulated by p (3×3) supercells, followed by full geometry relaxation, generating S(110), S(100) and S(111). Based on optimized surfaces, amorphous copper oxides have been model through random incorporation of oxygen atoms with an atomic ratio Cu:O = 1:1. To release strong strain introduced by oxygen, three steps have been carried out: (i) geometry optimization has been performed, with fixed dimension in a&b direction, which ensures the Li-coverage on crystal and amorphous surfaces is same; (ii) molecular dynamics (MD) simulation at the temperature T=300 K; and (iii) MD-obtained geometry has been further optimized until the force converged to 0.03 eV/Å, generating AS(110), AS(100) and AS(111). Step (i) and (ii) were performed using density-functional tight-binding (DFTB) approach,[15] as described below with more details. Li-adsorption has been investigated with single Li loaded on the above surfaces, with bottom four layers fixed during the calculations. Favorable adsorption site has been determined by two steps: (i) scanning potential adsorption with different initial geometries, followed by full relaxation; and (ii) two most favorable geometries have been used to calculate the adsorption energy $E_{ads}$ by $E_{ads}$ = E(Li-S)-E(Li)-E(S), where E(Li), E(S) and E(Li-S) are calculated total energies for single Li derived from bulk Li crystal, surface slab and Li adsorbed on the surface. Li-diffusion has been investigated through the searching of transition state when Li diffuses from most favorable (initial state, IS) to the 2nd favorable (final state, FS), generating the diffusion barrier $E_a$ = E(TS)-E(IS), where E(TS) and E(IS) are calculated total energies for TS and IS. TS searching is based on the protocol with the complete LST/QST method,[16] where LST and QST refer to linear and quadratic synchronous transition.

To further examine the dynamics of Li-diffusion from Cu surface to Li-layers, a cluster with layer step has been introduced to Cu surfaces, with a target Li-atom initially being bonded with Cu substrate but transferring to Li-step as the final state. TS has been further searched to illustrate the thermodynamics and kinetics associated with Li-clustering from full contact with Cu substrate.

Li behavior has been further investigated with 36 Li atoms being distributed over Cu or amorphous (CuO)x surfaces, focusing on the comparison of flat Li-layers (M1) and cluster morphologies (M2). Energy difference between M1 and M2 has been used to illustrate the energy preference for Li-distribution after full optimization. Moreover, ab initial molecular dynamics (AIMD) has been carried out at T=300 K with time step Δt=2.0 fs, simulating for 3.0 ps. After reaching the equilibrium state, we collected the energy profile for 1.0 ps for comparison. Interface interaction has been analyzed using bulk Li (bcc-phase, with two Li-atoms in the unit cell) and initial clean surfaces (S) as the reference by calculating the specific binding energy SBE = {E(Li36@S)-E(S)-(36/2) E(bulk Li)}/SA, where SA is the surface area of clear surfaces. Based on such definition, negative SBE indicates such interface is favorable.

DFTB calculations: Our calculations were carried out using the DFTB+ program, using Slater-Koster library, as embedded in Materials Studio suite. Specifically, the Coulombic interaction between partial atomic charges was determined under the scheme of the self-consistent charge formalism. During the optimization, the total energy and force convergence are set as $4.0\times10^{-4}$ eV and 0.04 eV/ Å. To generate amorphous layers, MD simulation was carried out with the temperature increased by three steps: (i) staring with T=100 K, MD simulation runs for total time of 2 ps; (ii) based on (i), another run with T=200 K has been performed for 2 ps; and (iii) based on (ii), full MD at T=300 K has been carried out for 10 ps. All final geometries have been examined with the calculations of radical distribution function, confirming that no long-ranged ordered structure as requested by amorphous nature.

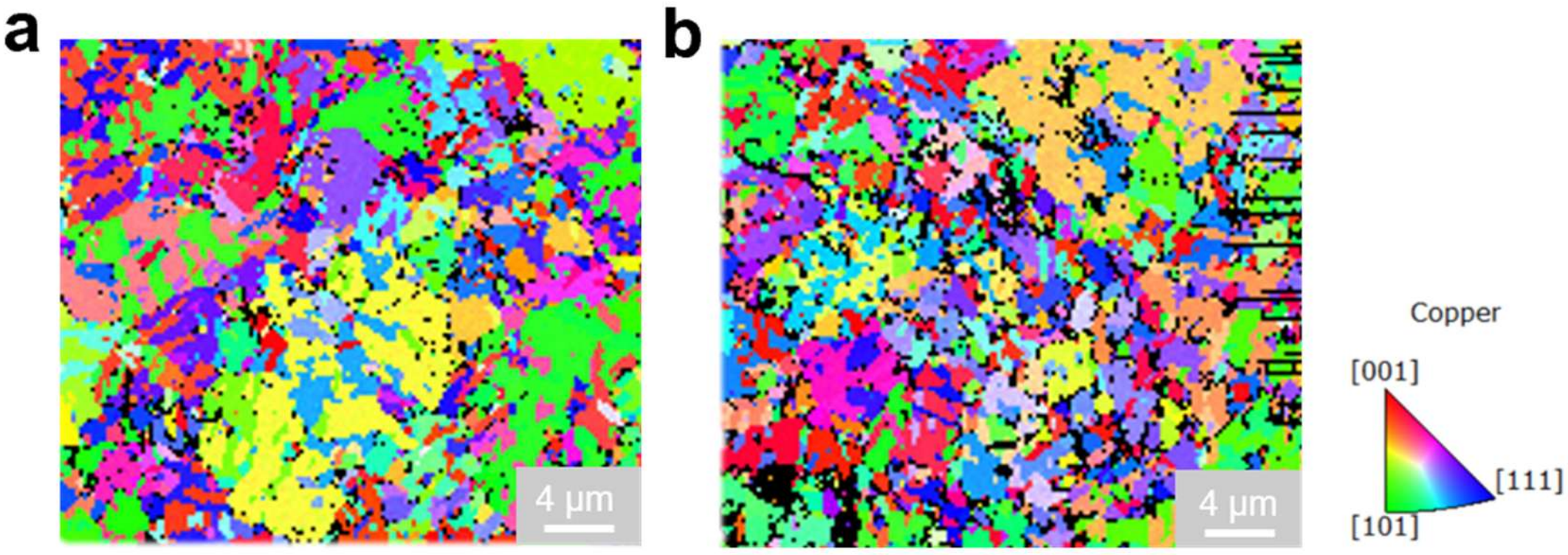

**Supplementary Figure 1.** EBSD spectra of Pristine-Cu (a) and Ion-Cu (b).

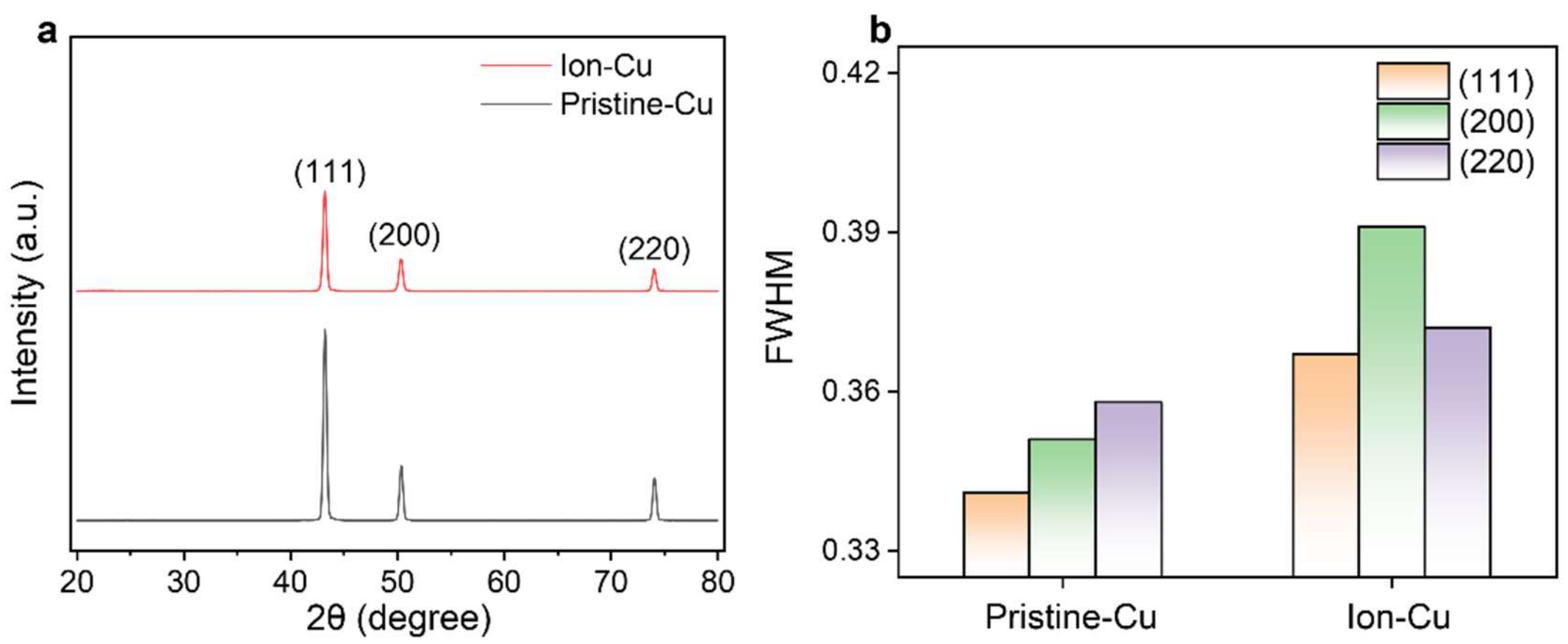

**Supplementary Figure 2.** Grazing incidence X-ray diffraction (GIXRD) analysis of Pristine-Cu and Ion-Cu. **a** GIXRD patterns of Pristine-Cu and Ion-Cu, measured at an incident angle of 0.3°, showing identical crystallographic orientations. **b** Statistical comparison of full width at half maximum (FWHM) for Cu(111), Cu(200), and Cu(220) peaks.

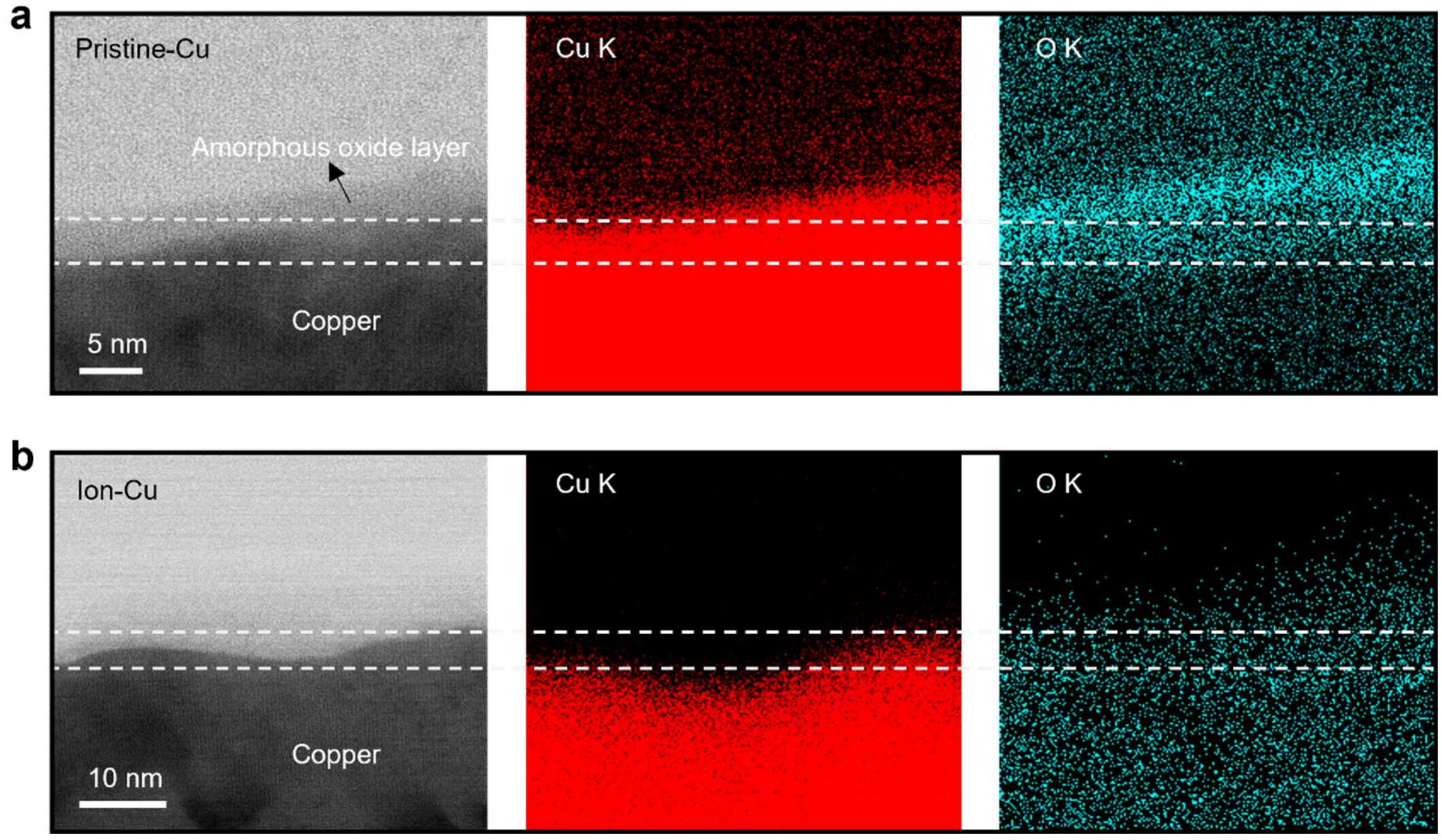

**Supplementary Figure 3.** Cross-sectional High Resolution Transmission Electron Microscopy (HRTEM) and Energy Dispersive Spectrometer (EDS) images of Pristine-Cu (a) and Ion-Cu (b).

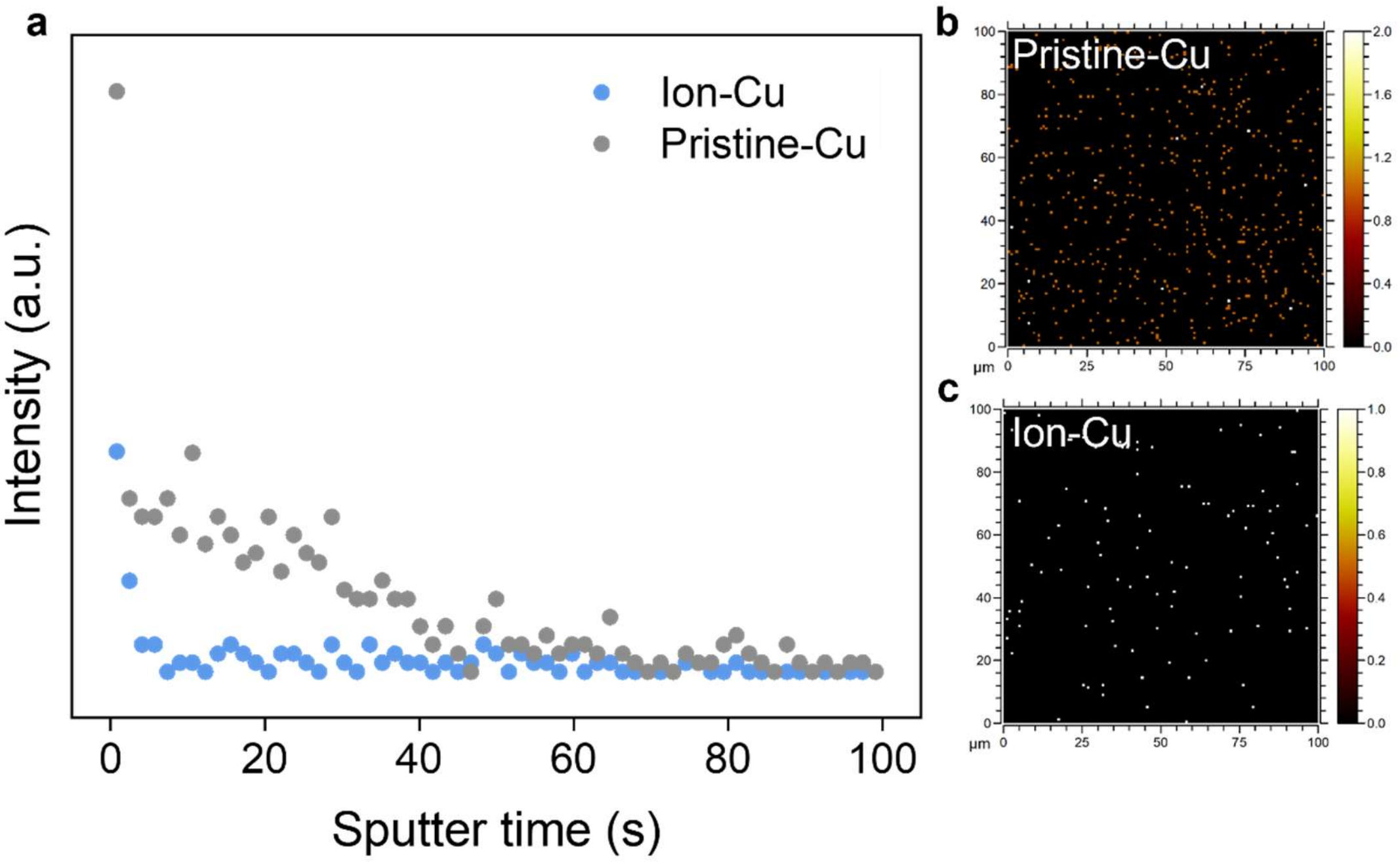

**Supplementary Figure 4.** TOF-SIMS of Ion-Cu and Pristine-Cu. **a** Intensity variation of $Cu_2(OH)_2CO_3$. **b-c** 2D planar mass spectrometry of $Cu_2(OH)_2CO_3$.

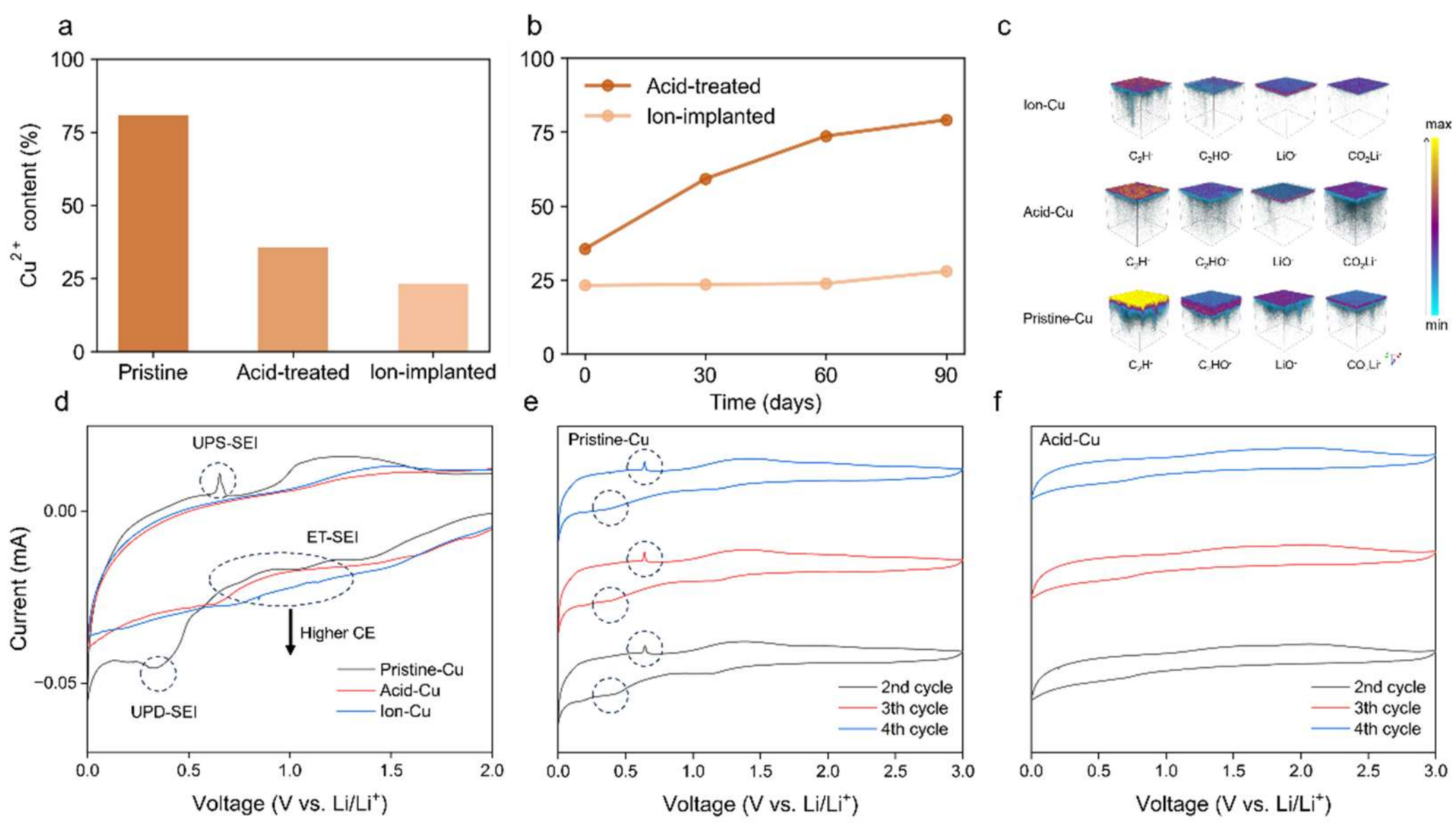

**Supplementary Figure 5.** Characterization of surface chemistry, SEI composition, and electrochemical behavior of Cu foils with different surface treatments. **a** Comparison of oxidation levels in pristine Cu, acid-treated Cu, and ion-implanted Cu based on XPS spectra analysis. **b** Long-term oxidation behavior of acid-treated and ion-implanted Cu foils (Ion-Cu). **c** 3D-rendered TOF-SIMS images of the SEI on Pristine-Cu, Acid-Cu, and Ion-Cu. **d-f** Cyclic voltammetry (CV) characterization of the effect of surface oxide on lithium underpotential deposition/stripping behavior on Cu foils. **(d)** First-cycle CV curves of Pristine-Cu, Acid-Cu, and Ion-Cu foils. **(e)** Multicycle CV curves of Pristine-Cu foil. **(f)** Multicycle CV curves of Acid-Cu foil. The lithium UPS peak persists in the 2nd, 3rd, and 4th cycles, demonstrating that the native oxide layer continuously disrupts Li deposition and stripping during long-term cycling.

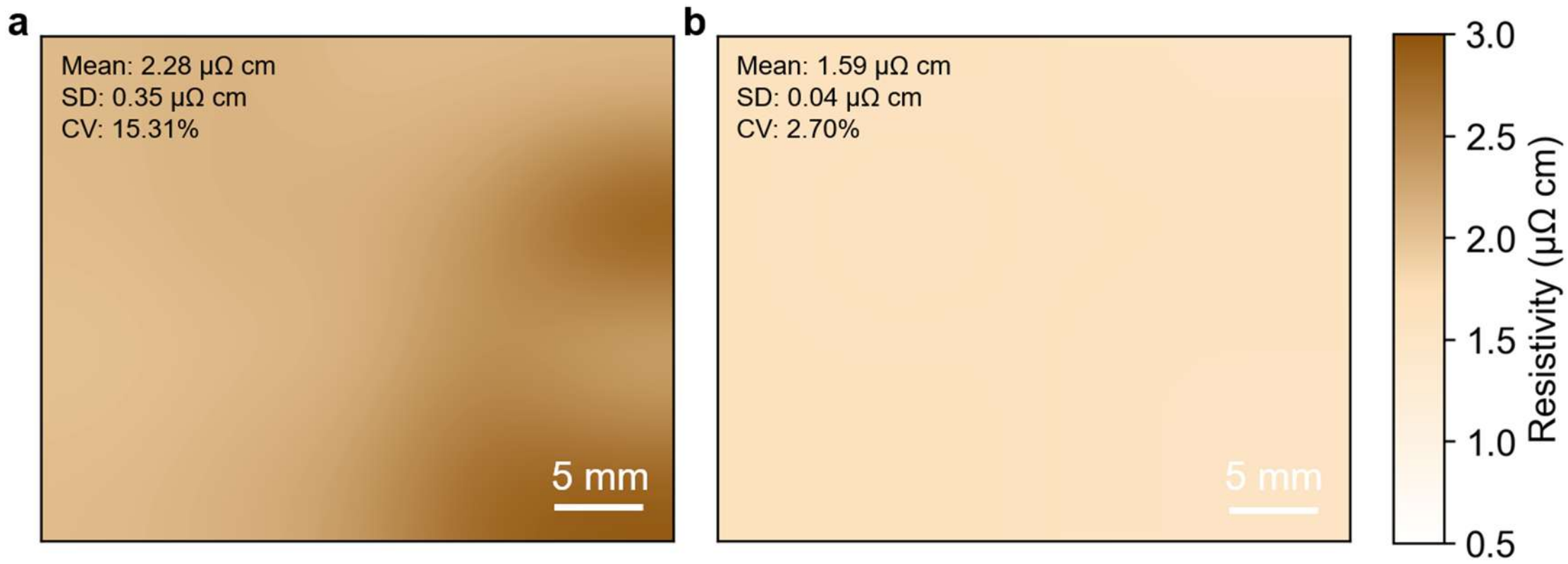

**Supplementary Figure 6.** Surface resistivity distribution map of Pristine-Cu (a) and Ion-Cu (b).

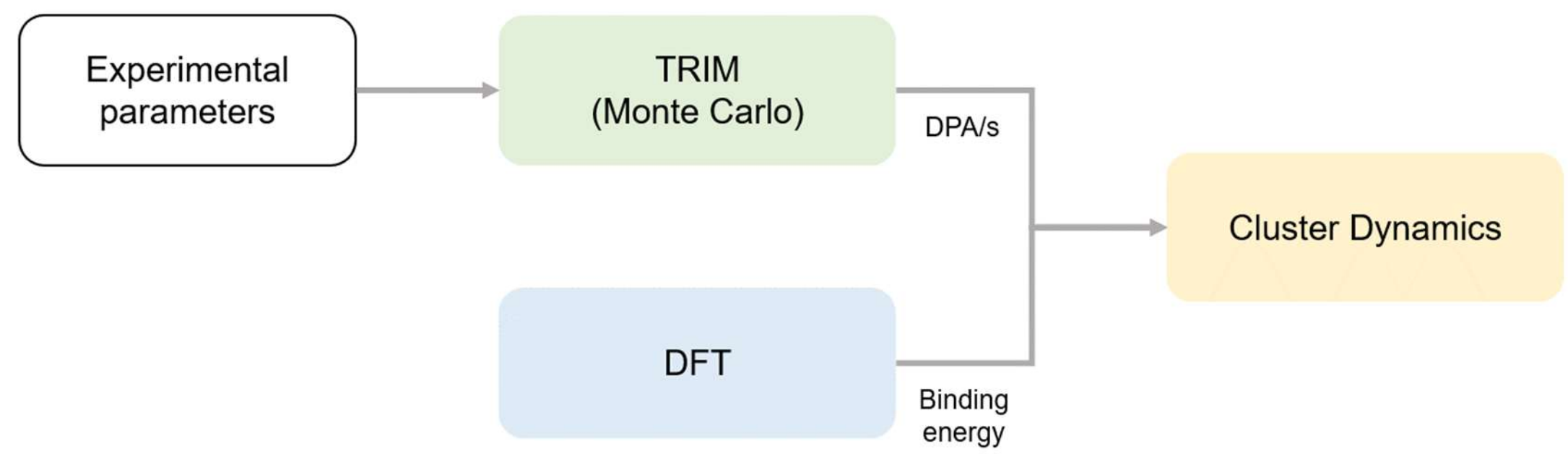

**Supplementary Figure 7.** Schematic summary of the modelling strategies.

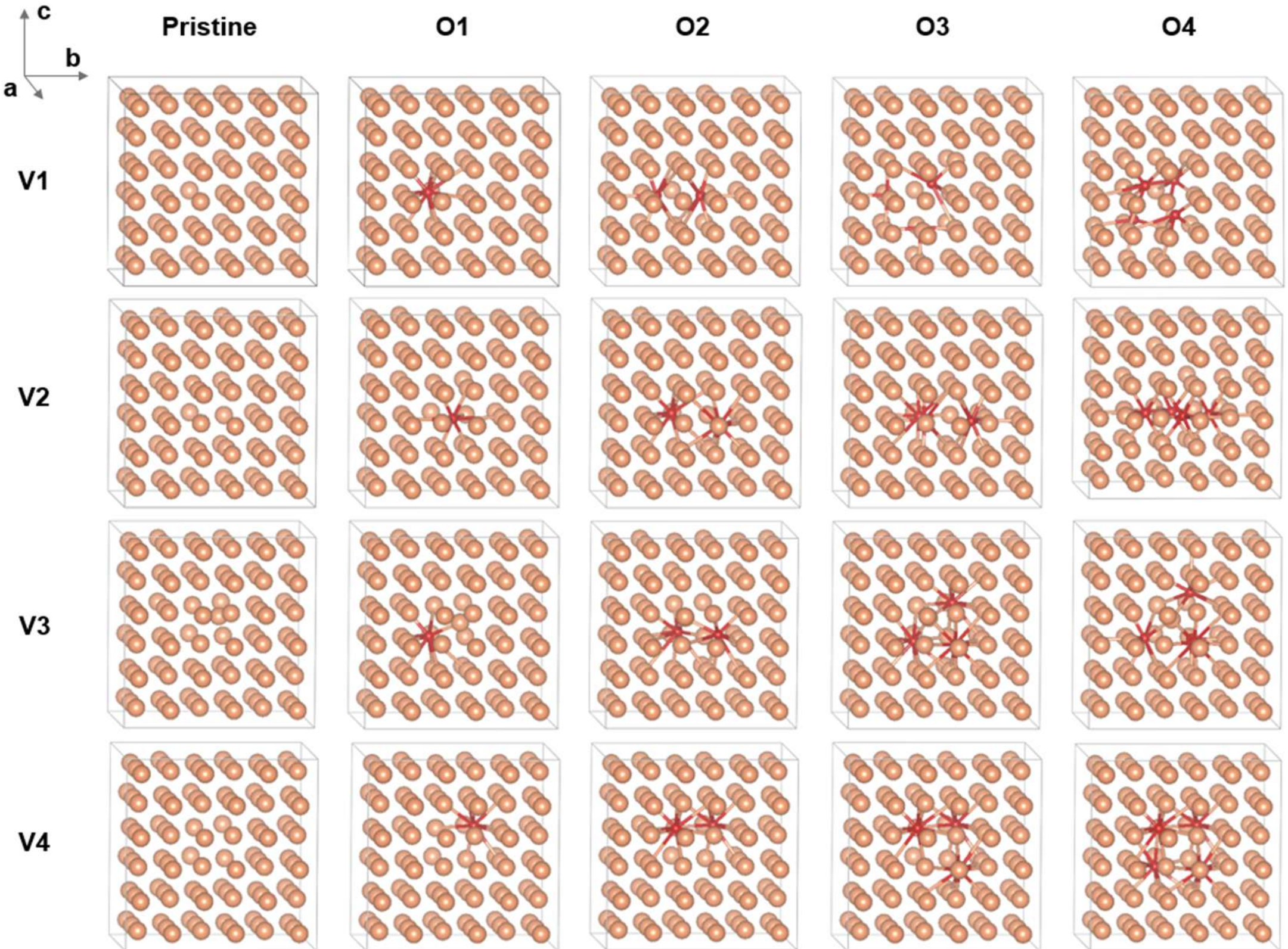

**Supplementary Figure 8.** Model construction of DFT calculations with varying vacancy and oxygen atom numbers.

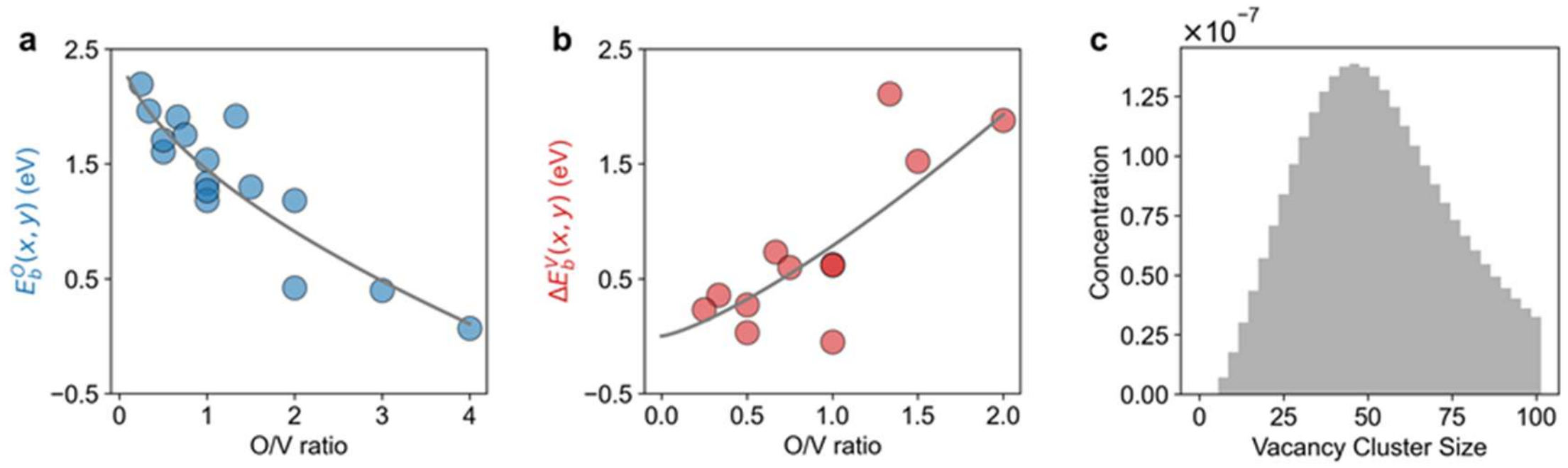

**Supplementary Figure 9.** DFT-calculated oxygen binding behavior and cluster dynamics simulation results. **a** Binding energy of a single oxygen atom with pre-existing vacancy clusters as a function of the O/V ratio. **b** Binding energy of a vacancy to an oxygen-containing cluster as a function of the O/V ratio. **c** Simulated size distribution of vacancy clusters based on cluster dynamics calculations under experimental ion implantation conditions.

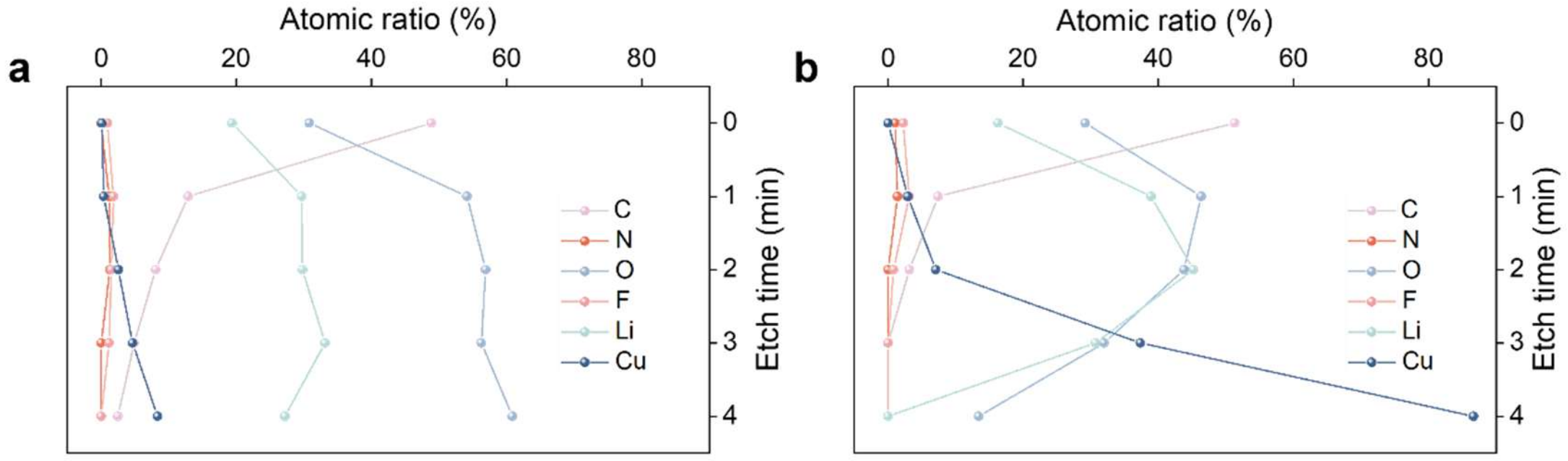

**Supplementary Figure 10.** XPS depth profiles with atomic concentration ratio of C, N, O, F, Li and Cu elements on Pristine-Cu (a) and Ion-Cu (b).

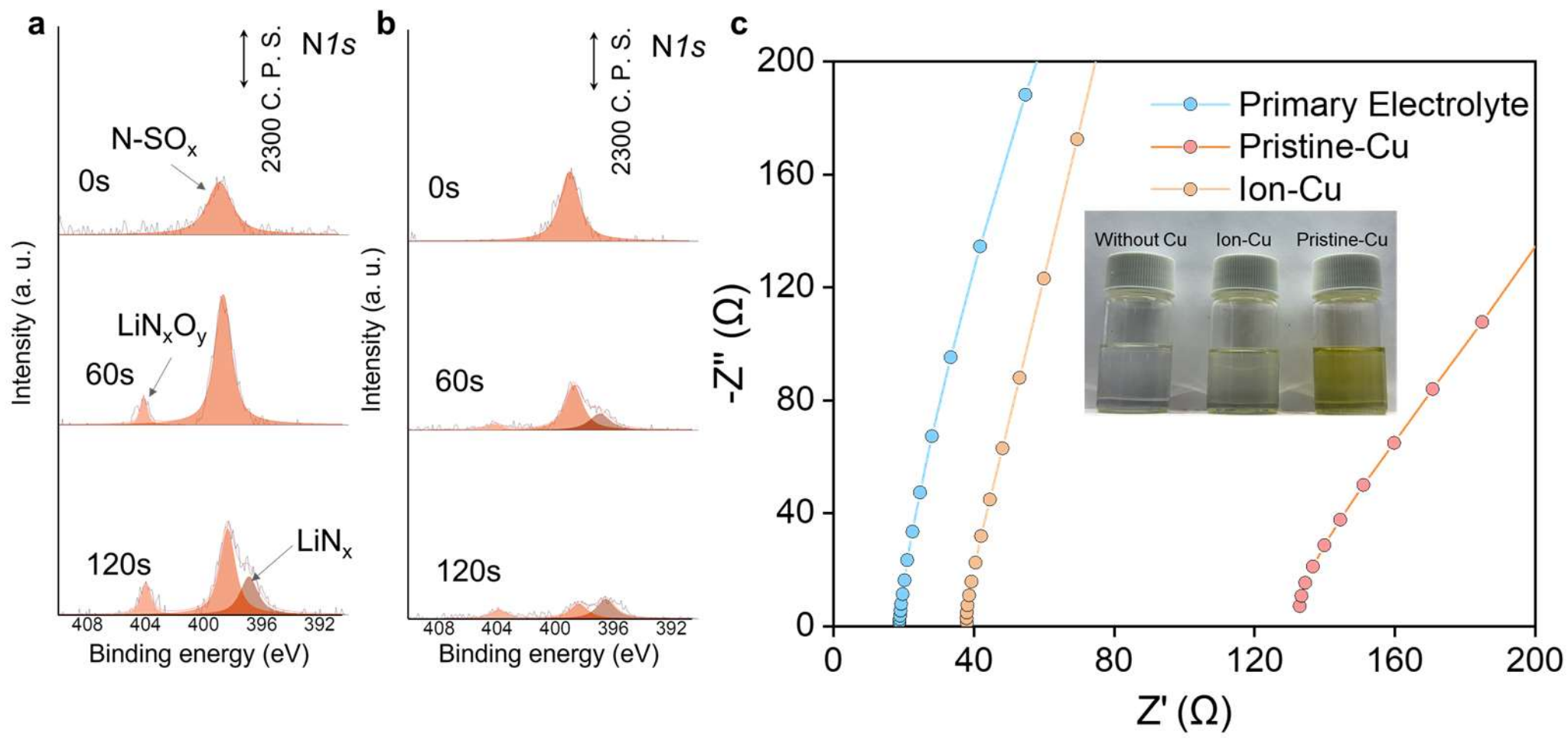

**Supplementary Figure 11.** Electrolyte Component Decomposition. **a-b** N1s XPS depth profile analysis of SEI components on Ion-Cu (a) and Pristine-Cu (b). **c** Color changes and Nyquist plots of electrolyte (1.0 M LiTFSI in a 1:1 vol/vol mixture of DOL/DME with 5% $LiNO_3$) solutions containing Pristine-Cu and Ion-Cu after immersion for 12 hours.

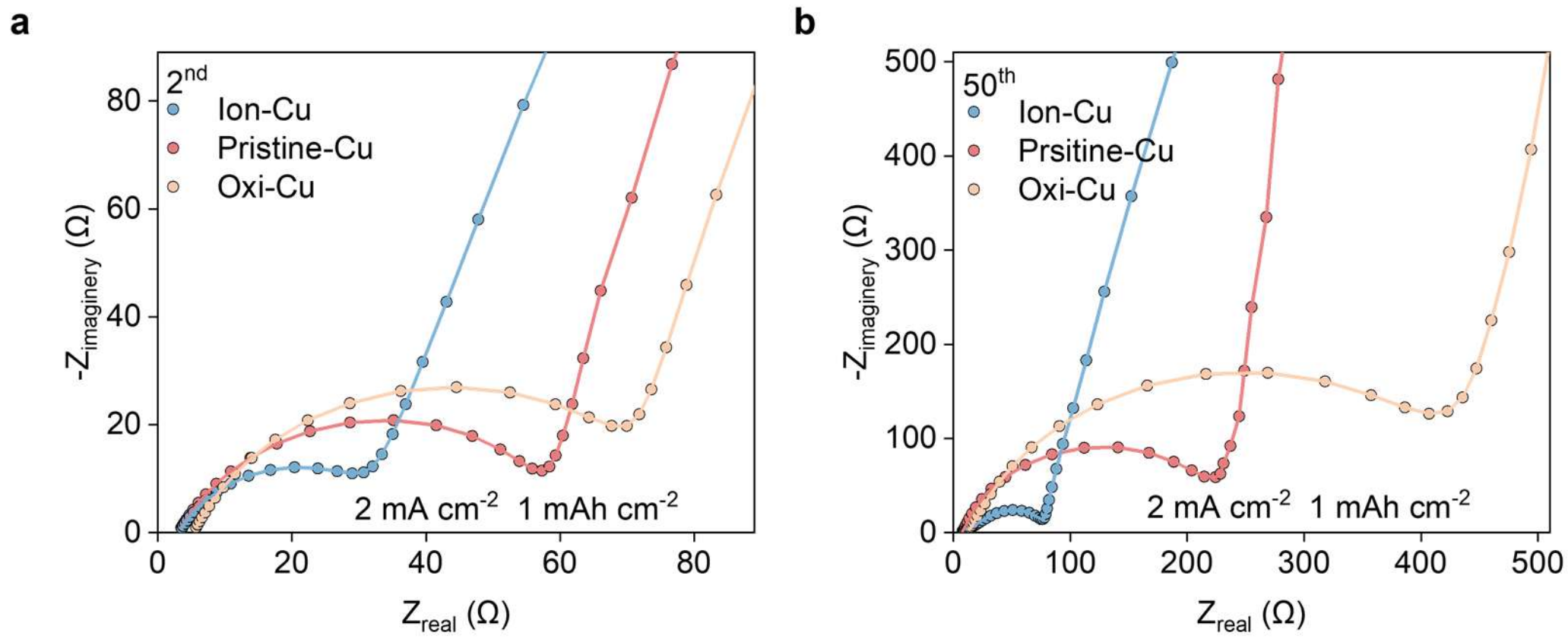

**Supplementary Figure 12. a-b** Nyquist plots of Li||Cu half-cells with different CuCCs at the 2nd (a) and 50th (b) cycle.

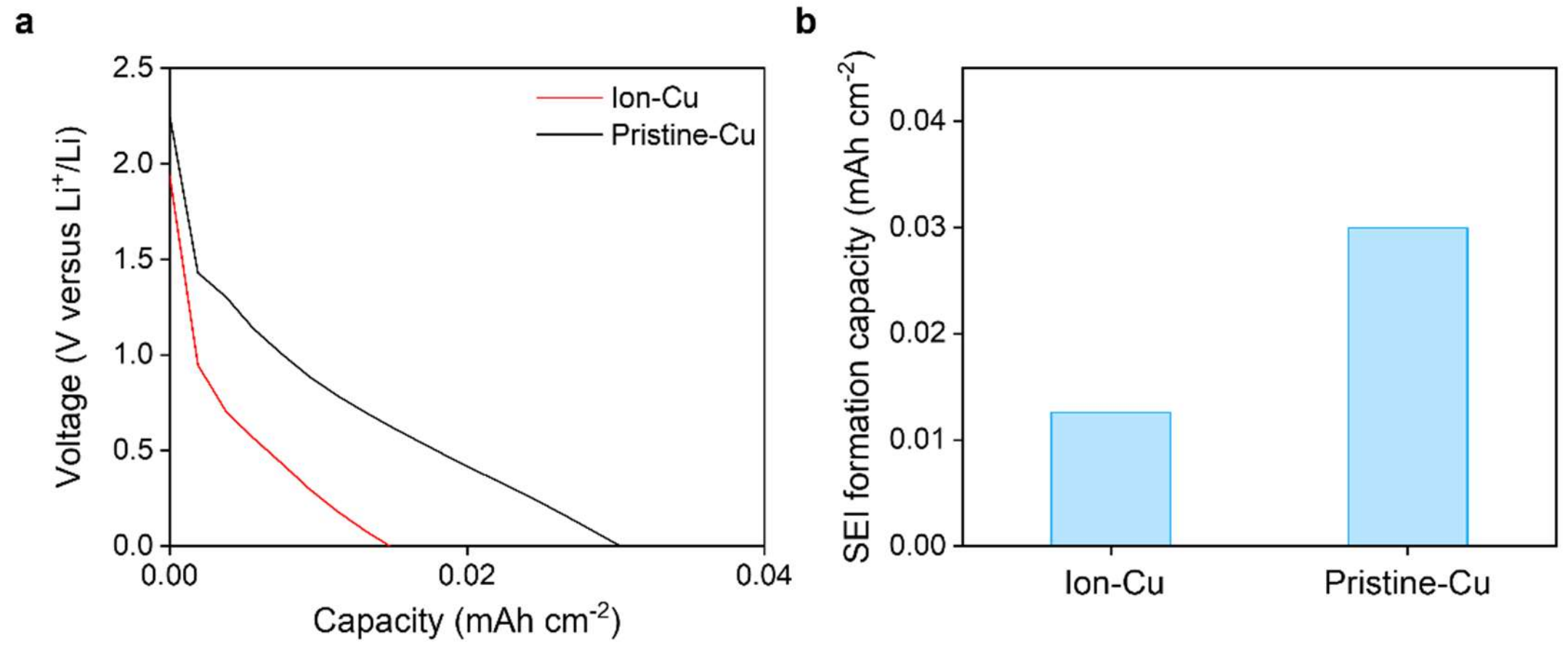

**Supplementary Figure 13.** Differences in SEI formation capacity between Ion-Cu and Pristine-Cu. **a** The initial galvanostatic discharge curves of different CuCCs at 0.5 mA $cm^{-2}$. **b** Capacity of SEI formed by Ion-Cu and Pristine-Cu at initial deposition potential of 0 V vs. Li.

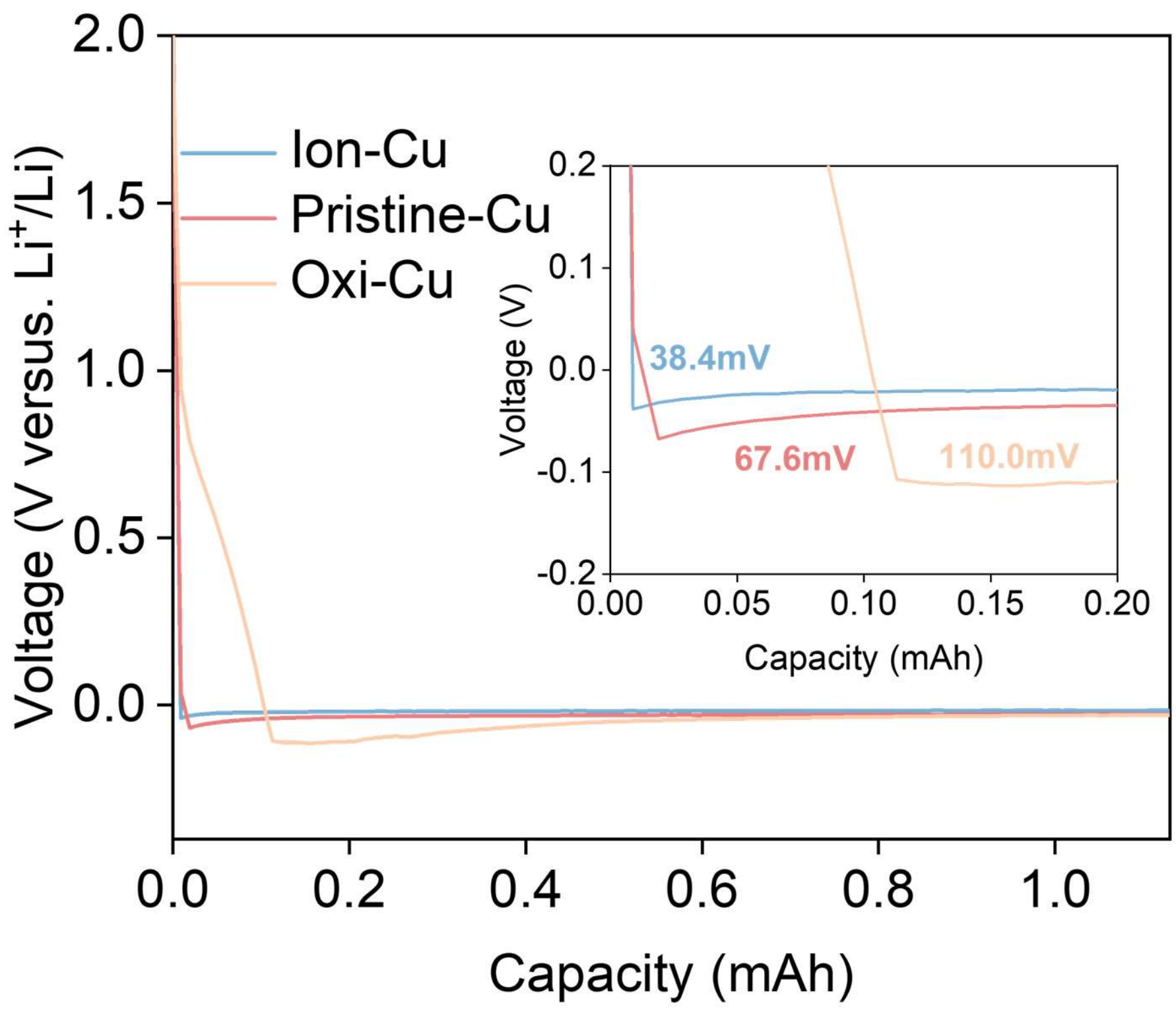

**Supplementary Figure 14.** Initial galvanostatic discharge curves of different CuCCs at 0.5 mA cm$^{-2}$.

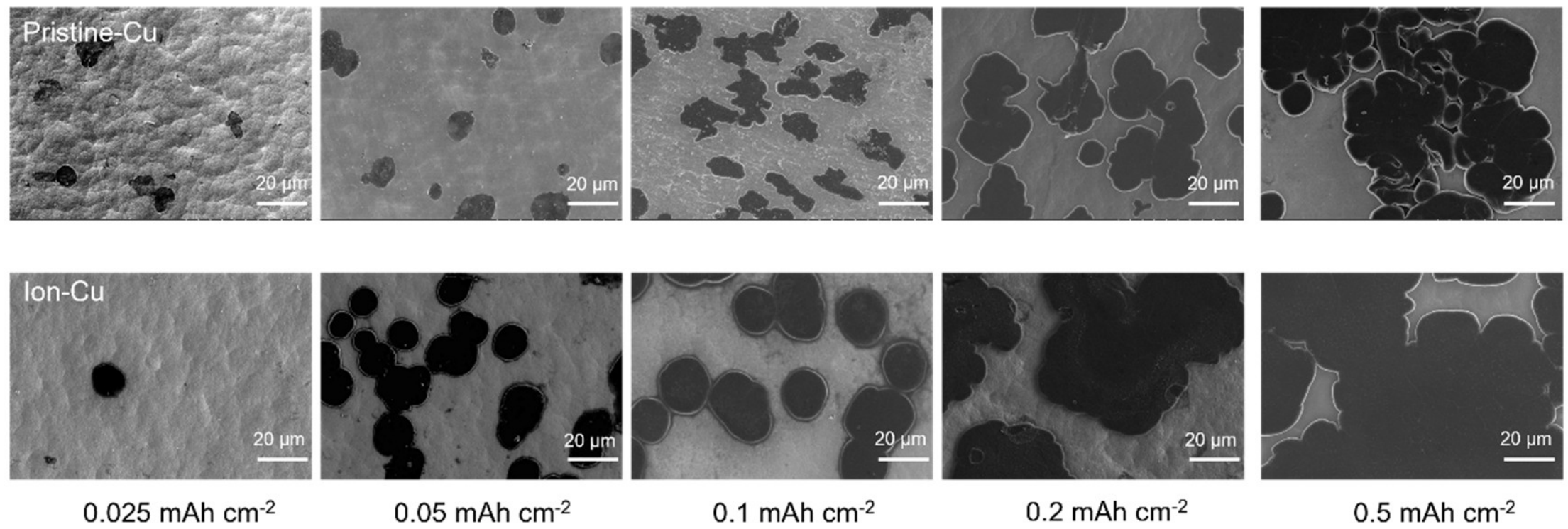

**Supplementary Figure 15.** SEM characterization of deposited Li. Morphological evolution of Li deposited on Pristine-Cu and Ion-Cu at 0.1 mA $cm^{-2}$ in an ether-based electrolyte (1.0 M LiTFSI in a 1:1 vol/vol mixture of DOL/DME with 5% $LiNO_3$).

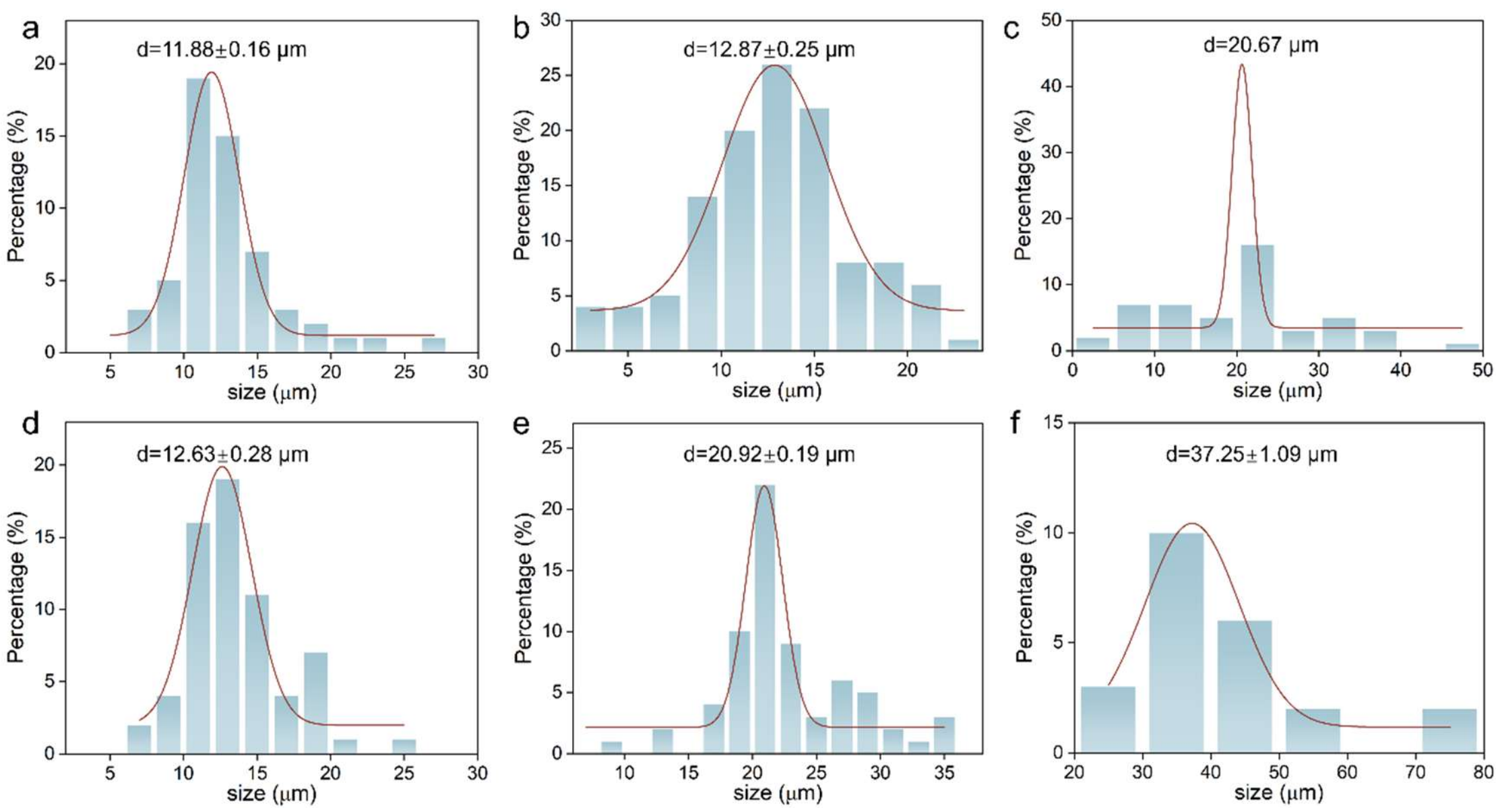

**Supplementary Figure 16.** Comparison of particle size distributions during early-stage Li nucleation on Pristine-Cu and Ion-Cu. **a-c** Particle size distribution of Li nuclei on Pristine-Cu at capacities of 0.05, 0.10, and 0.20 mAh $cm^{-2}$ under 0.1 mA $cm^{-2}$, respectively. **d-f** Corresponding distributions on Ion-Cu under identical deposition conditions. Ion-Cu consistently exhibits larger and more uniform nucleation radii compared to Pristine-Cu, indicating improved lateral growth behavior.

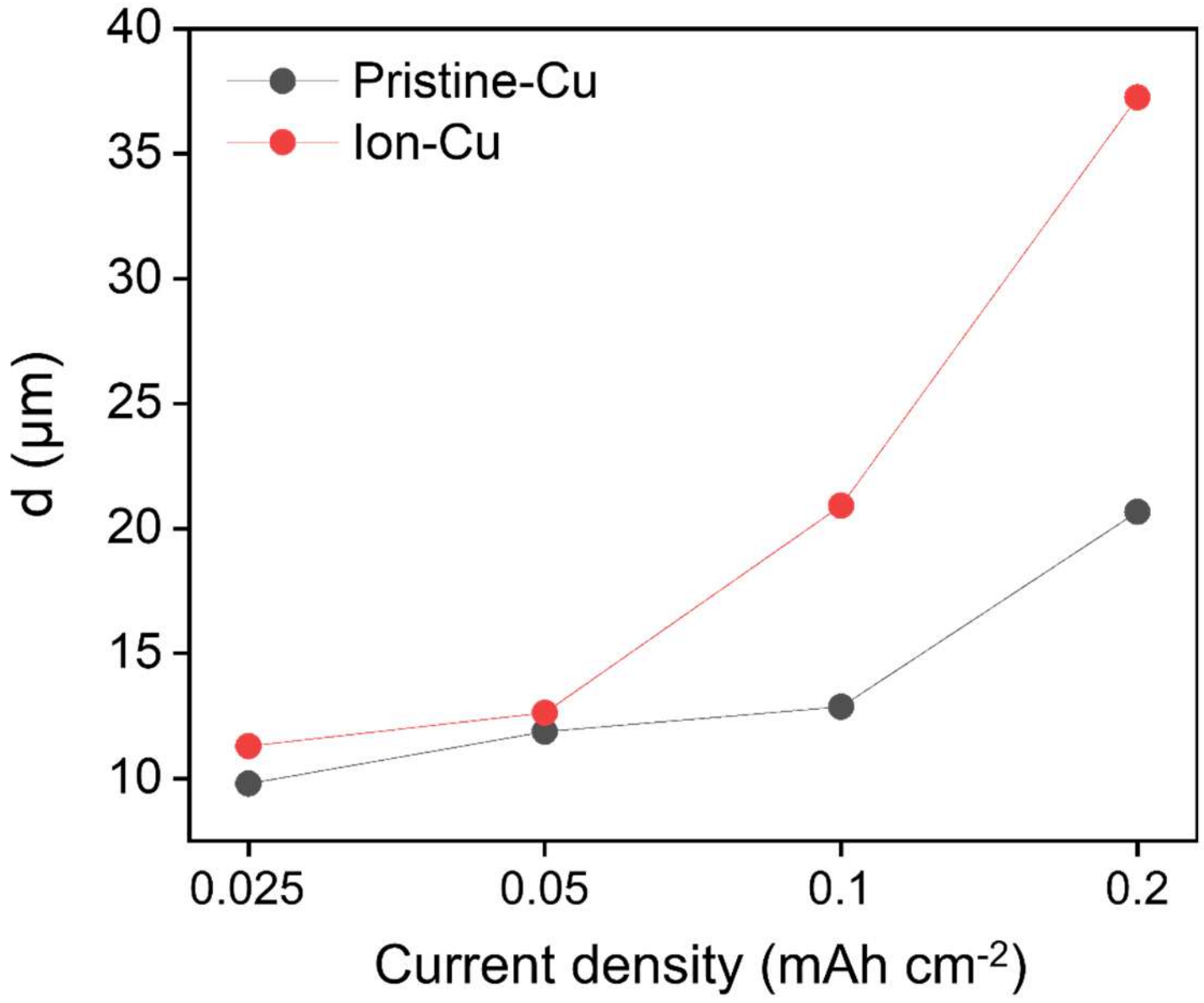

**Supplementary Figure 17.** Relationship between Li particle size and deposition amount on Ion-Cu and Pristine-Cu.

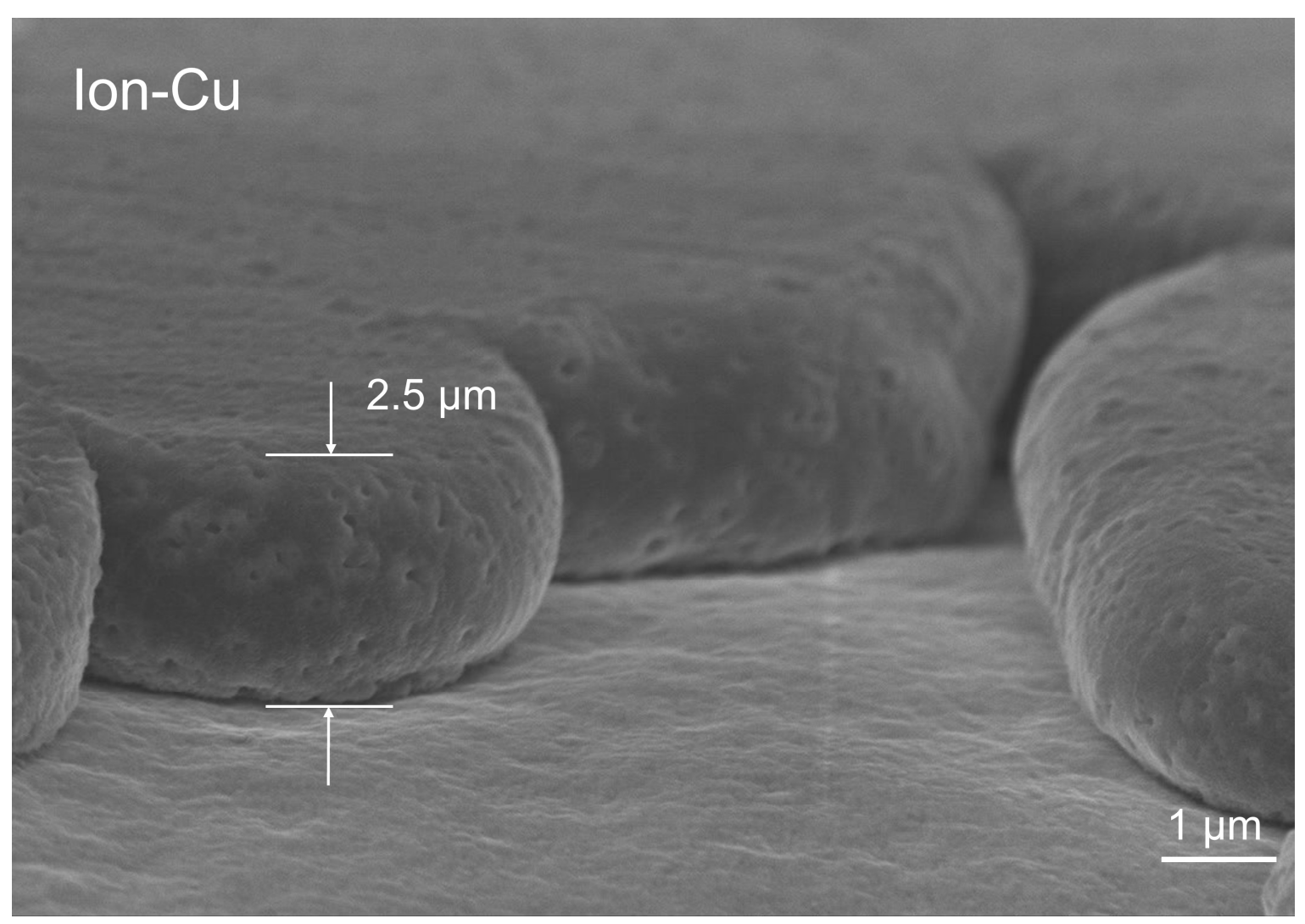

**Supplementary Figure 18.** Cross-sectional SEM image of Li metal deposition morphology on Ion-Cu. Obtained in ether-based electrolyte (1.0 M LiTFSI in a 1:1 vol/vol mixture of DOL/DME with 5% $LiNO_3$) with a capacity of 0.5 mAh $cm^{-2}$ under 0.5 mA $cm^{-2}$.

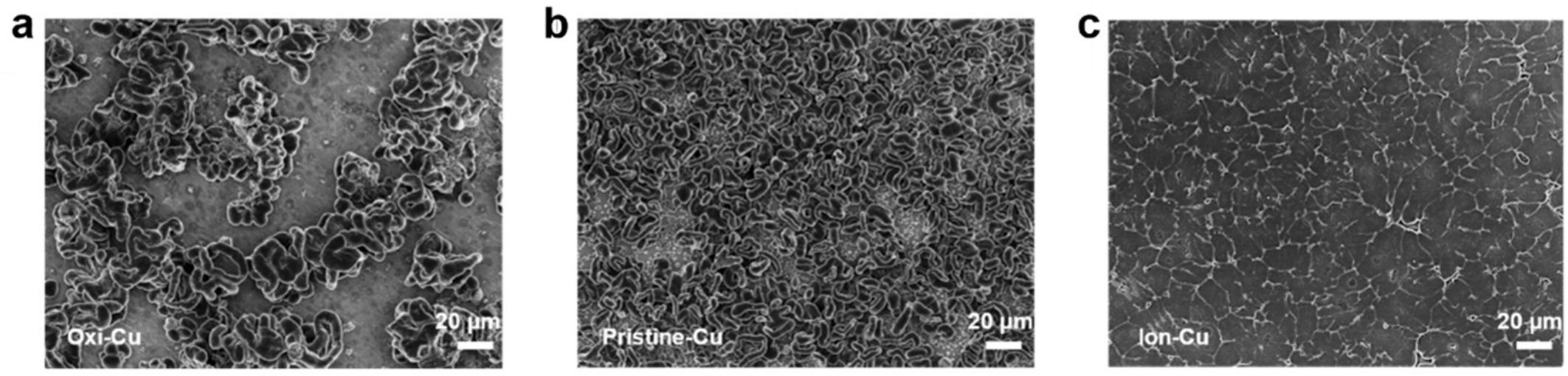

**Supplementary Figure 19.** SEM images of Li deposited on different Cu current collectors. **a** Oxi-Cu; **b** Pristine-Cu; **c**, Ion-Cu. All samples were tested at a deposition capacity of 1 mAh $cm^{-2}$ under a current density of 2 mA $cm^{-2}$.

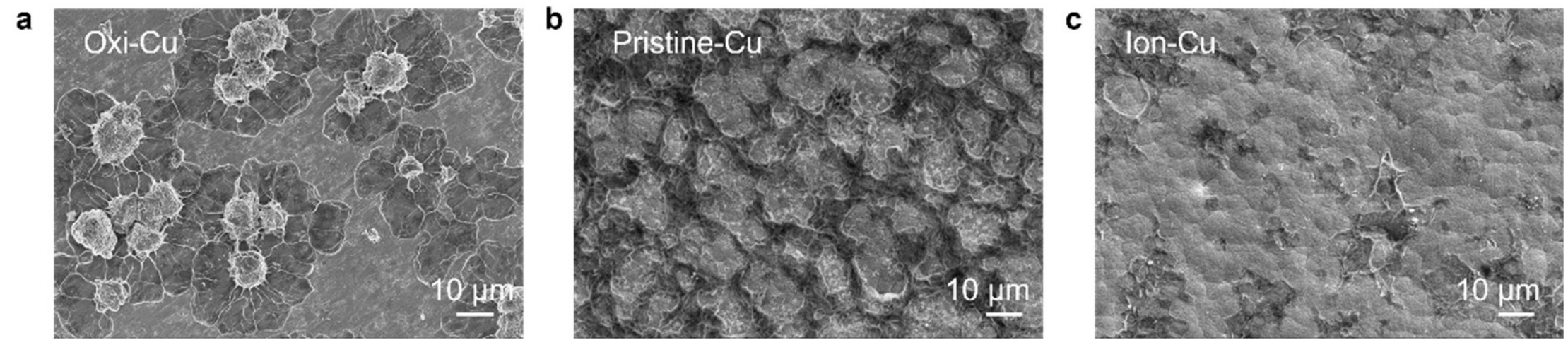

**Supplementary Figure 20.** SEM images of Li stripping morphology on different Cu current collectors. **a** Oxi-Cu; **b** Pristine-Cu; **c** Ion-Cu. All samples were tested after Li stripping in an ether-based electrolyte (1.0 M LiTFSI in a 1:1 vol/vol mixture of DOL/DME with 5% $LiNO_3$), with a capacity of 1 mAh $cm^{-2}$ under a current density of 2 mA $cm^{-2}$.

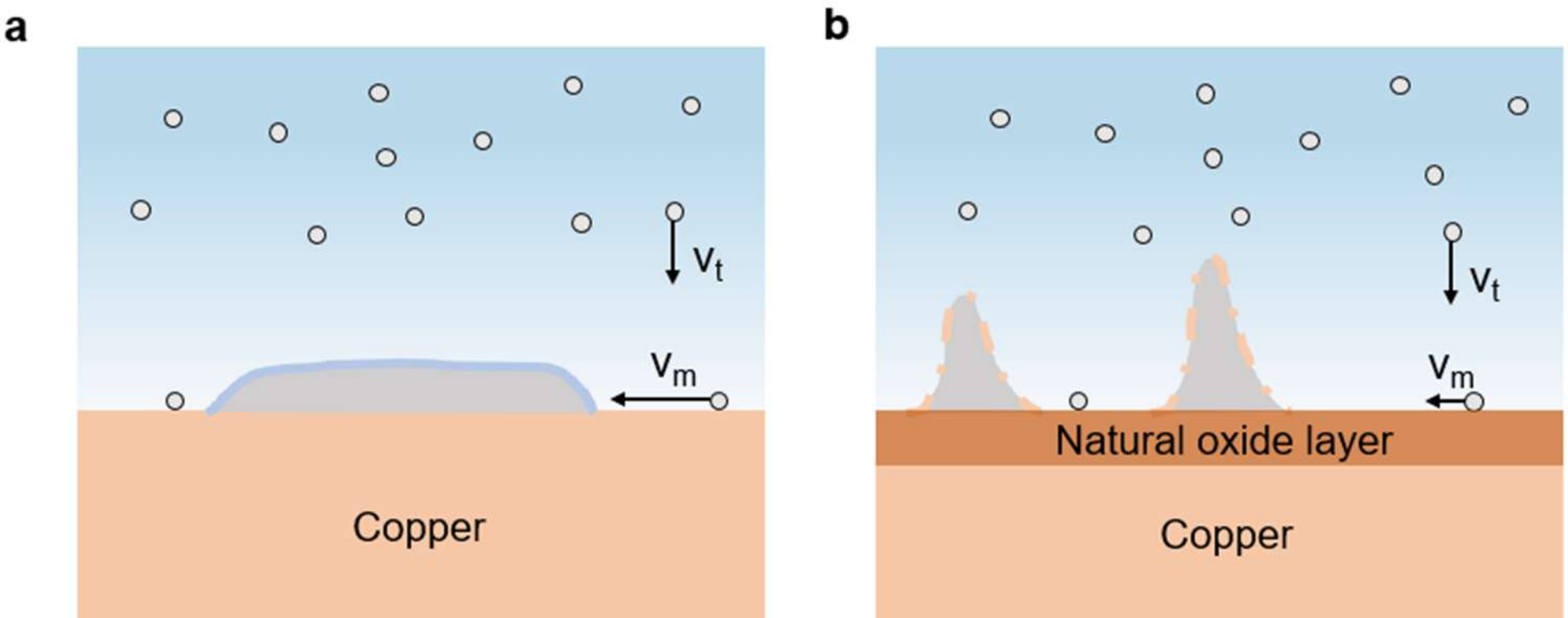

**Supplementary Figure 21.** Schematic illustrations of Li deposition behavior. On metallic Cu (a) and Cu with a natural oxide layer (b), highlighting differences in vertical ($v_t$) versus lateral ($v_m$) mobility.

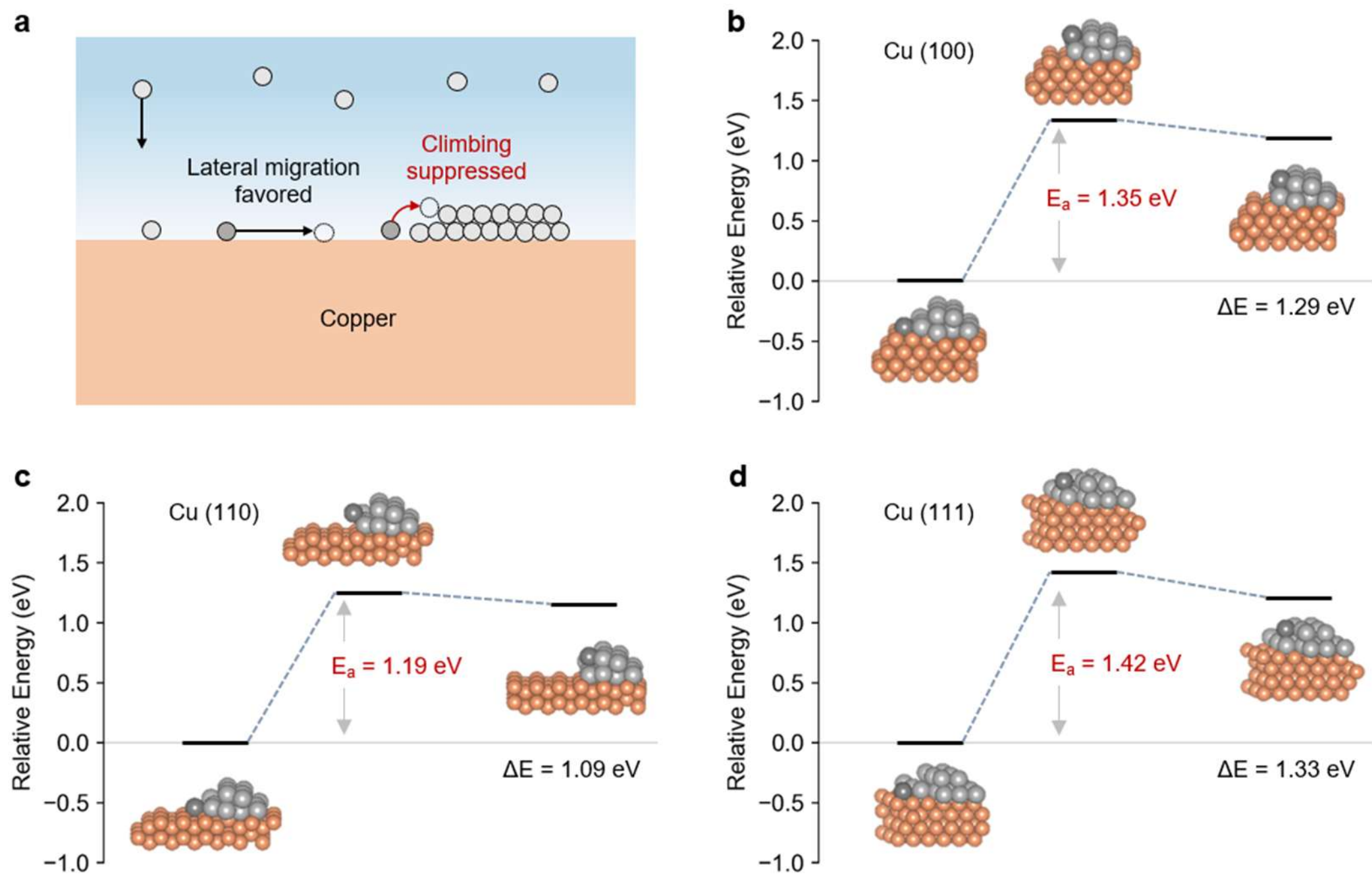

**Supplementary Figure 22.** Energy barriers for Li climbing on Cu surfaces evaluated via DFT. **a** Schematic illustration showing the competition between lateral migration and vertical climbing during Li deposition, where climbing is energetically suppressed. **b-d** Climbing energy profiles on Cu (100), (110), and (111) surfaces. These results confirm that Li atoms prefer to spread laterally rather than grow vertically on metallic Cu, favoring dense, film-like deposition morphology.

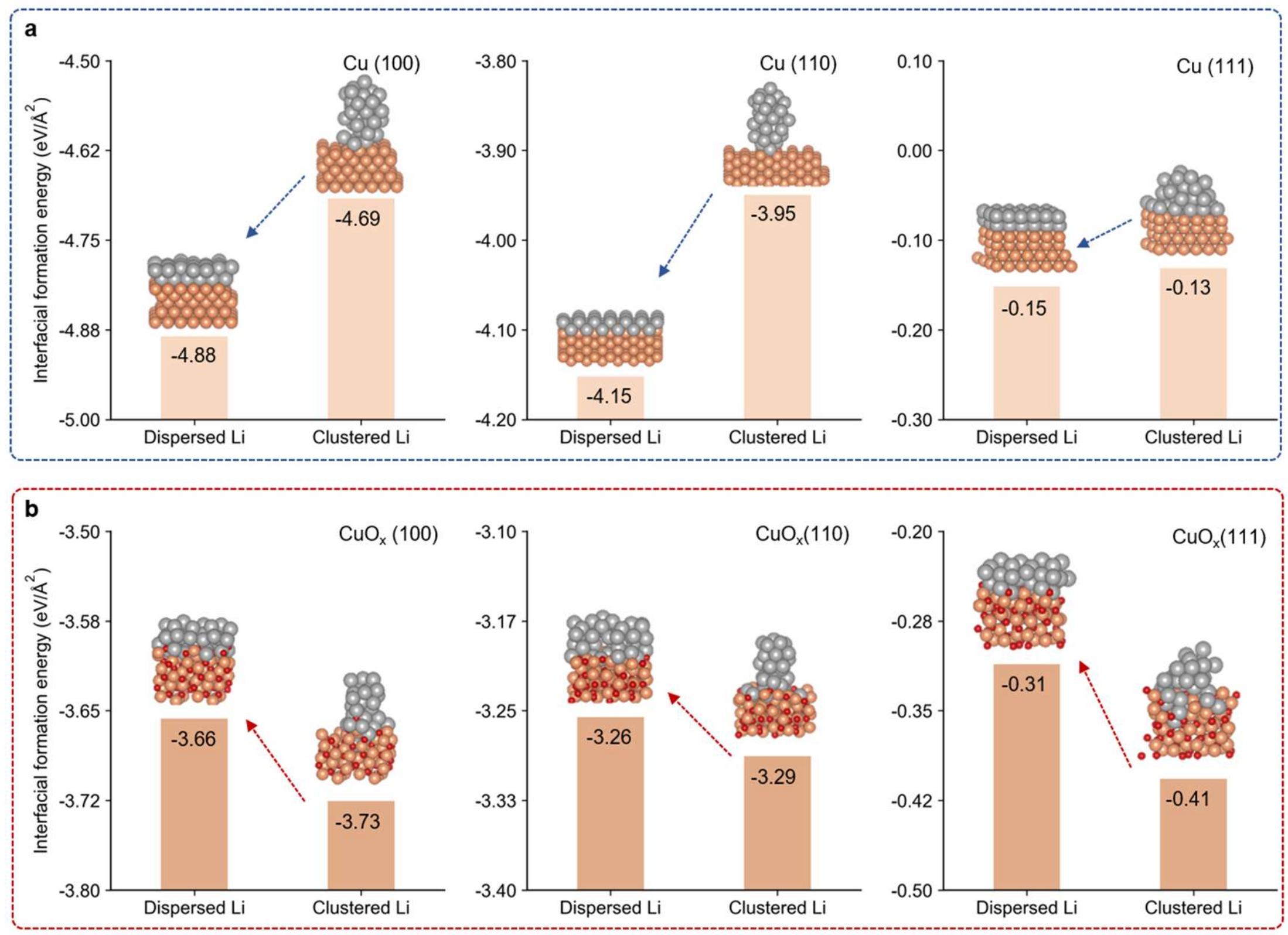

**Supplementary Figure 23.** Interfacial formation energies of dispersed versus clustered Li on Cu and $CuO_x$ surfaces. **a** Interfacial formation energies (eV/Å$^2$) of dispersed and clustered Li configurations on metallic Cu (100), (110), and (111) surfaces. On all facets, dispersed Li exhibits lower formation energy, indicating a thermodynamic preference for lateral growth. **b** Interfacial formation energies of Li on $CuO_x$ surfaces. In contrast to metallic Cu, clustered Li becomes more stable on oxidized surfaces ($CuO_x$ (100), (110), and (111)), suggesting a driving force for vertical stacking and aggregation. These results confirm that surface oxidation fundamentally alters the energy landscape of Li deposition, favoring clustering and dendritic growth on $CuO_x$ surfaces.

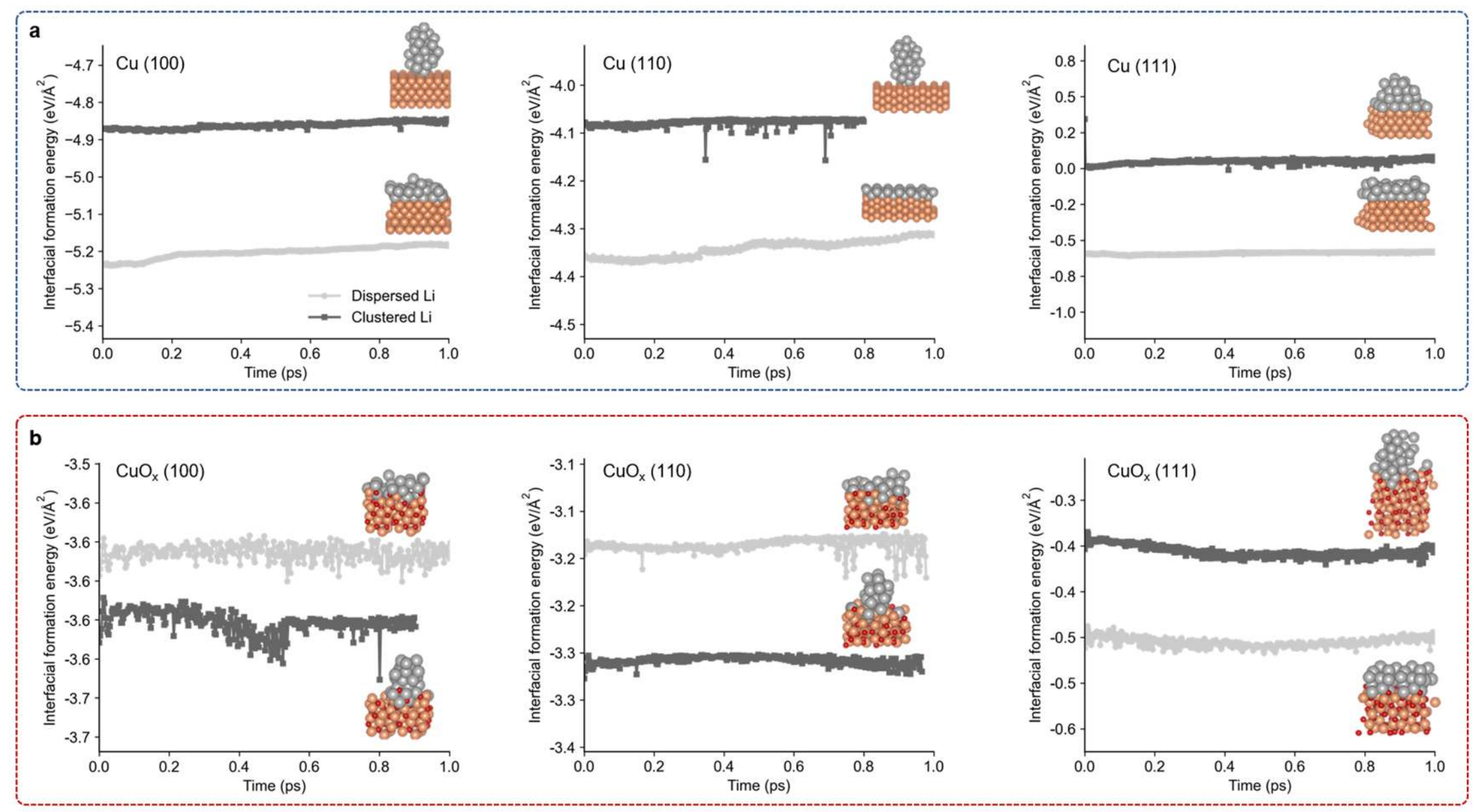

**Supplementary Figure 24.** AIMD simulations of interfacial formation energy evolution for dispersed and clustered Li on Cu and $CuO_x$ surfaces. **a** Time-resolved interfacial formation energy profiles of dispersed and clustered Li on Cu (100), (110), and (111) surfaces at 300 K. On all facets, dispersed Li configurations maintain lower interfacial energies, indicating enhanced thermodynamic stability and a preference for lateral growth. **b** Corresponding AIMD results for $CuO_x$ (100), (110), and (111) surfaces. In contrast to metallic Cu, clustered Li consistently exhibits lower energy (except for $CuO_x$ (111) surface), confirming its favorable aggregation behavior on oxidized surfaces. These results further validate the DFT-based energy landscape in Supplementary Figure 23 and demonstrate that oxidation critically alters Li deposition thermodynamics.

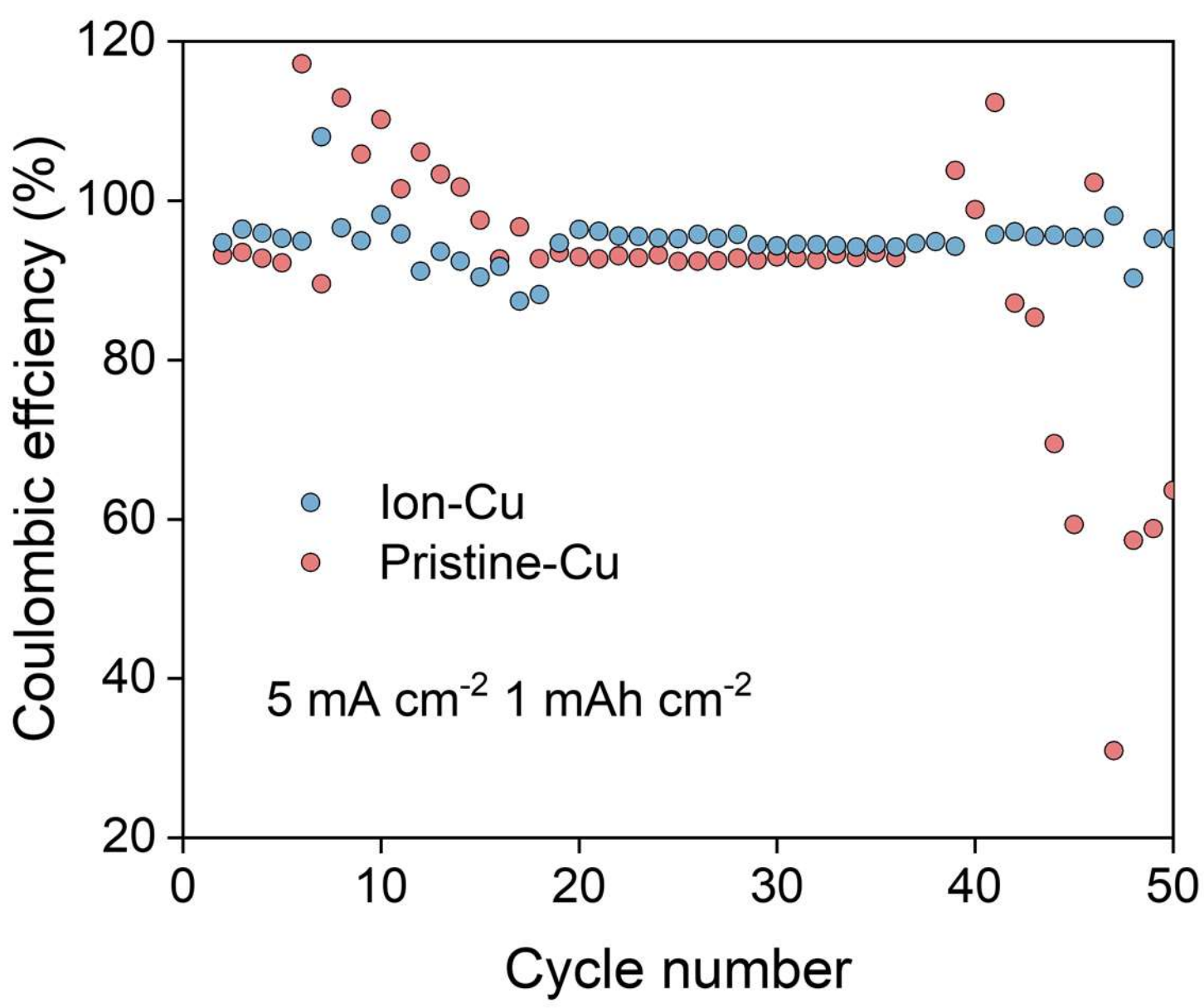

**Supplementary Figure 25.** Cycling test of Li||Cu half-cells with Ion-Cu and Pristine-Cu. Obtained under the current density of 5 mA $cm^{-2}$ and cycling areal capacity of 1.0 mAh $cm^{-2}$ in ether-based electrolyte (1.0 M LiTFSI in a 1:1 vol/vol mixture of DOL/DME with 5% $LiNO_3$) . At high current densities, pristine-Cu requires more activation cycles than Ion-Cu. The Coulombic efficiency of ion-Cu was 95.6%, while that of Pristine-Cu was 93.0%. Moreover, after 36 cycles, the Coulombic efficiency of Pristine-Cu dropped significantly, whereas after 50 cycles, the Coulombic efficiency of Ion-Cu remained above 95%.

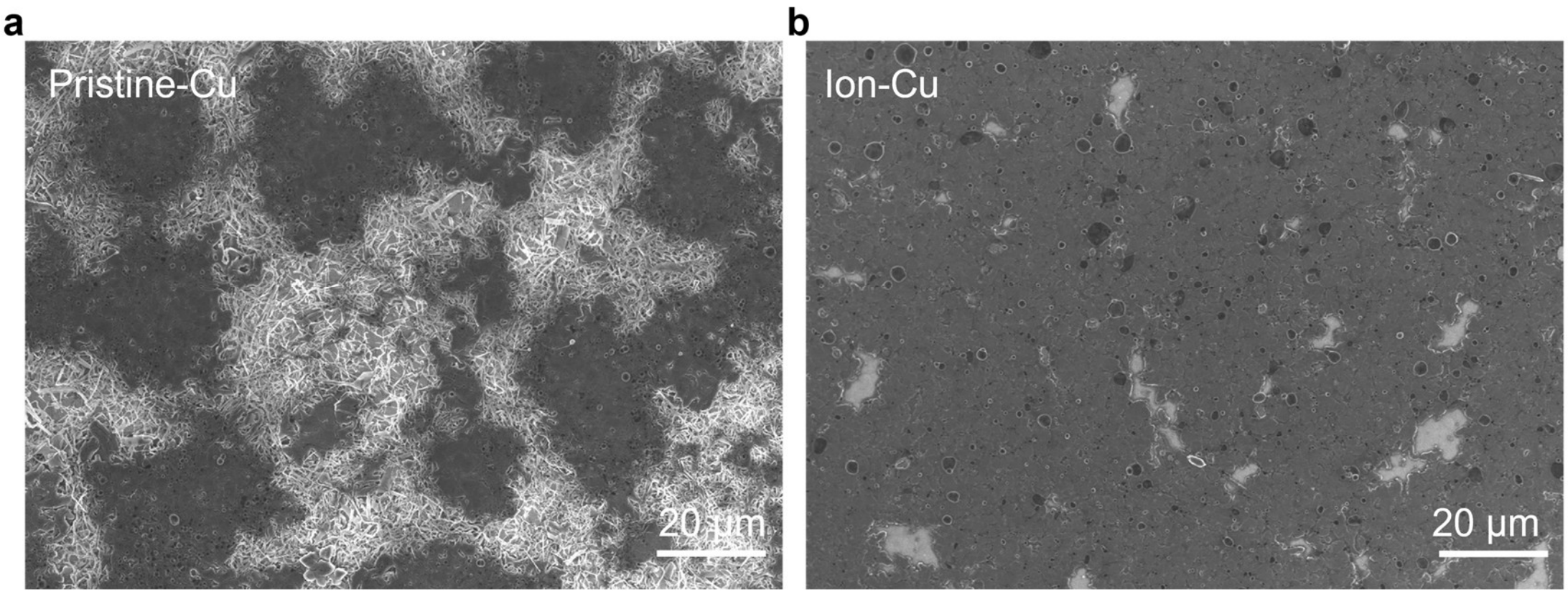

**Supplementary Figure 26.** SEM images of Li deposited on different Cu current collectors. **a** Pristine-Cu; **b** Ion-Cu. The deposition was performed in an ester-based electrolyte (1.0 M $LiPF_6$ in a 1:1 vol/vol mixture of EC/DEC with 10% FEC and 1% VC additives) with a capacity of 1 mAh $cm^{-2}$ under a current density of 1 mA $cm^{-2}$.

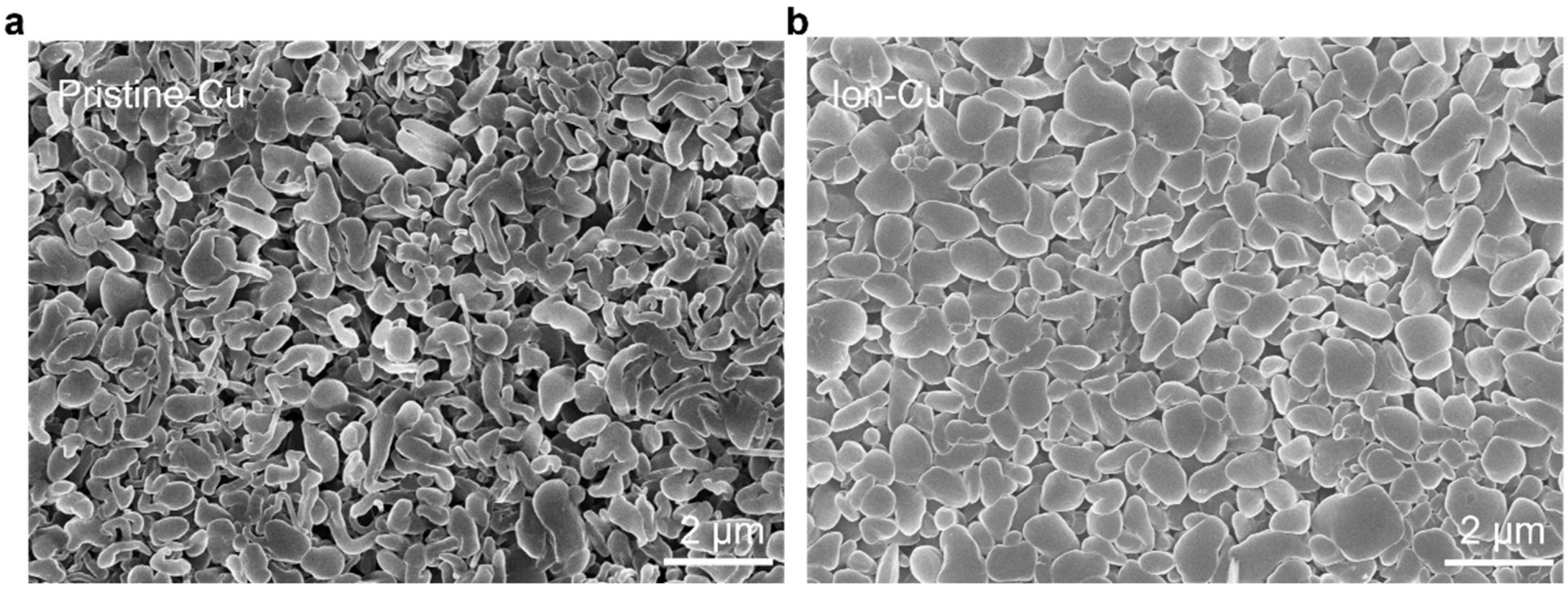

**Supplementary Figure 27.** SEM images of Li deposited on different Cu current collectors in a dual-salt electrolyte. **a** Pristine-Cu; **b** Ion-Cu. The deposition was conducted in a dual-salt electrolyte (0.6 M LiDFOB and 0.6 M $LiBF_4$ in a 2:6:2 wt/wt/wt mixture of FEC/FEMC/HFE) with a capacity of 1 mAh $cm^{-2}$ under a current density of 1 mA $cm^{-2}$.

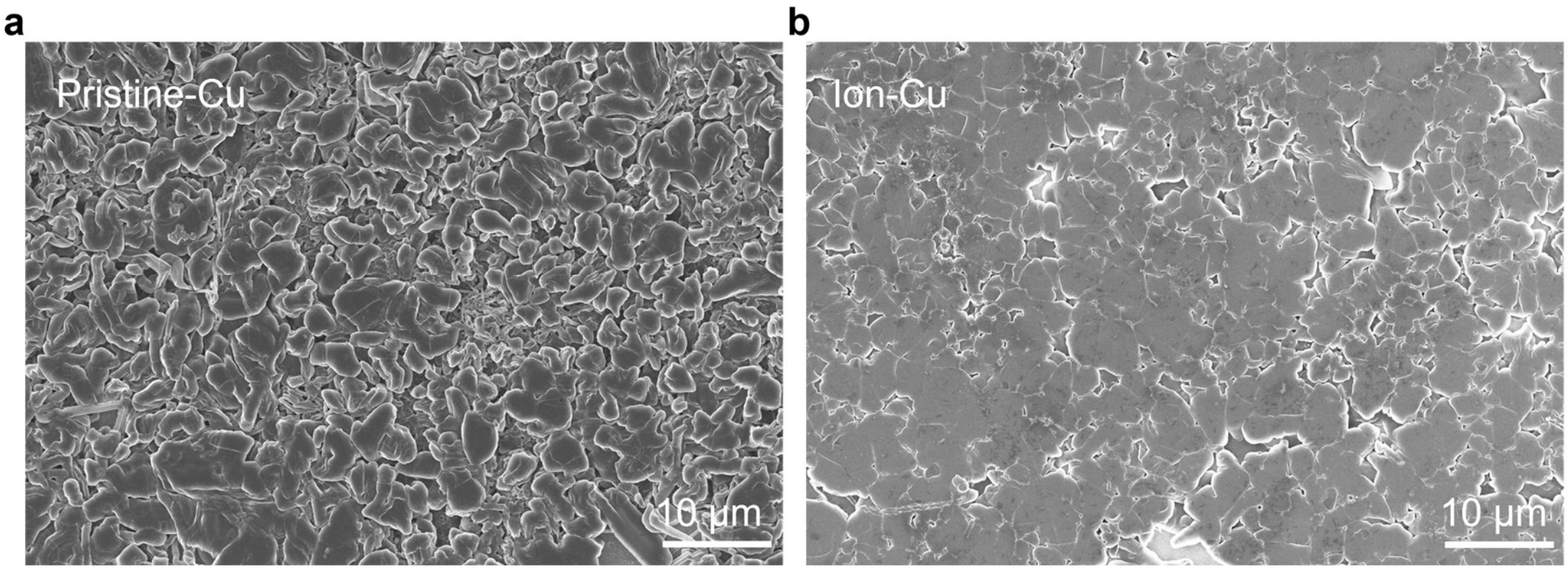

**Supplementary Figure 28.** SEM images of Li deposited on different Cu current collectors in a fluorinated ether electrolyte. **a** Pristine-Cu; **b** Ion-Cu. The deposition was carried out in an electrolyte consisting of a 1:1.8:2 mol/mol/mol mixture of LiFSI/DME/TTE with a capacity of 1 mAh $cm^{-2}$ under a current density of 1 mA $cm^{-2}$.

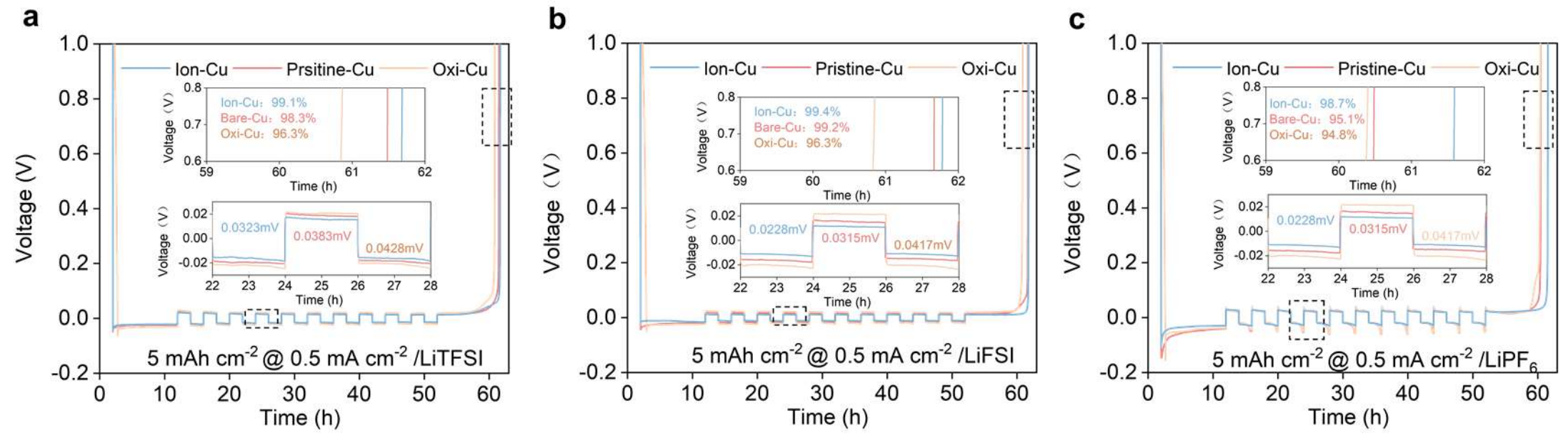

**Supplementary Figure 29.** Voltage profiles of Aurbach protocol in Li||Cu half-cells with different CuCCs in different electrolyte. **a** 1.0 M LiTFSI in a 1:1 vol/vol mixture of DOL/DME with 5% $LiNO_3$; **b** 1:1.8:2 mol/mol/mol mixture of LiFSI/DME/TTE; **c** 1.0 M $LiPF_6$ in a 1:1 vol/vol mixture of EC/DEC

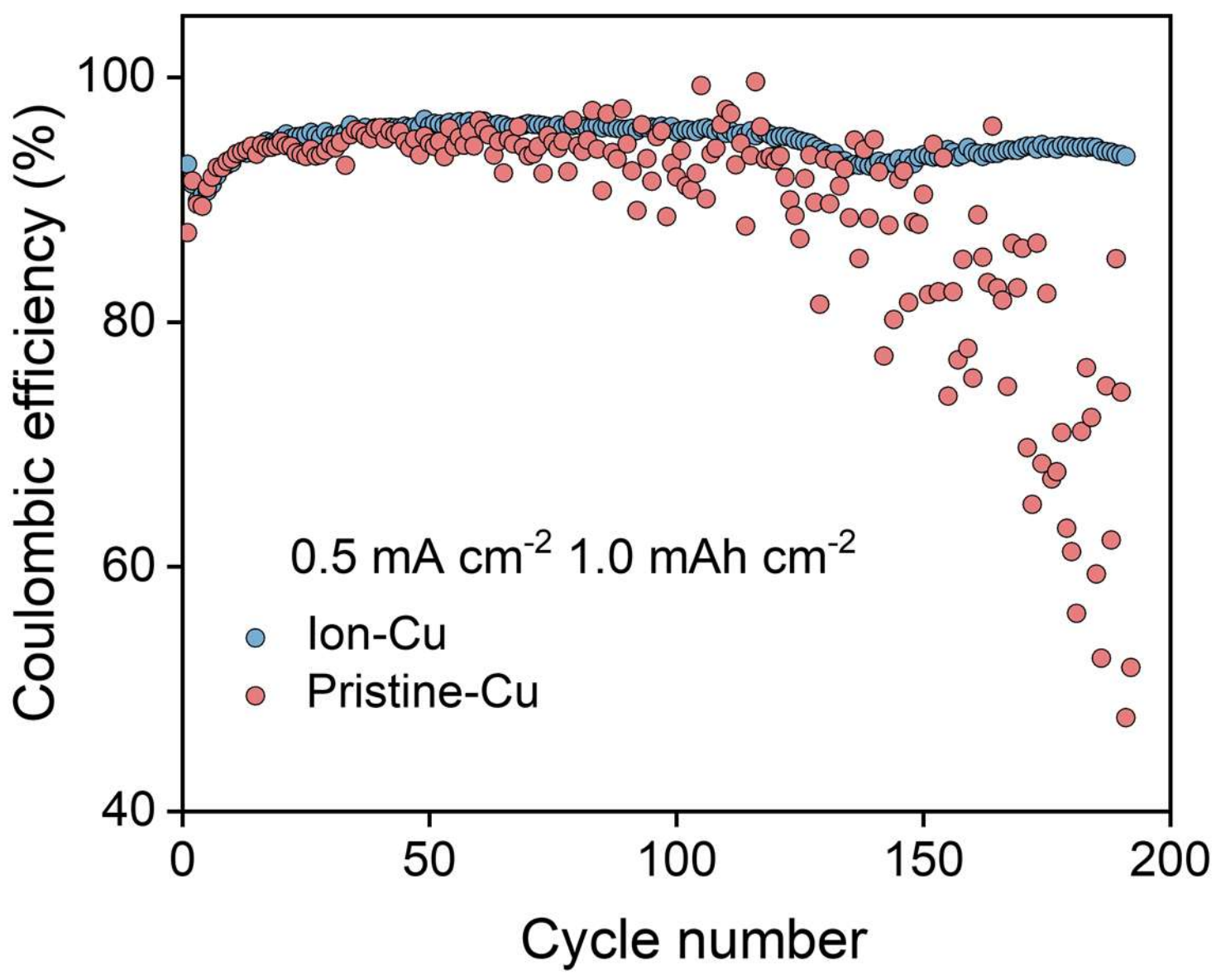

**Supplementary Figure 30.** Cycling test of Li||Cu half-cells with Ion-Cu and Pristine-Cu. Obtained under the current density of 0.5 mA cm$^{-2}$ and cycling areal capacity of 1.0 mAh cm$^{-2}$ in ester-based electrolyte (1.0 M $LiPF_6$ in a 1:1 vol/vol mixture of EC/DEC). Fluorine-free ester-based electrolytes are known for low Coulombic efficiency. The average Coulombic efficiency of pristine-Cu is 93.9%, while that of ion-Cu is 95.7%. The Coulombic efficiency of pristine-Cu drops below 90% after 100 cycles, whereas the Coulombic efficiency of ion-Cu remains above 94% even after 190 cycles.

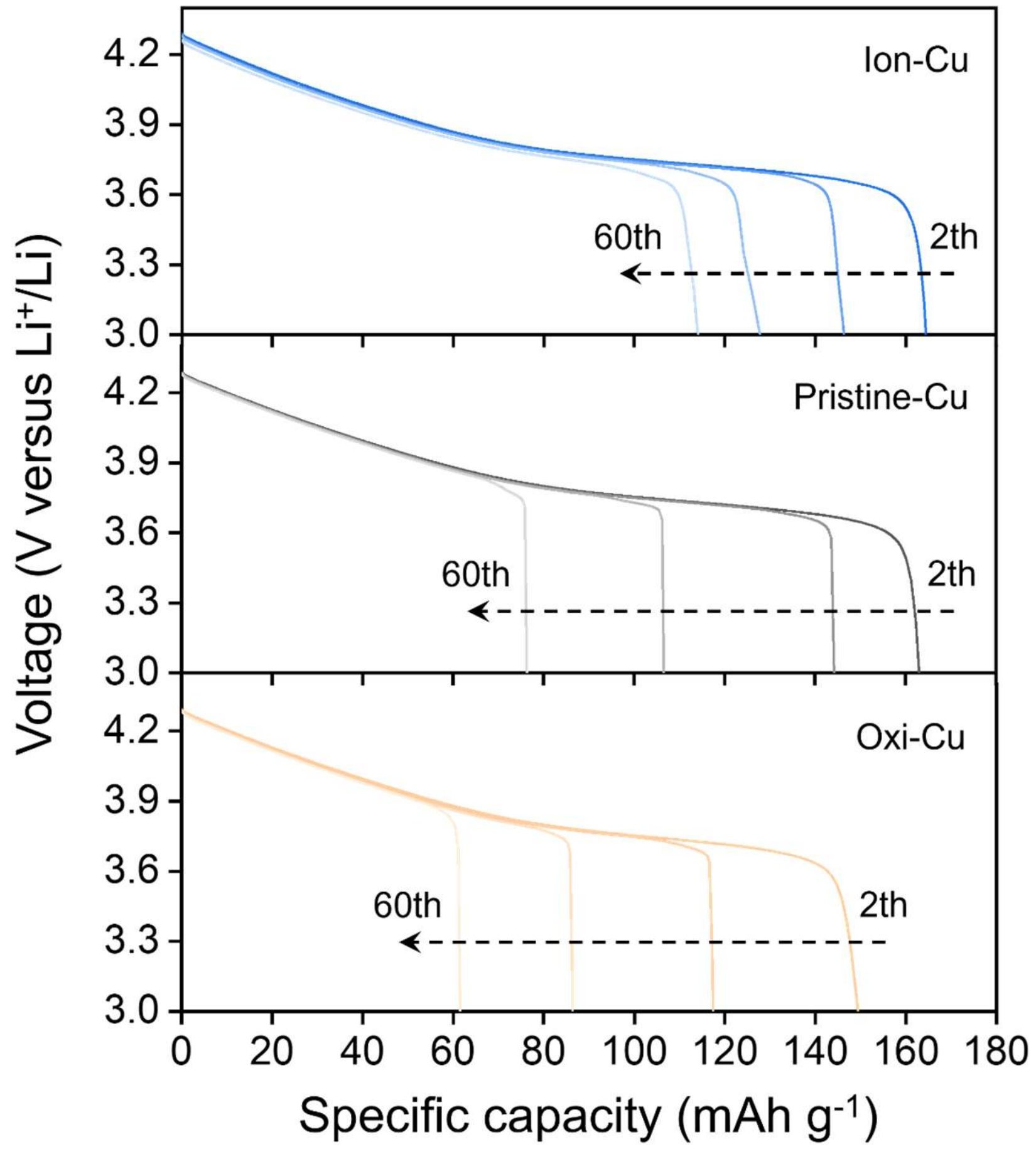

**Supplementary Figure 31.** The charge/discharge profiles of the AFLMB. Obtained at 2nd, 20th, 40th and 60th cycle using different CuCCs in ester-based electrolyte (1.0 M $LiPF_6$ in a 1:1 vol/vol mixture of EC/DEC with 10% FEC and 1% VC additives).

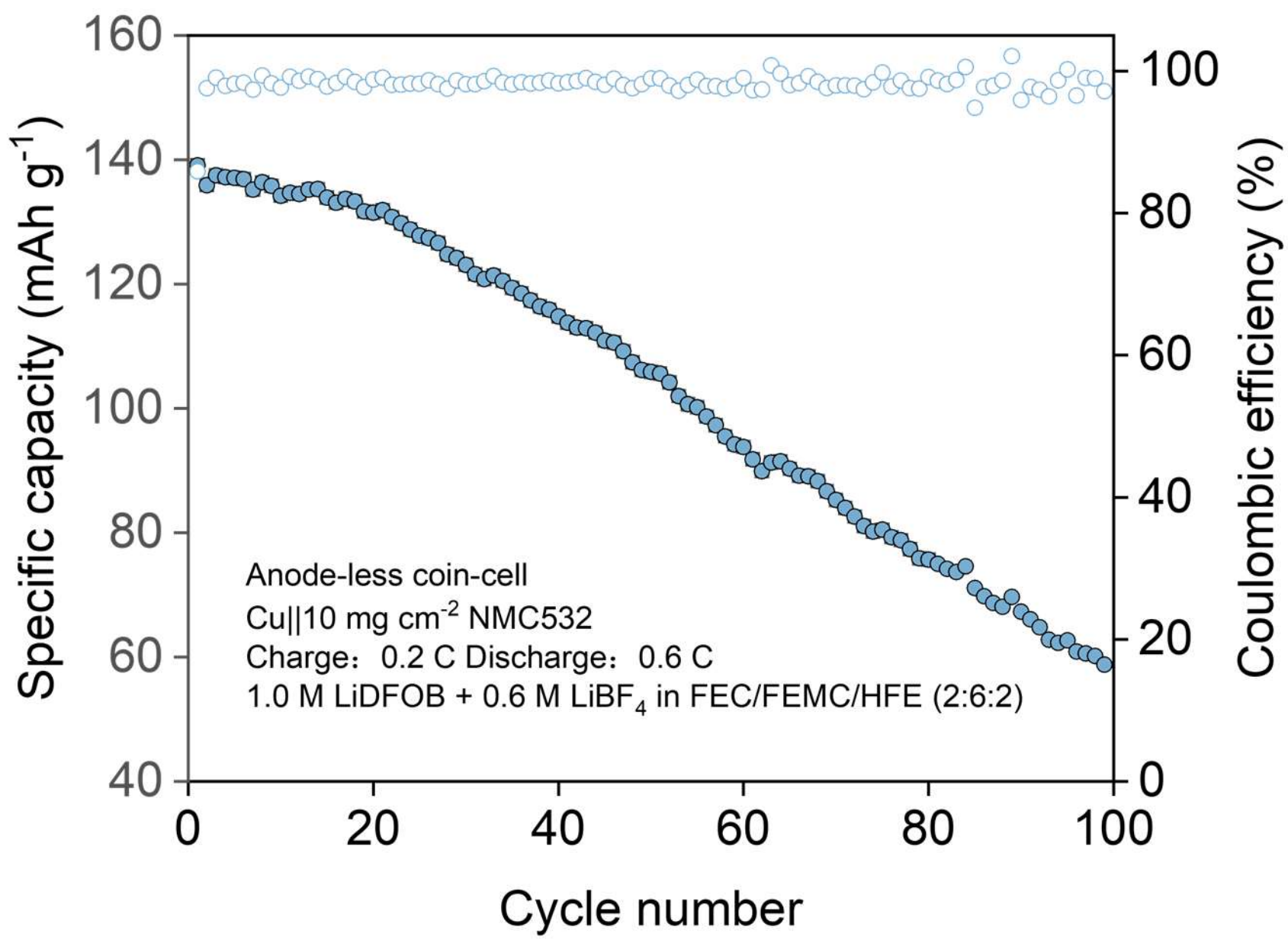

**Supplementary Figure 32.** Long-term cycling performances of anode-free Ion-Cu||NCM532 coin cells. Fabricated using dual-salt electrolyte (0.6 M LiDFOB and 0.6 M LiBF4 in a 2:6:2 wt/wt/wt mixture of FEC/FEMC/HFE) at charge/discharge current density of 0.2 C/0.6 C.

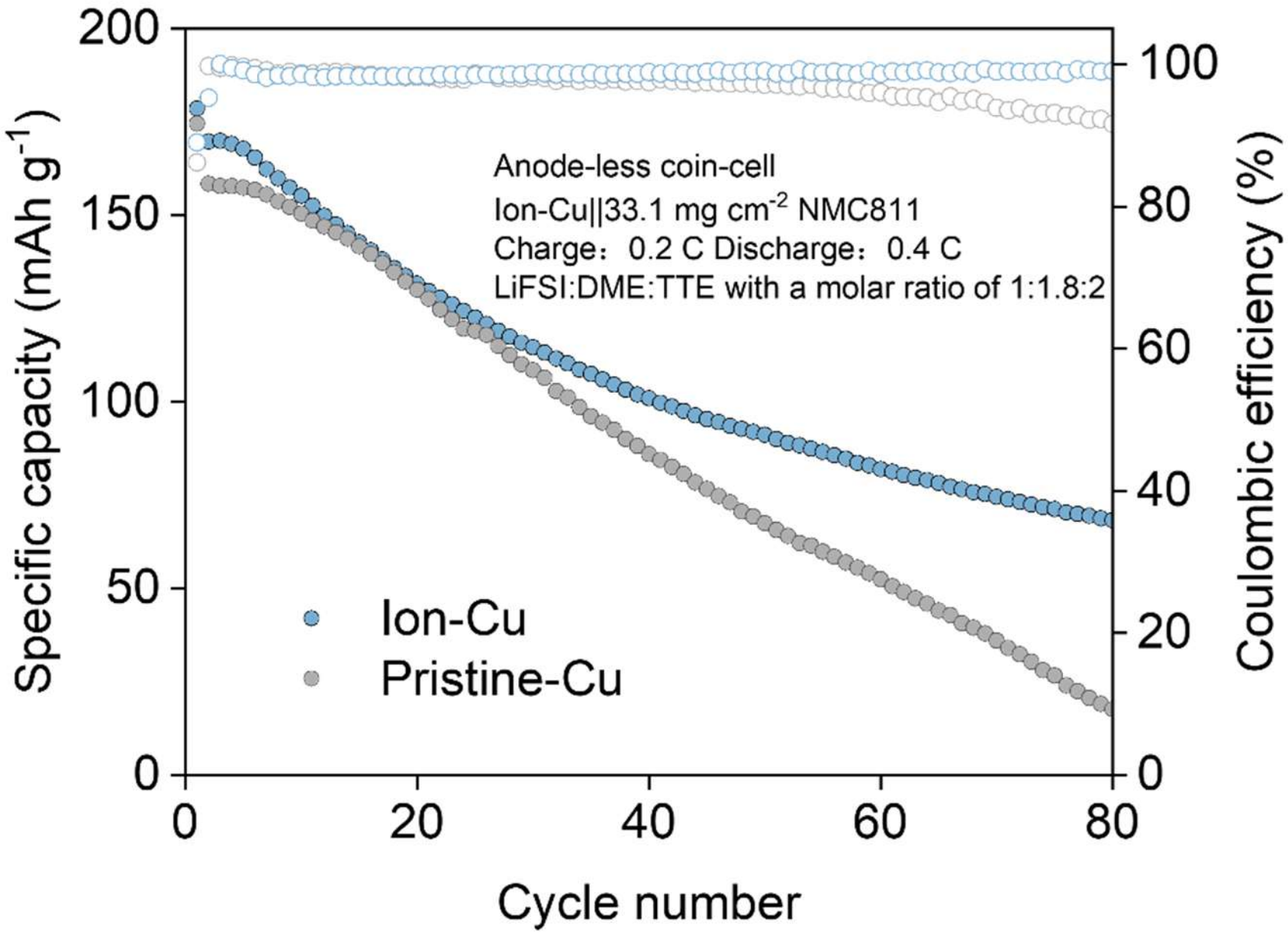

**Supplementary Figure 33.** Long-term cycling performance of anode-free Cu||High Loading NCM811 coin cells. Fabricated using a locally high-concentration electrolyte (LiFSI:DME:TTE at a molar ratio of 1:1.8:2) at charge/discharge current density of 0.2 C/0.4 C.

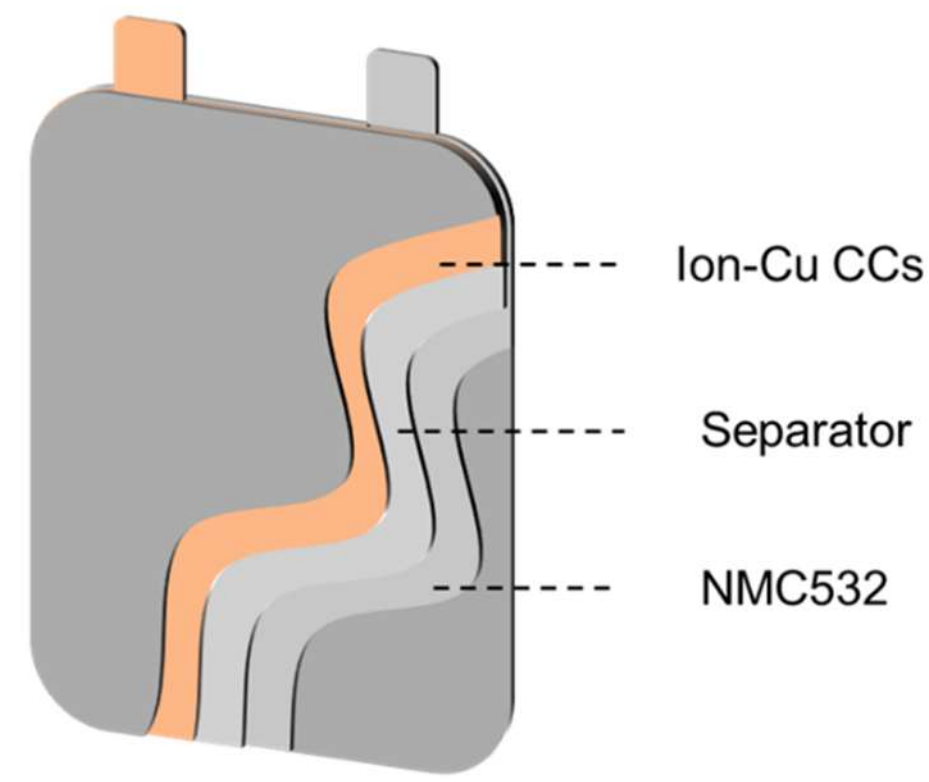

**Supplementary Figure 34.** Schematic of the pouch cell.

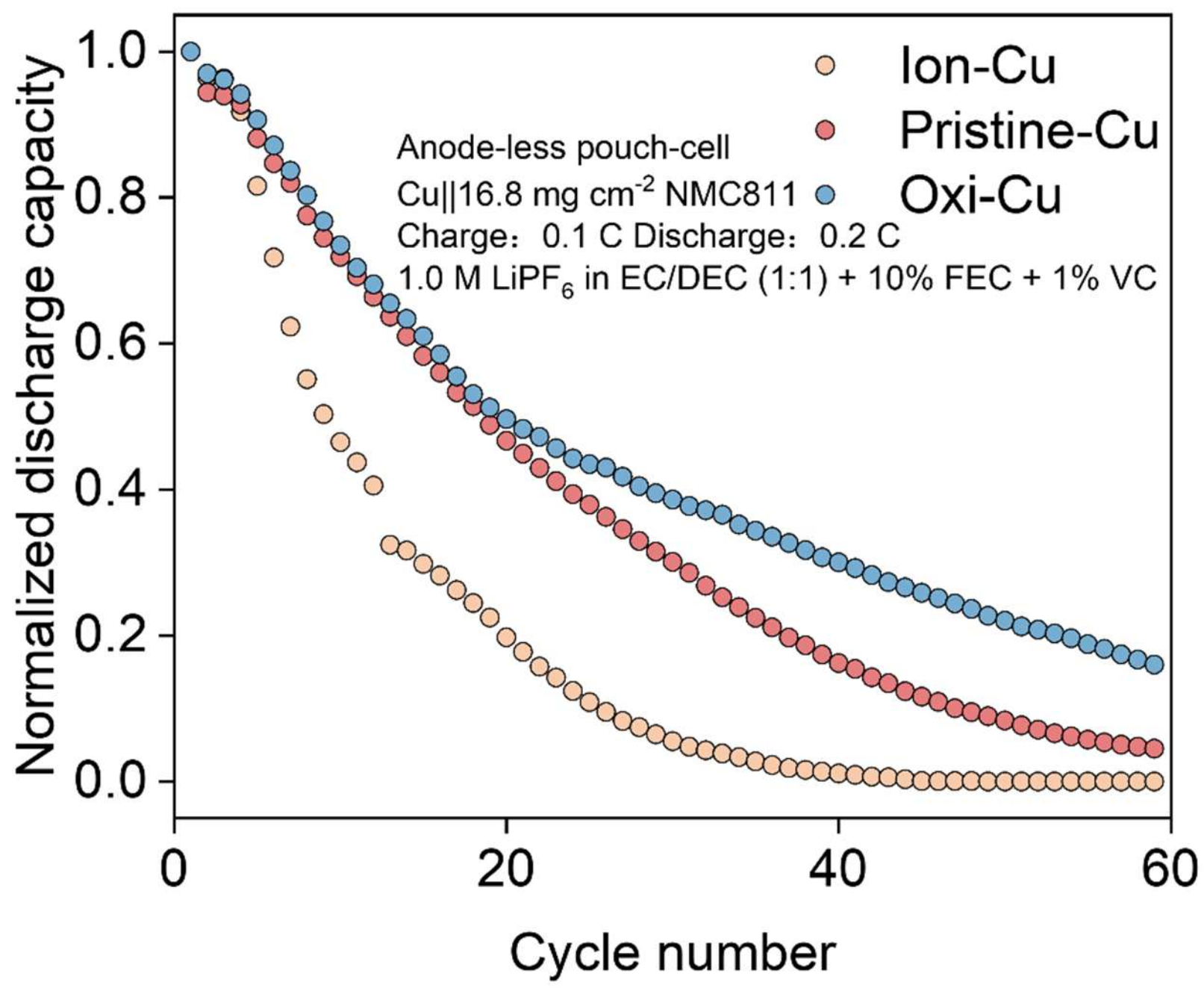

**Supplementary Figure 35.** Long-term cycling performances of anode-free Cu||NCM532 pouch-cell at charge/discharge current density of 0.1 C/0.2 C with different CuCCs.

**Supplementary Table S1.** Comparison of the coin cell performance using NMC cathode of our work with previously reported work.

| Samples | Areal capacity (mAh cm$^{-2}$) | Charge current density (mA cm$^{-2}$) | Cycle number | Capacity rentention (%) | Ref. |
|---|---|---|---|---|---|
| Ion-Cu | 2.9 | 1.05 | 100 | 59.2 | This work |
| CuCC-SSIC | 1.5 | 0.45 | 100 | 54.8 | [17] |
| Ag@PDA-GO | 2.2 | 0.5 | 40 | 71.4 | [18] |
| $Al_2O_3$-Cu | 2.5 | 0.5 | 100 | 60.0 | [19] |
| Concentrated $LiPF_6$ | 2 | 0.2 | 50 | 40.0 | [20] |
| Sn-Cu | 1 | 0.5 | 80 | 0 | [21] |
| Cu@GO | 1.76 | 0.2 | 50 | 44 | [22] |
| NPSFF | 1.5 | 0.45 | 50 | 57.3 | [23] |

**Supplementary Table S2.** Comparison of the coin cell performance using $LiFePO_4$ cathode of our work with previously reported work.

| Samples | Areal capacity (mAh $cm^{-2}$) | Charge current density (mA $cm^{-2}$) | Cycle number | Capacity rentention (%) | Ref. |
|---|---|---|---|---|---|
| Ion-Cu | 1.81 | 0.86 | 140 | 85.7 | This work |
| $Cu_3N$/Cu | 0.55 | 0.5 | 50 | 29.6 | [24] |
| CuCl-Cu | 1.05 | 1.05 | 50 | 78.4 | [25] |
| BTA-Cu | 1.5 | 0.75 | 50 | 73.3 | [26] |
| LiF@PVDF | 0.85 | 0.43 | 60 | 64.6 | [27] |
| $SiO_x$@Cu | 1.8 | 0.9 | 100 | 45.6 | [28] |
| Graphene@ Cu | 2.0 | 0.2 | 100 | 61.3 | [29] |
| DNA-Cu | 3.4 | 1.7 | 150 | 42.3 | [30] |
| $Cu_2S$ SN@Cu | 3.0 | 0.9 | 200 | 40.0 | [31] |
| Rested-3LiFSI | 1.6 | 0.2 | 50 | 63.8% | [32] |

## References

[1] M. Sharma, R. Singh, S. Dutt, P. Tomar, G. Trivedi, N. Robert, “Effect of absorbed dose on post-irradiation coloration and interpretation of polymerization reaction in the Gafchromic EBT3 film” *Radiat. Phys. Chem.* **2021**, *187*, 109569.

[2] J. F. Ziegler, M. D. Ziegler, J. P. Biersack, “SRIM – The stopping and range of ions in matter (2010)” *Nucl. Instrum. Methods Phys. Res. B: Beam Interact. Mater. At.* **2010**, *268*, 1818–1823.

[3] P. Zhao, Y. Shimomura, “Molecular dynamics calculations of properties of the self-interstitials in copper and nickel” *Computational Materials Science* **1999**, *14*, 84–90.

[4] C. Li, P. Zhang, J. Wang, J. A. Boscoboinik, G. Zhou, “Tuning the Deoxygenation of Bulk-Dissolved Oxygen in Copper” *J. Phys. Chem. C* **2018**, *122*, 8254–8261.

[5] V. Fotopoulos, D. Mora-Fonz, M. Kleinbichler, R. Bodlos, E. Kozeschnik, L. Romaner, A. L. Shluger, “Structure and Migration Mechanisms of Small Vacancy Clusters in Cu: A Combined EAM and DFT Study” *Nanomaterials* **2023**, *13*, 1464.

[6] C. S. Becquart, C. Domain, U. Sarkar, A. DeBacker, M. Hou, “Microstructural evolution of irradiated tungsten: *Ab initio* parameterisation of an OKMC model” *Journal of Nuclear Materials* **2010**, *403*, 75–88.

[7] G. Kresse, J. Furthmüller, “Efficient iterative schemes for *ab initio* total-energy calculations using a plane-wave basis set” *Phys. Rev. B* **1996**, *54*, 11169–11186.

[8] P. E. Blöchl, “Projector augmented-wave method” *Phys. Rev. B* **1994**, *50*, 17953–17979.

[9] J. P. Perdew, K. Burke, M. Ernzerhof, “Generalized Gradient Approximation Made Simple” *Phys. Rev. Lett.* **1996**, *77*, 3865–3868.

[10] S. J. Clark, M. D. Segall, C. J. Pickard, P. J. Hasnip, M. I. J. Probert, K. Refson, M. C. Payne, “First principles methods using CASTEP” *Zeitschrift für Kristallographie - Crystalline Materials* **2005**, *220*, 567–570.

[11] E. R. McNellis, J. Meyer, K. Reuter, “Azobenzene at coinage metal surfaces: Role of dispersive van der Waals interactions” *Phys. Rev. B* **2009**, *80*, 205414.

[12] J. P. Perdew, J. A. Chevary, S. H. Vosko, K. A. Jackson, M. R. Pederson, D. J. Singh, C. Fiolhais, “Atoms, molecules, solids, and surfaces: Applications of the generalized gradient approximation for exchange and correlation” *Phys. Rev. B* **1992**, *46*, 6671–6687.

[13] H. J. Monkhorst, J. D. Pack, “Special points for Brillouin-zone integrations” *Phys. Rev. B* **1976**, *13*, 5188–5192.

[14] S. Grimme, “Semiempirical GGA-type density functional constructed with a long-range dispersion correction” *J Comput Chem* **2006**, *27*, 1787–1799.

[15] B. Aradi, B. Hourahine, Th. Frauenheim, “DFTB+, a Sparse Matrix-Based Implementation of the DFTB Method” *J. Phys. Chem. A* **2007**, *111*, 5678–5684.

[16] T. A. Halgren, W. N. Lipscomb, “The synchronous-transit method for determining reaction pathways and locating molecular transition states” *Chemical Physics Letters* **1977**, *49*, 225–232.

[17] J. Zhan, L. Deng, Y. Liu, M. Hao, Z. Wang, L. Dong, Y. Yang, K. Song, D. Qi, J.

Wang, S. Wang, H. Liu, W. Zhou, H. Chen, "Self-Selective (220) Directional Grown Copper Current Collector Design for Cycling-Stable Anode-Less Lithium Metal Batteries" *Adv. Mater.* **2024**, 2413420.
[18] Z. T. Wondimkun, W. A. Tegegne, J. Shi-Kai, C.-J. Huang, N. A. Sahalie, M. A. Weret, J.-Y. Hsu, P.-L. Hsieh, Y.-S. Huang, S.-H. Wu, W.-N. Su, B. J. Hwang, "Highly-lithiophilic Ag@PDA-GO film to Suppress Dendrite Formation on Cu Substrate in Anode-free Lithium Metal Batteries" *Energy Storage Mater.* **2021**, *35*, 334–344.
[19] S. T. Oyakhire, W. Zhang, A. Shin, R. Xu, D. T. Boyle, Z. Yu, Y. Ye, Y. Yang, J. A. Raiford, W. Huang, J. R. Schneider, Y. Cui, S. F. Bent, "Electrical resistance of the current collector controls lithium morphology" *Nat. Commun.* **2022**, *13*, 3986.
[20] T. T. Hagos, B. Thirumalraj, C.-J. Huang, L. H. Abrha, T. M. Hagos, G. B. Berhe, H. K. Bezabh, J. Cherng, S.-F. Chiu, W.-N. Su, B.-J. Hwang, "Locally Concentrated $LiPF_6$ in a Carbonate-Based Electrolyte with Fluoroethylene Carbonate as a Diluent for Anode-Free Lithium Metal Batteries" *ACS Appl. Mater. Interfaces* **2019**, *11*, 9955–9963.
[21] S. S. Zhang, X. Fan, C. Wang, "A tin-plated copper substrate for efficient cycling of lithium metal in an anode-free rechargeable lithium battery" *Electrochim. Acta* **2017**, *258*, 1201–1207.
[22] Z. T. Wondimkun, T. T. Beyene, M. A. Weret, N. A. Sahalie, C.-J. Huang, B. Thirumalraj, B. A. Jote, D. Wang, W.-N. Su, C.-H. Wang, G. Brunklaus, M. Winter, B.-J. Hwang, "Binder-free ultra-thin graphene oxide as an artificial solid electrolyte interphase for anode-free rechargeable lithium metal batteries" *J. Power Sources* **2020**, *450*, 227589.
[23] P. Li, H. Zhang, J. Lu, G. Li, "Low Concentration Sulfolane-Based Electrolyte for High Voltage Lithium Metal Batteries" *Angew. Chem. Int. Ed.* **2023**, *62*, e202216312.
[24] Q. Li, H. Pan, W. Li, Y. Wang, J. Wang, J. Zheng, X. Yu, H. Li, L. Chen, "Homogeneous Interface Conductivity for Lithium Dendrite-Free Anode" *ACS Energy Lett.* **2018**, *3*, 2259–2266.
[25] Z. Li, X. Huang, L. Kong, N. Qin, Z. Wang, L. Yin, Y. Li, Q. Gan, K. Liao, S. Gu, T. Zhang, H. Huang, L. Wang, G. Luo, X. Cheng, Z. Lu, "Gradient nano-recipes to guide lithium deposition in a tunable reservoir for anode-free batteries" *Energy Storage Materials* **2022**, *45*, 40–47.
[26] T. Kang, J. Zhao, F. Guo, L. Zheng, Y. Mao, C. Wang, Y. Zhao, J. Zhu, Y. Qiu, Y. Shen, L. Chen, "Dendrite-Free Lithium Anodes Enabled by a Commonly Used Copper Antirusting Agent" *ACS Appl. Mater. Interfaces* **2020**, *12*, 8168–8175.
[27] O. Tamwattana, H. Park, J. Kim, I. Hwang, G. Yoon, T. Hwang, Y.-S. Kang, J. Park, N. Meethong, K. Kang, "High-Dielectric Polymer Coating for Uniform Lithium Deposition in Anode-Free Lithium Batteries" *ACS Energy Lett.* **2021**, *6*, 4416–4425.
[28] W. Chen, R. V. Salvatierra, M. Ren, J. Chen, M. G. Stanford, J. M. Tour, "Laser-Induced Silicon Oxide for Anode-Free Lithium Metal Batteries" *Advanced Materials* **2020**, *32*, 2002850.
[29] A. A. Assegie, C.-C. Chung, M.-C. Tsai, W.-N. Su, C.-W. Chen, B.-J. Hwang, "Multilayer-graphene-stabilized lithium deposition for anode-Free lithium-metal batteries" *Nanoscale* **2019**, *11*, 2710–2720.

[30] Z. Ouyang, Y. Wang, S. Wang, S. Geng, X. Zhao, X. Zhang, Q. Xu, B. Yuan, S. Tang, J. Li, F. Wang, G. Yao, H. Sun, "Programmable DNA Interphase Layers for High-Performance Anode-Free Lithium Metal Batteries" *Advanced Materials* **2024**, *36*, 2401114.
[31] Z. Yang, W. Liu, Q. Chen, X. Wang, W. Zhang, Q. Zhang, J. Zuo, Y. Yao, X. Gu, K. Si, K. Liu, J. Wang, Y. Gong, "Ultrasmooth and Dense Lithium Deposition Toward High-Performance Lithium-Metal Batteries" *Advanced Materials* **2023**, 2210130.
[32] T. T. Beyene, B. A. Jote, Z. T. Wondimkun, B. W. Olbassa, C.-J. Huang, B. Thirumalraj, C.-H. Wang, W.-N. Su, H. Dai, B.-J. Hwang, "Effects of Concentrated Salt and Resting Protocol on Solid Electrolyte Interface Formation for Improved Cycle Stability of Anode-Free Lithium Metal Batteries" *ACS Appl. Mater. Interfaces* **2019**, *11*, 31962–31971.